\documentclass[aps,10pt,prb,bibliography,citeautoscript,superscriptaddress,twocolumn]{revtex4-2}
\usepackage[T1]{fontenc}
\usepackage{graphicx}
\usepackage{dcolumn}
\usepackage{amsmath,amssymb,amsfonts,bm,dsfont,units}
\usepackage{hyperref}
\hypersetup{colorlinks=true,citecolor=blue,linkcolor=blue,urlcolor=blue,pdfstartview=FitH}
\usepackage{braket}
\usepackage{float,latexsym}
\usepackage{multirow}
\usepackage[normalem]{ulem} 
\usepackage[caption=false]{subfig} 

\usepackage{array}
\usepackage{xcolor}
\usepackage{graphicx} 
\usepackage{hhline,booktabs}
\usepackage{makecell}
\usepackage{enumitem}
\usepackage{tabularx}
\newcolumntype{Y}{>{\centering\arraybackslash}X}

\newcommand{\sfigure}[2]{Figure~\hyperref[#1]{\ref{#1}(#2)}}
\newcommand{\sfigref}[2]{Fig.~\hyperref[#1]{\ref{#1}(#2)}}

\usepackage{color}
\definecolor{dkgreen}{rgb}{0,0.5,0}
\definecolor{midnightblue}{rgb}{0.39,0.58,0.93}

\definecolor{kspink}{RGB}{200,0,200}
\definecolor{appendixgreen}{RGB}{9, 145, 58}

\newcommand{\comment}[1]{}{}

\usepackage{soul}

\newcommand{\mc}[1]{\mathcal{#1}}

\begin{document}

\title{Boundary transitions from a single round of measurements on gapless quantum states}

 \author{Yue Liu}
   \affiliation{Department of Physics, California Institute of Technology, Pasadena, CA 91125, USA}
   \affiliation{Institute for Quantum Information and Matter, California Institute of Technology, Pasadena, CA 91125, USA}

    \author{Sara Murciano}
    \affiliation{Department of Physics, California Institute of Technology, Pasadena, CA 91125, USA}
   \affiliation{Institute for Quantum Information and Matter, California Institute of Technology, Pasadena, CA 91125, USA}
    \affiliation{Walter Burke Institute for Theoretical Physics, California Institute of Technology, Pasadena, CA 91125, USA}

  \author{David F. Mross}
 \affiliation{Department of Condensed Matter Physics, Weizmann Institute of Science, Rehovot 7610001, Israel}
      \author{Jason Alicea}
   \affiliation{Department of Physics, California Institute of Technology, Pasadena, CA 91125, USA}
   \affiliation{Institute for Quantum Information and Matter, California Institute of Technology, Pasadena, CA 91125, USA}
  \affiliation{Walter Burke Institute for Theoretical Physics, California Institute of Technology, Pasadena, CA 91125, USA}

\date{\today}

	\begin{abstract}

Measurements can qualitatively alter correlations and entanglement emerging in gapless quantum matter.
We show how a single round of measurements on gapless quantum systems can, upon rotating the measurement basis, induce non-trivial transitions separating regimes displaying universal characteristics governed by distinct boundary conformal field theories. 
We develop the theory of such `measurement-induced boundary transitions' by investigating a gapless parent of the one-dimensional cluster state, obtained by appropriately symmetrizing a commuting projector Hamiltonian for the latter.  Projective measurements on the cluster state are known to convert the wavefunction, after post-selection or decoding, into a long-range-ordered Greenberger-Horne-Zeilinger (GHZ) state.  Similar measurements applied to the gapless parent (i) generate long-range order coexisting with power-law correlations when post-selecting for uniform outcomes, and (ii) yield power-law correlations distinct from those in the pre-measurement state upon decoding.  In the post-selection scenario, rotating the measurement basis preserves long-range order up until a critical tilt angle marking a measurement-induced boundary transition to a power-law-ordered regime.  Such a transition---which does not exist in the descendant cluster state---establishes new connections between measurement effects on many-body states and non-trivial renormalization-group flows.  
We extend our analysis to tricritical Ising and three-state Potts critical theories, which also display measurement-induced boundary transitions, and propose general criteria for their existence in other settings. 

	\end{abstract}
	\maketitle

\section{Introduction}

Quantum mechanics permits two qualitatively different ways of modifying wavefunctions: unitarily through Hamiltonian dynamics and non-unitarily through measurements. 
Local unitaries preserve real-space entanglement, and may change the detailed manifestation of correlations but not their fundamental character.  Conversely, local measurements can dramatically alter both entanglement and correlation functions---e.g., via wavefunction collapse or amplitude restructuring under weak measurements.  Extensive work has capitalized on such nontrivial measurement effects to explore a wide variety of novel many-body phenomena.  On one hand, the interplay between random unitary dynamics and monitoring brought forth the paradigm of measurement-induced phase transitions \cite{Li2018,Skinner2019,Fisher2023}.
On the other, local measurements on \emph{static} many-body systems potentially offer a fruitful new knob for controlling quantum matter, for instance to produce behavior beyond that possible through Hamiltonian engineering alone. 

\begin{figure}[h!]
    \centering
    \includegraphics[width=\linewidth]{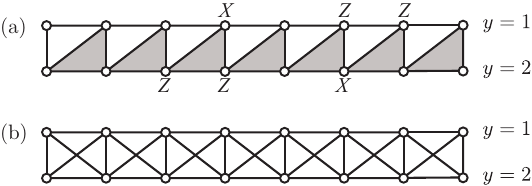}
    \caption{(a) Canonical cluster state model. The Hamiltonian consists of $ZXZ$ terms on all the white and grey triangles, manifestly breaking interchain reflection symmetry. (b) A gapless, reflection-symmetric extension of the model is constructed by completing $ZXZ$ terms on all the triangles.}
    \label{fig:cluster}
\end{figure}

The one-dimensional (1D) cluster state provides a classic setting for nontrivial measurement effects on many-body wavefunctions.  First introduced as a resource for measurement-based quantum computation \cite{Raussendorf01}, the cluster state represents a gapped $\mathbb{Z}_2 \times \mathbb{Z}_2$ symmetry-protected topological phases realizable with an exactly solvable Hamiltonian [Eq.~\eqref{eq:CanonicalCluster}] consisting of commuting `$ZXZ$' three-spin interactions.  For our purposes it proves convenient to view the model as living on a two-chain square ladder---see Fig.~\ref{fig:cluster}---such that each $\mathbb{Z}_2$ symmetry acts on a different chain.  Upon projectively measuring Pauli $X$ operators on one chain and either performing an outcome-dependent decoding unitary or post-selecting for a uniform outcome, the wavefunction converts into a 
Greenberger-Horne-Zeilinger (GHZ) state displaying long-range $Z$ order on the other chain. The resulting GHZ is fragile, however, in the sense that rotating the measurement basis \cite{Lee2022}, or weakening the measurement, immediately destroys the long-range order.  

Quantum critical systems exhibit an arguably even richer interplay with measurements. Gaplessness renders their universal long-distance properties innately sensitive to perturbations---including from weak measurements that reveal arbitrarily small amounts of information \cite{Garratt2023}. 
Weak measurement effects have been investigated in various critical systems including Luttinger liquids \cite{Garratt2023,Sun2023,Ashida2023,tang2024}, transverse-field Ising chains and related minimal models \cite{Yang2023,Weinstein2023,usmeasurementaltered,Sala2024,Paviglianiti2023,patilLud2024}, and $(2+1)$-dimensional quantum critical points such as $O(N)$ models \cite{Lee2023} (see also \cite{Zou2023,leequantum2023,ma2023tmy,garratt2024probe,McGinley2024,McGinleyLett2023,Sala2024} and~\cite{vicari,Lu22,Chan2019,deluca,Li2021,Friedman,Hsieh1,Rajabpour_2016,Rajabpour_2015} for earlier works on the fate of criticality under `imperfections').  In these models, measurements qualitatively alter correlations and entanglement in a manner dependent on the measurement basis, on the outcome, and of course on the underlying critical theory.  More technically, weak measurements can drive a renormalization-group flow to new fixed points exhibiting modified power-laws and entanglement scaling.  

Motivated by these developments, we explore connections between the  fixed points induced by measurements on 1D critical states and their field theory description, specifically as boundary conformal field theories (BCFTs) \cite{Cardy1984}.  A BCFT is a conformal field theory with certain boundary conditions along a line in Euclidean spacetime, imposed for instance by an impurity, a local field, or, as in our problem, measurements.   We pay special attention to cases where different measurement bases can generate \emph{multiple stable fixed points} characterized by distinct boundary conditions and hence distinct physical properties.  For such cases, we show that rotating the measurement basis can yield novel `measurement-induced boundary transitions' that intervene between stable measurement-driven fixed points.  These transitions differ fundamentally from those obtained by combining projective measurements and random unitary dynamics \cite{Li2018,Skinner2019,Fisher2023,Dongheng2024,yoshida2023,nahum2023,bao2020,Zabalo2020,chaoming2020,Lavasani2020xea,turkeshi2023,Turkeshi2020}, since they arise after a single round of measurements on the ground state of a critical system.  Reference~\onlinecite{Lee2022} identified phase transitions tuned by rotating the measurement basis for a $D$-dimensional cluster state with $D \geq 2$.  These transitions also differ from those captured here: we focus on $D = 1$, our measurement-induced boundary transitions are governed by theories distinct from the conformal quantum critical points studied in Ref.~\onlinecite{Lee2022}, and our transitions separate regimes exhibiting power-law correlations on both sides.

We primarily develop the theory of measurement-induced boundary transitions in the context of a gapless parent of the 1D cluster state.  We introduce this gapless state---which we hope will interest readers in its own right---by judiciously symmetrizing three-spin $ZXZ$ terms on the square ladder in a way that yields an inter-chain reflection symmetry.  Remarkably, under a `Kennedy-Tasaki' duality transformation \cite{Kennedy1992,Kennedy1992_2,Li23,Li23_2}, the resulting model exactly maps to two decoupled $\mathcal{X}\mathcal{Y}$ spin chains that realize Luttinger liquids with interactions that can be tuned by adding symmetry-allowed terms.  The descendant gapped cluster-state SPT emerges from the gapless parent upon explicitly breaking inter-chain reflection symmetry.  

The well-understood measurement-induced phenomena for the cluster state provides a useful reference point for investigating the fate of the gapless parent under measurements.  What kind of spin correlations (e.g., GHZ-like vs power-law) emerge from measurement in the latter?  Can one design a decoding protocol to reveal nontrivial measurement-induced correlations without post-selection? How do correlations vary with the choice of measurement basis?  And can this system harbor measurement-induced boundary transitions separating distinct stable fixed points?

We address these questions using a field-theory analysis 
and density matrix renormalization group (DMRG) simulations \cite{white1992}.  We find that upon measuring one chain of the square ladder in the $X$ basis and post-selecting for the most likely outcome, the other chain exhibits GHZ-like 
area-law entanglement and long-range $Z$ order, but coexisting with power-law correlations that reflect gaplessness of the pre-measurement state.  
Upon tilting the measurement basis away from $X$, long-range $Z$ order persists until a critical tilt angle (unlike the cluster-state SPT), after which power-law-decaying $Z$ correlations emerge.  In other words, rotating the measurement basis indeed generates a measurement-induced boundary transition.  We find evidence that the intermediate fixed point exhibits logarithmic entanglement scaling, contrary to the area-law behavior arising in the adjacent stable fixed points.  

Non-trivial measurement effects on the gapless parent also appear without post-selection.  For this purpose, we cannot naively average physical observables over measurement outcomes weighted by Born probabilities, as doing so simply returns  pre-measurement expectation values.  For the case of $X$ measurement, we show that the same decoding protocol used for the cluster state (see also \cite{Lu2023}) outputs a mixed state exhibiting measurement-altered power-law spin correlations with a larger or smaller exponent depending on the Luttinger parameter.  For tilted measurement bases, we show that weighting outcomes instead by the \emph{square} of the Born probabilities captures a measurement-induced boundary transition similar to that arising with post-selection.  

Finally, we provide a broader perspective on measurement-induced boundary transitions by proposing a general criterion for their emergence and studying additional gapless quantum states---in particular, a tricritical Ising spin chain and the three-state Potts model. 
In both cases, we identify a measurement-induced boundary transition characterized by modified power-law correlations.

The paper is organized as follows. Section~\ref{sec:canonical_cluster} reviews the canonical cluster state, while Sec.~\ref{sec:symcluster} introduces our gapless parent.  By exploiting the duality mapping to independent Luttinger liquids, we study uniform weak $X$- or $Z$-basis measurements on the gapless parent state in Sec.~\ref{sec:uniform}.  Section~\ref{sec:decoding} discusses a decoding protocol to obtain non-trivial behavior from $X$-basis measurements after averaging over outcomes.   Tilting the measurement angle in the $XZ$-plane yields non-trivial measurement-induced boundary transitions described in Sec.~\ref{sec:tiltmeas}. Section~\ref{sec:minimal} explores measurement-induced boundary transitions in the tricritical Ising and three-state Potts models, and Sec.~\ref{sec:conclusion} provides a discussion and outlook.  Extensive appendices provide technical details and additional numerical data that complement the main text.

\section{Canonical cluster state review}\label{sec:canonical_cluster}

To establish a baseline, we review known features of the canonical 1D $\mathbb{Z}_2\times \mathbb{Z}_2$ cluster state SPT---including the structure of the ground state \cite{Verresen17} as well as the influence of measurements~\cite{Raussendorf01,Lee2022}. 
As noted in the introduction, we view the model as defined on a square ladder shown in 
Fig.~\ref{fig:cluster}(a), such that the two $\mathbb{Z}_2$ symmetries correspond to flipping spins on one leg or the other.   In the zero-correlation-length limit, the cluster-state Hamiltonian on this geometry reads
\begin{equation}
    H_{\rm cluster} = -\sum_{j=1}^{N-1}(Z_{j,1} X_{j,2} Z_{j+1,1} + Z_{j,2} X_{j+1,1} Z_{j+1,2}),
    \label{eq:CanonicalCluster}
\end{equation}
where $X_{j,y}, Z_{j,y}$ are  Pauli operators acting on the site $j=1,\ldots,N$ in chain $y=1,2$. 
The first and second terms respectively represent three-spin $ZXZ$ interactions on the white and grey triangles in Fig.~\ref{fig:cluster}(a).  Notice that $H_{\rm cluster}$ violates interchain reflection symmetry, since the orientation of these triangles flips upon swapping the legs.  The two $\mathbb{Z}_2$ symmetries protecting SPT order in the ground state are generated by
\begin{equation}
    G_1 = \prod_{j=1}^N X_{j,1} \quad \text{and} \quad G_2 = \prod_{j=1}^{N} X_{j,2}.
    \label{Z2generators}
\end{equation}

Exact solvability of Eq.~\eqref{eq:CanonicalCluster} descends from the fact that all $ZXZ$ terms commute with one another and square to the identity.  Consequently, the ground state satisfies $Z_{j,1} X_{j,2} Z_{j+1,1} = Z_{j,2} X_{j+1,1} Z_{j+1,2}=+1$ for all $j$. Multiplying $ZXZ$ terms on consecutive triangles with the same color in Fig.~\ref{fig:cluster}(a) defines $\mathbb{Z}_2 \times \mathbb{Z}_2$ SPT string order parameters, which are equal to one in the ground state of the above cluster state model.  Taking the product on white triangles, for instance, yields
\begin{equation}\label{eq:sop}
\mathcal{S}_{j,k} \equiv \left[{
    \begin{matrix}
        &X_{j+1} & \cdots & X_{k-1}&X_{k}\\
         Z_j&  & & & Z_{k} \\
    \end{matrix}
    }\right]  = 1
\end{equation}
for any $k>j$.  
On the left-hand side, all operators are multiplied, with the top and bottom rows corresponding to the upper and lower chain of the ladder.
A second string order parameter similarly follows from the grey triangles.  Beyond the zero-correlation-length limit, these string order parameters tend to a non-zero constant at $|j-k| \rightarrow \infty$ in the SPT phase. 

Interestingly, measurements together with decoding convert the zero-correlation-length SPT ground state $\ket{\psi_0}$ into a GHZ state $\ket{\rm GHZ} = \frac{1}{\sqrt{2}}(\ket{\uparrow\cdots\uparrow} + \ket{\downarrow\cdots\downarrow})$.  
Suppose that one projectively measures the local operators $\{X_{j,1}\}$ for all $j$ on the upper chain, and denotes the outcome as $\bm{s} \equiv \{s_{j}\}$ and the corresponding Born probability as $p_{\bm{s}}$. The post-measurement state (up to normalization) is then
\begin{equation}\label{eq:projective}
    \ket{\psi}_{\bm{s}} \propto \mathcal{P}_{\bm{s}} \ket{\psi_0}\equiv\prod_{j=1}^N \left(\frac{1+s_j X_{j,1}}{2}\right) \ket{\psi_0},
\end{equation}
where $\mathcal{P}_{\bm{s}}$ is the projector corresponding to the measurement.  In the state $\ket{\psi}_{\bm{s}}$, two-point $ZZ$ correlators for the unmeasured bottom chain readily evaluate to
\begin{align}
    \langle Z_{j,2} Z_{k,2}\rangle_{\bm{s}} = \frac{\bra{\psi_0} \mathcal{P}_{\bm{s}} Z_{j,2} Z_{k,2} \mathcal{P}_{\bm{s}} \mathcal{S}_{j,k} \ket{\psi_0}}{\braket{\psi_0 |\mathcal{P}_{\bm{s}}|\psi_0}} = s_{j+1}\cdots s_{k}.
    \label{ZZcorr}
\end{align}
[Using Eq.~\eqref{eq:sop}, in the middle, we benignly inserted an $\mathcal{S}_{j,k}$ factor to immediately obtain the result.] 
The above correlator remains non-zero even at $|j-k|\rightarrow \infty$---though the value depends on the measurement outcome.  Correspondingly, performing a standard average over measurement outcomes gives
\begin{equation}
    \sum_{\bm{s}} p_{\bm{s}} \braket{Z_{j,2} Z_{k,2}}_{\bm{s}} = \braket{Z_{j,2} Z_{k,2}} = 0
    \label{trivial_ave}
\end{equation}
for any $k\neq j$.  We see here that Born-rule averaging effectively erases the measurement, yielding a two-point $ZZ$ correlator for the pre-measurement state that naturally becomes trivial in the SPT phase.  

One can nevertheless extract correlations characteristic of a GHZ state in several ways.  By post-selecting for a uniform measurement outcome with all $s_j = +1$, Eq.~\eqref{ZZcorr} returns standard long-range ordered correlations characteristic of the wavefunction $\ket{\rm GHZ}$.  Since the probability for the target measurement outcome decays exponentially with system size, however, this approach is non-ideal---but fortunately unnecessary.  A post-measurement 
state $\ket{\psi}_{\bm{s}}$ associated with an arbitrary outcome $\bm{s}$ can be converted into a GHZ state via controlled unitary feedback: namely, $U_{\bm{s}}\ket{\psi}_{\bm{s}} = \ket{\rm{GHZ}}$ with $U_{\bm{s}} = \prod_j X_{j,2}^{(1-p_j)/2}$ a unitary dependent on the measurement outcome through $p_j = \cdots s_{j-2} s_{j-1} s_{j}$~\cite{Lu2023}.  An alternative, and closely related, perspective is to consider
\begin{equation}
    \sum_{\bm{s}} p_{\bm{s}} \braket{Z_{j,2} Z_{k,2}}_{\bm{s}}s_{j+1}\cdots s_{k} = 1.
    \label{nontrivial_ave}
\end{equation}
The left side averages the $ZZ$ correlator weighted by a measurement-outcome-dependent sign structure that cancels the signs in Eq.~\eqref{ZZcorr}; hence averaging over measurement outcomes returns 1 [instead of 0 as in Eq.~\eqref{trivial_ave}].  Equation~\eqref{nontrivial_ave} is equivalent to what one would obtain by evaluating $ZZ$ in the unitarily modified state $U_{\bm s} \ket{\psi}_{\bm s}$ and then averaging over measurement outcomes.  For a more general discussion of unitary feedback in related settings, see Ref.~\onlinecite{Lu2023}.  

Appendix~\ref{app:moreproperties} discusses additional properties of the cluster state including the effects of finite correlation length, enacting weak vs projective measurements, and tilting the measurement basis away from $X$.  As reviewed there, the emergence of a GHZ state requires both strict projective measurements \emph{and} pristinely measuring $X$ for the lower chain ~\cite{Lee2022}.  Weakening the measurement or tilting the measurement basis---even by arbitrarily small amounts---produces exponentially decaying $ZZ$ correlations in the post-measurement state, rather than long-range-ordered GHZ-like correlations arising in the ideal case.  We will later uncover qualitatively different resilience to such modifications in the context of the gapless parent of the 1D cluster state that we introduce next.

\section{Gapless parent of the cluster state}
\label{sec:symcluster}

\begin{figure}[h]
    \centering
    \includegraphics[width=\linewidth]{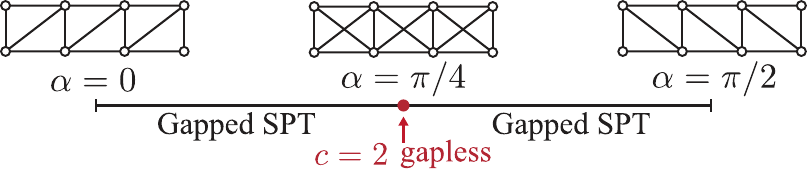}
    \caption{Phase diagram of Eq.~\eqref{eq:SymmetricCluster}. For $0\leq \alpha <\pi/4$ and $\pi/4 < \alpha \leq \pi/2$, two cluster-state SPT regimes emerge that are related by interchain reflection symmetry. At $\alpha=\pi/4$, the system is gapless with central charge $c=2$.}
    \label{fig:cluster_transition}
\end{figure}

We introduce the gapless parent of the cluster state SPT by first considering the generalized Hamiltonian 
\begin{align}
    H = -\sum_{j=1}^{N-1}[\cos\alpha (Z_{j,1} X_{j,2} Z_{j+1,1} &+  Z_{j,2} X_{j+1,1} Z_{j+1,2}) \nonumber \\
    +\sin\alpha(Z_{j,2} X_{j,1} Z_{j+1,2} &+ Z_{j,1} X_{j+1,2} Z_{j+1,1})].
\label{eq:SymmetricCluster}
\end{align}
The $ZXZ$ terms in the first line, which are the same as those in Eq.~\eqref{eq:CanonicalCluster}, correspond to the triangles in Fig.~\ref{fig:cluster}(a).  The second line includes a new set of $ZXZ$ terms on `flipped' triangles.  When $\alpha = 0$, $H$ simply recovers the cluster state SPT reviewed in Sec.~\ref{sec:canonical_cluster}; $\alpha = \pi/2$ corresponds to an alternative cluster state SPT realization that relates to the former by interchain reflection.  At $\alpha = \pi/4$ the Hamiltonian uniquely preserves interchain reflection symmetry [see Fig.~\ref{fig:cluster}(b) for a schematic].  We will show below that this limit realizes a gapless state that intervenes between the two gapped cluster state SPT regimes highlighted above.  Figure~\ref{fig:cluster_transition} illustrates the phase diagram as a function of $\alpha$ for this model.  We stress, however, that $\alpha <\pi/4$ and $\alpha > \pi/4$ do not represent distinct phases.  In fact the cluster state SPT can harmoniously coexist with interchain reflection symmetry as shown explicitly in Ref.~\onlinecite{Liu2023}.  A reflection-symmetric SPT can also emerge as a descendent of our gapless parent state; see Appendix~\ref{app:sptwithreflection} for details.  

\begin{table*}[ht]
    \centering
        \caption{Dictionary between select operators in the original basis and in the new basis specified in Eq.~(\ref{eq:map}).}
    \setlength{\extrarowheight}{2pt}
    \begin{tabular*}{\linewidth}{@{\extracolsep{\fill}} |c|c|c|c|c|c|c|c|c|}
    \hline
         Original &$\bigg[\begin{matrix}
             Z_j& Z_{j+1} \\
             X_{j} &\\
         \end{matrix}\bigg]$& $\bigg[\begin{matrix}
             Z_j& Z_{j+1} \\
             & X_{j+1} \\
         \end{matrix}\bigg]$ & $\bigg[\begin{matrix}
             X_{j} &\\
             Z_j& Z_{j+1} \\
         \end{matrix}\bigg]$ & $\bigg[\begin{matrix}
              & X_{j+1}\\
             Z_j& Z_{j+1} \\
         \end{matrix}\bigg]$ & $\bigg[\begin{matrix}
             Y_j& Z_{j+1} \\
             Y_j& Z_{j+1} \\
         \end{matrix}\bigg]$ & $\bigg[\begin{matrix}
             Z_j& Y_{j+1} \\
             Z_j& Y_{j+1} \\
         \end{matrix}\bigg]$ & $\bigg[\begin{matrix}
             X_j \\
              \\
         \end{matrix}\bigg]$ & $\bigg[\begin{matrix}
              \\
             X_j \\
         \end{matrix}\bigg]$ \\
         \hline
         New & 
         $\bigg[\begin{matrix}
             \mathcal{Y}_j & \mathcal{Y}_{j+1} \\
              & \\
         \end{matrix}\bigg]$ &
         $\bigg[\begin{matrix}
              & \\
              \mathcal{\tilde Y}_j & \mathcal{\tilde Y}_{j+1} \\
         \end{matrix}\bigg]$ &
         $\bigg[\begin{matrix}
             \mathcal{X}_j & \mathcal{X}_{j+1} \\
              & \\
         \end{matrix}\bigg]$ & 
         $\bigg[\begin{matrix}
              & \\
              \mathcal{\tilde X}_j & \mathcal{\tilde X}_{j+1} \\
         \end{matrix}\bigg]$ & 
         $-\bigg[\begin{matrix}
             \mathcal{Z}_j & \mathcal{Z}_{j+1} \\
              & \\
         \end{matrix}\bigg]$ & 
         $-\bigg[\begin{matrix}
              & \\
             \mathcal{\tilde Z}_j & \mathcal{\tilde Z}_{j+1} \\
         \end{matrix}\bigg]$ & 
         $\bigg[\begin{matrix}
              \mathcal{X}_j \\
              \mathcal{\tilde X}_j \\
         \end{matrix}\bigg]$ & 
         $\bigg[\begin{matrix}
              \mathcal{Y}_j \\
              \mathcal{\tilde Y}_j \\
         \end{matrix}\bigg]$ \\
         \hline
    \end{tabular*}
    \label{tab:ZXZtoXY}
\end{table*}

Gaplessness at $\alpha = \pi/4$ becomes manifest under a non-local mapping---equivalent to a Kennedy-Tasaki transformation \cite{Kennedy1992,Kennedy1992_2,Li23,Li23_2}---to Pauli operators,
\begin{equation}\label{eq:map}
\begin{aligned}
    \mathcal{X}_j &= \left[{
    \begin{matrix}
        &\cdots & X_{j-2}  & X_{j-1}  & & \\
        & &  &  & Z_{j} &\\
    \end{matrix}
    }\right],    \\
    \mathcal{\tilde X}_j &= \left[{
    \begin{matrix}
         & \cdots & X_{j-2}  & X_{j-1}  & X_{j} & \\
        &  &  &  & Z_{j} & \\
    \end{matrix}
    }\right], \\
    \mathcal{Y}_j &= \left[{
    \begin{matrix}
        &Z_{j} &  &  &  &  \\
        &X_{j} & X_{j+1} & X_{j+2} & \cdots & \\
    \end{matrix}
    }\right],    \\
    \mathcal{\tilde Y}_j &= \left[{
    \begin{matrix}
        &Z_{j} &  &  &  &  \\
        &         & X_{j+1} & X_{j+2} & \cdots &  \\
    \end{matrix}
    }\right].    \\
\end{aligned}
\end{equation}
[Operators in each bracket are multiplied. The upper (lower) row of each bracket refers to the first (second) chain.]
The structure of this transformation---attaching a string of $X$ operators in one chain to a $Z$ operator in the opposite chain---closely resembles the SPT string order parameters in Eq.~\eqref{eq:sop}. The new variables satisfy the usual Pauli algebra with $\mathcal{Z}_j = -i \mathcal{X}_j \mathcal{Y}_j$ and similarly for $\tilde{\mathcal{Z}}_j$. 
Table~\ref{tab:ZXZtoXY} provides a dictionary between select operators in the original basis and in the new basis above.  In particular, the constituent $ZXZ$ operators from Eq.~\eqref{eq:SymmetricCluster} map to nearest-neighbor bilinears of the new Pauli operators in Eq.~\eqref{eq:map}, yielding (after some rearrangement) 
\begin{align}
    H = -\sum_{j=1}^{N-1}(\sin\alpha \mathcal{X}_j \mathcal{X}_{j+1} &+ \cos\alpha \mathcal{Y}_j \mathcal{Y}_{j+1} \nonumber \\
    +\cos\alpha \tilde{\mathcal{X}}_j \tilde{\mathcal{X}}_{j+1} &+ \sin\alpha\tilde{\mathcal{Y}}_j \tilde{\mathcal{Y}}_{j+1}).
\label{eq:SymmetricClusterNew}
\end{align}
For general $\alpha$, Eq.~\eqref{eq:SymmetricClusterNew} describes two decoupled $\mathcal{X}\mathcal{Y}$ models, each of which is gapped due to anisotropy in the couplings. The limits $\alpha=0$ and $\alpha=\pi/2$ correspond to two decoupled Ising chains in the spontaneous symmetry breaking phase. The Kennedy-Tasaki transformation mapping the cluster state SPT \eqref{eq:CanonicalCluster} to this symmetry-broken phase has been studied in \cite{Li23}.
The anisotropy disappears at $\alpha = \pi/4$---where each $\mathcal{X}\mathcal{Y}$ model manifestly exhibits an exact U(1) symmetry.  Here the system is gapless with central charge $c = 2$ and admits the usual free-fermion representation.  

Adding local, symmetry-preserving terms to Eq.~\eqref{eq:SymmetricCluster} generates $\mathcal{Z} \mathcal{Z}$ terms (see Table~\ref{tab:ZXZtoXY}) that lead to interactions in the fermion picture and turn the system into a two-channel Luttinger liquid with nontrivial Luttinger parameters.  Taking $\alpha = \pi/4$, dropping a grotesque factor of $\sqrt{2}$, and allowing for such interactions leads to the following Hamiltonian that we will study extensively in the remainder of this paper:
\begin{equation}\label{eq:2XXZ}
    \begin{aligned}
        H_\Delta=- \sum_j\bigg( & \mathcal{X}_j \mathcal{X}_{j+1}+\mathcal{Y}_j \mathcal{Y}_{j+1}+\Delta\mathcal{Z}_j \mathcal{Z}_{j+1}\\
+&\tilde{\mathcal{X}}_j \tilde{\mathcal{X}}_{j+1}+\tilde{\mathcal{Y}}_j \tilde{\mathcal{Y}}_{j+1}+\Delta\tilde{\mathcal{Z}}_j \tilde{\mathcal{Z}}_{j+1}\bigg),\\
    \end{aligned}
\end{equation}
which describes two decoupled $\mathcal{X}\mathcal{X}\mathcal{Z}$ models.  Since we are interested in the gapless phase, we restrict our analysis to $\Delta \in [-1,1)$ throughout.
Next, we will study the effect of weak measurements on the gapless parent state realized by Eq.~\eqref{eq:2XXZ}.

\section{Uniform weak measurements on the gapless parent state}\label{sec:uniform}

\subsection{Weak measurement on a Tomonaga-Luttinger liquid}\label{sec:LL}

As a primer let us discuss the impact of weak measurement on a single $\mathcal{XXZ}$ spin chain---described by just the first line of Eq.~\eqref{eq:2XXZ}---mainly following Refs.~\onlinecite{Garratt2023,Sun2023,Ashida2023} but also adding some new insights. 
The $\mathcal{XXZ}$ model realizes a Tomonaga-Luttinger liquid (TLL) governed by the low-energy fixed-point action
\begin{equation}\label{eq:TLLaction}
\begin{aligned}
    S_{\rm TLL} &= \frac{1}{2\pi } \int_{x,\tau}\left[2i \partial_\tau \phi \partial_x \theta+ K^{-1}(\partial_x \phi)^2+K(\partial_x \theta)^2 \right].
\end{aligned}
\end{equation}
Here $\phi(x,\tau),\theta(x,\tau)$  are slowly varying bosonic fields dependent on a coarse-grained position $x$ and imaginary time $\tau$, while $K=\frac{\pi}{2\arccos{\Delta}}$ is the Luttinger parameter determined by the microscopic $\mathcal{ZZ}$ coupling $\Delta$.  In particular, since we consider only $\Delta \in [-1,1)$, we always have $K \geq 1/2$. 
In the low-energy, long-distance limit, the Pauli operators at site $j$ relate to the bosonic fields through \cite{Giamarchi2003} 
\begin{equation}\label{eq:dictionary}
    \begin{aligned}
 \mathcal{X}_j -i\mathcal{Y}_j &\sim  
e^{i \theta}+ i c_1 (-1)^j\left[e^{i(\theta + 2\phi)} -e^{i(\theta-2\phi)}\right] 
 \\
        \mathcal{Z}_j &\sim -\frac{2}{\pi}\partial_x \phi + c_2(-1)^j \cos{(2\phi)},\\
    \end{aligned}
\end{equation}
where $c_1,c_2 \in \mathbb{R}$ are non-universal coefficients.  (In the first line, we do not factor out an overall $e^{i \theta}$ factor to avoid subtle issues regarding commutation of $\theta$ and $\phi$ evaluated at microscopically separated positions.  Symmetries are more readily accounted for in the form above.)
Equation~\eqref{eq:TLLaction} implies equal-time correlators  $\braket{[\phi(x)-\phi(0)]^2} = K \log(x)$ and $\braket{[\theta(x)-\theta(0)]^2} = \frac{1}{K} \log(x)$. Two-point correlators of the lattice Pauli operators follow as 
\begin{align}
    \braket{\mathcal{X}_j \mathcal{X}_0} &= \braket{\mathcal{Y}_j \mathcal{Y}_0} \sim C_0 |j|^{-\frac{1}{2K}}+C_1(-1)^{j} |j|^{-\frac{1}{2K}-2K}, \nonumber\\
    \braket{\mathcal{Z}_j \mathcal{Z}_0} &\sim C_2 |j|^{-2} + C_3 (-1)^{j} |j|^{-2K}\label{eq:premeas}
\end{align}
with $C_i$ some other non-universal constants.

The weak measurements that we consider here and below are a softened version of the standard projective measurements used, e.g., in Eq.~\eqref{eq:projective}.  In physical setups, weak measurements
can be achieved by entangling the many-body wavefunction
to ancillary degrees of freedom and then projectively measuring the latter.  References~\onlinecite{Garratt2023,Sun2023} previously investigated $\mathcal{Z}$-basis measurements in a Luttinger liquid; see especially the seminal study of Ref.~\onlinecite{Garratt2023} that established a connection to the classic Kane-Fisher impurity problem~\cite{Kane1992}.  We will consider $\mathcal{X}$-basis weak measurements, as they illustrate the main points in a way that closely connects to measurements of interest in our gapless parent of the cluster state. 

Upon weakly measuring all sites, the ground state $\ket{\psi_c}$ of the $\mathcal{XXZ}$ spin chain becomes 
\begin{equation}\label{eq:post_meas}
    \ket{\psi_{\bm{s}}}=\frac{1}{\mathcal{N}}e^{\beta\sum_js_j\mathcal{X}_j}\ket{\psi_c}
\end{equation}
where $\bm{s}=\{s_j=\pm 1\}$ encodes the set of measurement outcomes (e.g., in a scheme utilizing ancillas), $\mathcal{N}$ is an overall normalization factor, and $\beta$ sets the measurement strength~\cite{Sun2023,Yang2023,Weinstein2023,Lee2023}.  In particular, when acting on some initial state, the on-site operator $e^{\beta s_j\mathcal{X}_j}$ amplifies the contribution with $\mathcal{X}_j = s_j$ and suppresses the contribution with $\mathcal{X}_j = -s_j$; $\beta\to \infty$ recovers the projective limit where the latter is killed.

In the following, we post-select for the uniform outcome where $s_j=1$ for all $j$.  Using Eq.~\eqref{eq:post_meas} and the bosonized expression in Eq.~\eqref{eq:dictionary}, such a post-selected weak measurement manifests as a boundary term acting at all spatial points but only at imaginary time $\tau=0$:
\begin{equation}\label{eq:action_imp}
    S' = S_{\rm TLL} + \beta b_1\int_{x,\tau} \delta(\tau)\cos{\theta}, 
\end{equation}
where $b_1$ is a non-universal factor. 
We observe that the scaling dimension of $\cos{\theta}$ is $\Delta_{\cos\theta}=1/(4K)$---implying that the measurement-induced boundary perturbation is marginal for $K=1/4$, relevant for larger $K$ and irrelevant for smaller $K$. Weak measurements do not qualitatively impact universal long-distance properties in the latter case. 

In our setup, the Luttinger parameter is constrained by $K\geq 1/2$, and the $\tau=0$ perturbation is always relevant. Even arbitrarily weak measurements will therefore eventually pin $\theta$. Specifically, the flow of $\beta$ implies a length scale $L_\beta(\tau) \sim\ell_\tau\beta^{-1/\Delta_{\cos\theta}}$ with $\ell_0$ the UV cutoff and 
$\ell_\tau \approx \tau$ at large $\tau$.  
Equal-time correlations at distances at $x \gtrsim L_\beta(\tau)$ feel the full impact of the measurement while those at $x \lesssim L_\beta(\tau)$ are only affected perturbatively. In this paper we are always interested in correlations of microscopic operators that are \emph{not} time-evolved, which in the bosonized theory follow from correlations of fields evaluated at $\tau = 0$.  For such cases the crossover length $L_\beta(0)$ depends only on $\beta$ (and a UV cutoff).  

We use BCFT formalism to efficiently compute various correlation functions. Specifically, we encode pinning of $\theta$ in the IR by a Dirichlet boundary condition (DBC) $\theta(\tau^*)=0$ at all $x$ and at a $\beta$-dependent time $\tau^*$. The dual field, $\phi$, consequently obeys free or Neumann boundary conditions (NBc), i.e., $\partial_\tau\phi |_{\tau=\tau^*}= 0$. The BCFT exhibits a crossover scale $L^\text{BCFT}_{\tau^*}(\tau)$ which depends on the distance $\tau-\tau^*$, and in the limit $\tau-\tau^*\rightarrow 0$ is set by a UV cutoff. 
To determine $\tau^*(\beta)$, we match the crossover length scales of Eq.~\eqref{eq:post_meas} and the BCFT at $\tau=0$, i.e., we require $L^\text{BCFT}_{\tau^*}= L_\beta(0)$. In particular, the correlations of Eq.~\eqref{eq:post_meas} for a given $\beta$ correspond to BCFT correlations evaluated at a specific distance from the boundary.  

Combining relevance of the measurement-induced perturbation and the flow to a BCFT with DBC for $\theta$, Appendix~\ref{app:correlators} evaluates two-point correlation functions $\braket{e^{ia\theta(0,\tau)} e^{-ib\theta(x,\tau)}}_{\rm DBC}$ and $\braket{e^{ia\phi(0,\tau)} e^{-ib\phi(x,\tau)}}_{\rm NBC}$ for general $x,\tau$.  The dictionary in Eq.~\eqref{eq:dictionary} allows us to leverage this computation to evaluate two-point correlations of microscopic spin operators at spatial separation $x$ but at the same imaginary time $\tau$.  
To extract the characteristic scale $\tau^*$, we need to also examine two-point spin correlations in an alternative way that directly incorporates the measurement strength $\beta$.  We will do so by expanding around the projective measurement limit and  using scaling arguments.  

Consider first nearly projective measurements ($\beta \gg 1$); here correlations are expected to be governed by a scale $\tau^*$ close to 0. 
By exploiting the results reviewed in Appendix~\ref{app:correlators}, in this regime we obtain up to $O(\tau^{*4}/x^4)$
\begin{align}\label{eq:bcft_tt}
    &\frac{\braket{e^{ia\theta(0,0)} e^{-ib\theta(x,0)}}_{\rm DBC}}{\braket{e^{ia\theta(0,0)}}_{\rm DBC} \braket{e^{-ib\theta(x,0)}}_{\rm DBC}} = 1+\frac{a b}{K} \left(\frac{\tau^*}{x}\right)^2 
\end{align}
at long distances $x \gg \tau^*$.
Using Eq.~\eqref{eq:dictionary}, two-point $\mathcal{Y}$ correlators ---which suffice for the present aims---are then given in the same regime by
\begin{equation}\label{eq:YY2}
\frac{\braket{\mathcal{Y}_0 \mathcal{Y}_x}_{\rm uni}}{\braket{\mathcal{X}_0 \mathcal{X}_x}_{\rm uni}}\sim \frac{\braket{\sin[\theta(0)] \sin[\theta(x)]}_{\rm DBC}}{\braket{\cos[\theta(0)] \cos[\theta(x)]}_{\rm DBC}}\sim \left(\frac{\tau^*}{x}\right)^2.
\end{equation}
Above we normalized the two-point correlator $\braket{\mathcal{Y}_0 \mathcal{Y}_x}_{\rm uni}$ by $\braket{\mathcal{X}_0 \mathcal{X}_x}_{\rm uni}$ to eliminate dependence on the UV cutoff of the field theory.
Appendix~\ref{app:pert} alternatively finds 
\begin{equation}\label{eq:YY1}
\frac{\braket{\mathcal{Y}_0 \mathcal{Y}_x}_{\rm uni}}{\braket{\mathcal{X}_0 \mathcal{X}_x}_{\rm uni}}\sim \left(\frac{e^{-2\beta} }{x}\right)^2
\end{equation}   
by examining the ground-state wavefunction near the projective-measurement limit.  Comparing the preceding two equations yields an exponentially small characteristic scale $\tau^*\sim e^{-2\beta}$ in the $\beta \gg 1$ limit.  

For $\beta \ll 1$, we do not know how to explicitly compute the $\beta$ dependence of the asymptotic long-distance correlations---but can nevertheless proceed using scaling arguments.   In particular, prior to performing the weak measurement, 
the scaling dimension of $\beta$ in Eq.~\eqref{eq:action_imp} is $[\beta] = -1 + \frac{1}{4K}$, such that the overall action is dimensionless. Dimensional analysis then gives $\beta\sim (\tau^*)^{-1+1/(4K)}$, and hence long-distance correlations at $x \gg \tau^*$ should be obtained with $\tau^* \sim \beta^{4K/(1-4K)}$ in the $\beta \ll 1$ limit.

\subsection{$X$-basis measurements}\label{sub:xbasis}
Analyzing the effects of measurements on a single-channel Luttinger liquid serves as a useful warm-up to study their consequences in the gapless parent of the cluster state.  Let 
\begin{equation}
    \ket{\psi_\Delta} = |\psi_c\rangle|\tilde{\psi}_c\rangle
\end{equation} 
denote the ground state of Eq.~\eqref{eq:2XXZ}; on the right side, $\ket{\psi_c}$ and $\ket{\tilde\psi_c}$ denote the ground states of the two decoupled $\mathcal{XXZ}$ spin chains that arise under the Kennedy-Tasaki mapping from Eq.~\eqref{eq:map}.  Weakly measuring all $X_{j,1} = \mathcal{X}_j \mathcal{\tilde X}_j$ operators from the upper chain and post-selecting for the uniform outcome $\boldsymbol{s}=\{ s_j=+1\}$ modifies the wavefunction to $\ket{\psi}_{\rm uni} = \frac{1}{\mathcal{N}} M_X \ket{\psi_\Delta}$, where 
\begin{align}
    M_X = e^{\beta \sum_j X_{j,1}} = e^{\beta \sum_j \mathcal{X}_j \mathcal{\tilde X}_j}
    \label{modified_psi}
\end{align}
is the associated non-unitary measurement operator.

Following the previous subsection, long-distance properties of the weakly measured wavefunction can be extracted from the action
\begin{equation}\label{eq:action_2chain}
    \begin{aligned}
        S = S_{\rm TLL}[\theta,\phi] + S_{\rm TLL}[\tilde \theta,\tilde \phi] + \delta S_{\rm meas}[\theta, \tilde\theta].
    \end{aligned}
\end{equation}
Here $S_{\rm TLL}$ is given in Eq.~\eqref{eq:TLLaction} and 
\begin{equation}\label{eq:pert}
    \delta S_{\rm meas} \propto \beta \int_{x,\tau} \delta(\tau) \cos\theta \cos\tilde \theta
\end{equation}
encodes the measurement-induced perturbation that couples the Luttinger liquids at all $x$ but only at $\tau = 0$.  
It is useful to introduce symmetric and antisymmetric combinations of the bosonic fields via 
$\theta_{\pm} \equiv \theta \pm \tilde \theta$ and $\phi_\pm = (\phi \pm \tilde \phi)/2$. 
\cite{Ashida2023}. In this basis the problem maps onto two independent Luttinger liquids for the $+$ and $-$ fields that remain decoupled for $\beta \neq 0$; indeed Eq.~\eqref{eq:pert} becomes 
\begin{equation}\label{eq:pert2}
    \delta S_{\rm meas} \propto \beta \int_{x,\tau} \delta(\tau) (\cos\theta_+ + \cos\theta_-).
\end{equation}
The scaling dimension of $\cos\theta_\pm$ (and equivalently of $\cos\theta \cos\tilde\theta$) is $1/(2K)$. Consequently, for $K>1/2$ any nonzero measurement strength generates a relevant perturbation that, at long distances, imposes DBC's that pin $\theta_+$ and $\theta_-$ as described earlier. 
We assume this regime in what follows. Given the decoupling between the $\theta_\pm$ sectors, we can use boundary CFT calculations just like those in the previous section to extract correlation functions in the post-measurement wavefunction. 

Inspecting the right side of Eq.~\eqref{modified_psi}, the post-measurement wavefunction clearly breaks the two U(1) symmetries associated with each $\mathcal{XXZ}$ chain.  A subgroup that sends $\mathcal{X}_j \rightarrow - \mathcal{X}_j$ \emph{and} $\mathcal{\tilde X}_j \rightarrow - \mathcal{\tilde X}_j$ is, however, preserved---implying that $\langle \mathcal{X}_j \rangle = \langle \tilde{\mathcal{X}}_j \rangle = 0$.   In the bosonized description, one can account for the vanishing of these expectation values by observing that Eq.~\eqref{eq:pert} exhibits two possible DBC's: either $\theta=\tilde{\theta}=0$ or $\theta=\tilde{\theta}=\pi$~\cite{Ashida2023}. Democratically sampling both boundary conditions leads to 
$\langle\cos\theta\rangle=\langle\cos\tilde{\theta}\rangle=0$ as dictated by symmetry.  The measurement nevertheless catalyzes long-range order in both $\mathcal{X}_j$ and $\tilde{\mathcal{X}}_j$. 
For instance, for $\mathcal{X}$ we find that
\begin{equation}\label{eq:2XY_xx}
    \begin{aligned}
        &\langle\mathcal{X}_j \mathcal{X}_k\rangle_{\bm{s}} \\&\sim \left\langle \cos\left(\frac{\theta_{+}(j)+\theta_{-}(j)}{2} \right) \cos\left(\frac{\theta_{+}(k)+\theta_{-}(k)}{2} \right) \right\rangle_{\rm DBC}
    \end{aligned}
\end{equation}
tends to a non-zero constant at $|j-k| \rightarrow \infty$ due to pinning of $\theta_\pm$.  

\begin{figure}[ht]
    \centering
    \includegraphics[width=\linewidth]{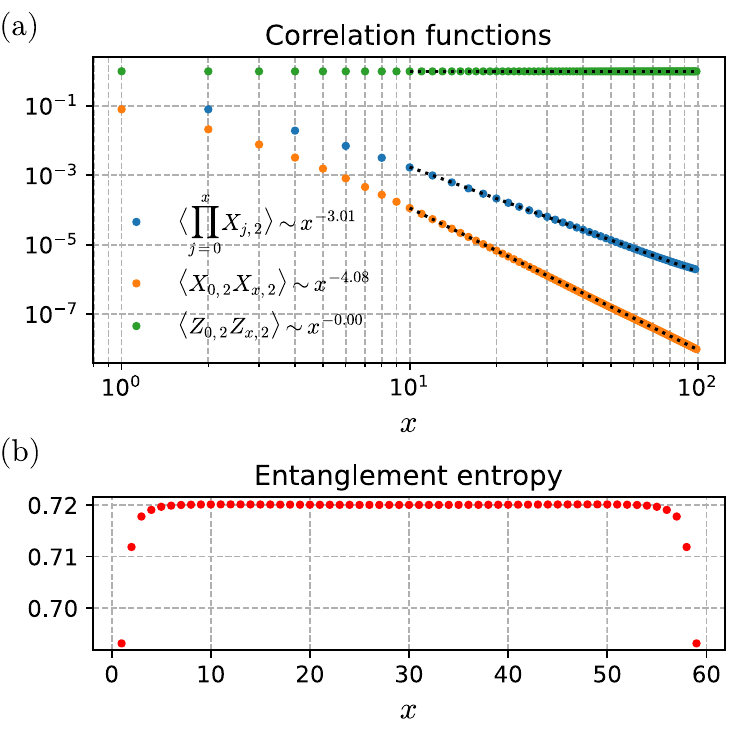}
    \caption{(a) Correlation functions and (b) entanglement entropy of the gapless parent of the cluster state with $K = 1.5$ after projective measurement of $X$ in the upper chain with post-selection for a uniform outcome. 
 The gapless parent state is obtained with bond dimension $\chi = 2000$ via iDMRG in (a) and via finite DMRG with $L = 60$ and periodic boundary conditions in (b).  Fitted power-laws in (a)---see black dotted lines and the legend---agree well with analytical predictions, e.g., from Eqs.~\eqref{eq:2XYcorrelations_1} and \eqref{eq:prod_X}. 
 In (b) we show the entanglement entropy of the post-measurement wavefunction between subsystems $A = [0,x]$ and $B = [x+1, L]$, supporting area-law behavior. 
 For both (a) and (b), we have numerically checked that the results for finite measurement strength $\beta = 0.3, 0.5, 1$ are similar to the projective limit.}
    \label{fig:K=1.5}
\end{figure}

Consider next the two-point $Z$ correlator
\begin{equation}\label{eq:ZZ}
    \begin{aligned}
        \braket{{Z}_{j,2} {Z}_{k,2}}_{\rm uni} &= \left\langle{
        \begin{matrix}
             & \mathcal{X}_{j+1} & \mathcal{X}_{j+2} & \cdots& \mathcal{X}_{k} \\
            \mathcal{\tilde X}_j&\mathcal{\tilde X}_{j+1} &\cdots&\mathcal{\tilde X}_{k-1}&\\
        \end{matrix}       }\right\rangle_{\rm uni} . 
    \end{aligned}
\end{equation}
In the projective measurement limit $\beta \rightarrow \infty$, we can use the post-selected measurement result $X_{j,1} = \mathcal{X}_j\tilde{\mathcal{X}}_j=1$ for all sites to write
\begin{equation}\label{eq:ZZ2}
    \begin{aligned}
        \braket{{Z}_{j,2} {Z}_{k,2}}_{\rm uni} &= \braket{\mathcal{X}_j \mathcal{X}_k}_{\rm uni},
    \end{aligned}
\end{equation}
i.e., in this limit long-range $\mathcal{X}$ order implies long-range $Z$ order for the bottom chain.  DMRG simulations presented in Fig.~\ref{fig:K=1.5}(a) (green points) confirm this prediction.  Importantly, even though Eq.~\eqref{eq:ZZ2} is valid only in the projective-measurement limit, our DMRG results (not shown) reveal that long-range $Z$ order persists for any finite $\beta$.  
According to Eq.~\eqref{eq:map}, long-range order in $\mathcal{X}$ \emph{also} implies a non-zero expectation value for two $Z$ operators on the unmeasured bottom chain linked by a non-local product of $X$ operators on the weakly measured top chain.  
We therefore conclude that the weak $X$ measurement catalyzes long-range $Z$ order in the bottom chain \emph{and} long-range order in the disorder operator $\mu_{j,1} = \prod_{k\leq j} X_{k,1}$ for the top chain.  For reference, both quantities exhibit only power-law order in the pre-measurement state.  Recall also that in the gapped descendant cluster-state SPT, long-range $Z$ order emerges only with strict projective measurements.  

Using Table~\ref{tab:ZXZtoXY} as well as calculations detailed in Appendix~\ref{app:correlators}, we can further evaluate $X$ correlations in the unmeasured bottom chain:
\begin{equation}\label{eq:2XYcorrelations_1}
    \braket{{X}_{j,2} {X}_{k,2}}_{\rm uni} = \braket{\mathcal{Y}_j \mathcal{\tilde Y}_j \mathcal{Y}_k \mathcal{\tilde Y}_k }_{\rm uni} \sim |j-k|^{- \min{(4, 4K)}}
\end{equation}
for any finite measurement strength $\beta$. (Obtaining the above result requires incorporating irrelevant tunneling processes between saddle points with $\theta = \tilde \theta= 0$ and $\pi$ into the boundary CFT analysis.) See \sfigref{fig:K=1.5}{a} for numerical support of Eq.~\eqref{eq:2XYcorrelations_1} at $\beta = \infty$. 
We are also interested in the behavior of the string operator $\prod_{i=j}^k X_{i,2}$. Our DMRG simulations reported in Appendix~\ref{app:2K} and \sfigref{fig:K=1.5}{a} suggest that, in the projective-measurement limit, the two-point correlations of the  disorder operator $\mu_{j,2} = \prod_{k\leq j} X_{k,2}$ are
\begin{equation}\label{eq:prod_X}
\langle \mu_{j,2}\mu_{k,2}\rangle \sim |j-k|^{-2K}.
\end{equation}
Correlations that decay to zero with separation are natural given the long-range $Z$ correlations identified above. The particular \emph{power-law} decay, however, evident in Eq.~\eqref{eq:prod_X} is nontrivial and reflects the gaplessness of the pre-measurement state. 
We can analytically recover this scaling relation by first using the post-selected projective measurement outcome to write 
\begin{equation}
    X_{j,2} = \mathcal{Y}_j \mathcal{\tilde Y}_j = -\mathcal{X}_j \mathcal{\tilde X}_j  \mathcal{Z}_j \mathcal{\tilde Z}_j \xrightarrow[]{\mathcal{X}_j \mathcal{\tilde X}_j = 1} e^{i \frac{\pi}{2}(\mathcal{Z}_j + \mathcal{\tilde Z}_j)}.
\end{equation}
Further applying the bosonization dictionary from Eq.~\eqref{eq:dictionary} yields 
$\langle\prod_{i=j}^k X_{i,2}\rangle_{\rm uni} \sim  \braket{e^{-i \int_{j}^{k} dx (\partial \phi + \partial \tilde{\phi})}}_{\rm NBC} \sim \langle e^{i [\phi(j)-\phi(k)]} e^{i [\tilde\phi(j)-\tilde\phi(k)]} \rangle_{\rm NBC}$.  Techniques from Appendix~\ref{app:correlators} then recover Eq.~\eqref{eq:prod_X}. For the weak measurement case with finite $\beta$, it is natural to expect the string operator to continue displaying power-law correlations, presumably with the same exponent.

Another important tool for characterizing many-body quantum systems is the entanglement entropy. Given a pure state, $\ket{\psi}$, and a bipartite system $A\cup B$, the crucial object to compute is the reduced density matrix $\rho_A=\mathrm{Tr}_B\ket{\psi}\bra{\psi}$. The entanglement entropy between $A$ and $B$ is given by $S=-\mathrm{Tr}(\rho_A\log \rho_A)$.
In \sfigref{fig:K=1.5}{b}, we present numerical evidence that the post-measurement state exhibits area-law entanglement, 
as expected in the presence of a relevant perturbation~\cite{Sun2023,Weinstein2023,Yang2023}. Weak $X$ measurements therefore generate a state exhibiting a curious coexistence of long-range order, power-law correlations, and area-law entanglement.

\subsection{$Z$-basis measurement}\label{sub:zbasis}

Next we explore $Z$-basis weak measurements.  Individual $Z$ operators map to non-local combinations of the operators defined in Eq.~\eqref{eq:map},
\begin{equation}\label{eq:Zj1toY}
    Z_{j,1} = \left[{
    \begin{matrix}
         & \mc{Y}_{j+1} & \mc{Y}_{j+2}  & \cdots \\
        \mc{\tilde Y}_j& \mc{\tilde Y}_{j+1} &\mc{\tilde Y}_{j+2}  & \cdots \\
    \end{matrix}
    }\right],
\end{equation}
which makes it difficult to apply bosonization techniques.
Therefore, it proves enlightening to first examine measurement of nearest-neighbor products $Z_j Z_{j+1}$---which according to Table~\ref{tab:ZXZtoXY} simply map to the local operators $\mathcal{\tilde{Y}}_j \mathcal{Y}_{j+1}$.  Once again post-selecting for a uniform measurement outcome, we specifically consider the post-measurement wavefunction $\ket{\psi}_{\rm uni} = \frac{1}{\mathcal{N}}M_{ZZ}\ket{\psi_{\Delta}}$ with 
\begin{equation}\label{eq:ZjZj_meas}
    M_{ZZ} = e^{\beta\sum_j Z_{j,1} Z_{j+1,1}} = e^{\beta\sum_j \mathcal{\tilde{Y}}_j \mathcal{Y}_{j+1}}.  
\end{equation}
In the $\beta \rightarrow \infty$ limit, the above non-unitary operator projects onto configurations for which the spins in the measured top chain point either all `up' or all `down'.  Projectively measuring single-body $Z$ operators (which we examine later) manifestly breaks the spin-flip symmetry and selects one of those two polarizations.  It is thus plausible that the two types of measurements yield similar consequences---as we will indeed confirm.  

Using Eq.~\eqref{eq:dictionary}, weak measurements encoded by Eq.~\eqref{eq:ZjZj_meas} perturb the Luttinger liquid action via \begin{equation}\label{eq:pertZZ}
    \delta S_{\rm meas}\propto \beta \int_{x,\tau} \delta(\tau)\sin\theta\sin\tilde{\theta},
\end{equation}
which has a very similar form to Eq.~\eqref{eq:pert} and is also relevant for $K>1/2$ (as usual, assumed hereafter).  The microscopic Hamiltonian in Eq.~\eqref{eq:SymmetricCluster} is invariant under the duality transformation 
\begin{equation}\label{eq:duality}
    X_{j,y} \to Z'_{j,y} Z'_{j+1,y}, \quad Z_{j-1,y} Z_{j,y} \to X'_{j,y}.
\end{equation}
Self-duality also persists when $\Delta \neq 0$ in Eq.~\eqref{eq:2XXZ}.
Consequently, measuring $Z_{j,1}Z_{j+1,1}$ is dual to measuring $X_{j,1}$, allowing us to adapt results for $X$-basis  measurements in Sec.~\ref{sub:xbasis} to the present case.  In particular, we immediately deduce the following properties of the post-measurement state in Eq.~\eqref{eq:ZjZj_meas}:
\begin{enumerate}[label=(\roman*)]
    \item The measured top chain exhibits long-range $Z$ order.
    \item The unmeasured bottom chain exhibits long-range order in the disorder operator $\mu_{j,2} = \prod_{k \leq j} X_{k,2}$.
    \item The bottom chain  exhibits power-law $Z$ correlations,
    \begin{equation}\label{eq:measZ_corr1}
        \langle Z_{j,2} Z_{k,2}\rangle_{\rm uni}\sim |j-k|^{-2K}
    \end{equation}
and
    \begin{equation}\label{eq:measZ_corr2}
        \langle Z_{j,2} Z_{j+1,2} Z_{k,2} Z_{k+1,2}\rangle_{\rm uni} \sim |j-k|^{-\min(4,4K)}.
    \end{equation}
    \item The post-measurement state exhibits area-law entanglement. 
\end{enumerate}

Now we turn to single-$Z$ weak measurements and consider the state 
\begin{equation}\label{eq:Zj_meas}
    \ket{\psi}_{\rm{uni}} = \frac{1}{\mathcal{N}}e^{\beta\sum_jZ_{j,1} }\ket{\psi_{\Delta}}.
\end{equation}
Despite the non-locality of the right side of Eq.~\eqref{eq:Zj1toY}, we can make progress via numerics, analytics in limiting cases, and intuition from our results for $ZZ$ measurements [i.e., Eq.~\eqref{eq:ZjZj_meas}]. Appendix~\ref{app:Z1scalingdim} reports DMRG simulations indicating that the scaling dimension of $Z_{j,1}$ is
\begin{equation} \label{eq:dim_Z}
    [Z_{j,1}] = \frac{1}{4},
\end{equation}
for any $K \geq 1/2$; there we also provide an analytical derivation in the two special cases $K=1/2$ and $1$.  Consequently, single-$Z$ weak measurement in the uniform post-selection sector always comprises a strongly relevant perturbation in the allowed range of $K$ for our setup.  

\begin{figure}[h]
    \centering
    \includegraphics[width=\linewidth]{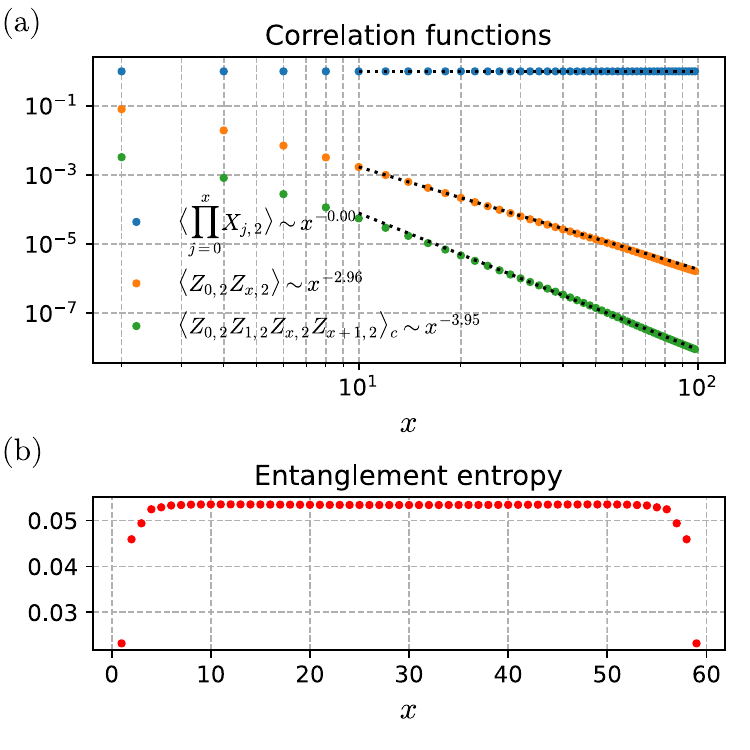}
    \caption{Same as Fig.~\ref{fig:K=1.5}, but after projective measurement of $Z$ in the upper chain. The fitted power-laws in (a) agree well with analytical predictions in Eqs.~\eqref{eq:measZ_corr1} and \eqref{eq:measZ_corr2}. For both (a) and (b), we have numerically checked that the results for finite measurement strength $\beta = 0.3, 0.5, 1$ are similar to the projective limit.}
    \label{fig:Zmeas}
\end{figure}

As noted earlier, it is reasonable to anticipate related properties emerging at the fixed points generated by single-$Z$ versus $ZZ$ measurements, given the similar spin configurations promoted in the two cases (e.g., $\ket{\uparrow\dots\uparrow}$ in the former and $\{\ket{\uparrow\dots\uparrow}, \ket{\downarrow\dots\downarrow}\}$ in the latter). Explicit symmetry breaking by the single-$Z$ measurement clearly induces $\langle Z_{j,1} \rangle \neq 0$---implying long-range $Z$ correlations in the measured top chain as in property (\romannumeral 1)~enumerated for the $ZZ$ case.  
In the projective measurement limit, we can establish a stronger result.  There, the post-$Z$-measurement state is $\ket{\psi^{\text{post-}Z}} = \mathcal{P}_{\uparrow\cdots\uparrow} \ket{\psi_\Delta}$, while the post-$ZZ$-measurement wavefunction can be written as $\ket{\psi^{\text{post-}ZZ}} = \frac{1}{\sqrt{2}} (\mathcal{P}_{\uparrow\cdots\uparrow} \ket{\psi_\Delta} + \mathcal{P}_{\downarrow\cdots\downarrow} \ket{\psi_\Delta} )$. Notice that $\mathcal{P}_{\uparrow\cdots\uparrow} \ket{\psi_\Delta}$ and $\mathcal{P}_{\downarrow\cdots\downarrow} \ket{\psi_\Delta}$ represent the same wavefunction for the lower-chain degrees of freedom, because neither the pre-measurement state $\ket{\psi_\Delta}$ nor the $ZZ$ measurement breaks the $\mathbb{Z}_2$ spin-flip symmetry in the upper chain. Thus the expectation value of any operator $\mathcal{O}_2$ in the lower chain is the same for the two post-measurement wavefunctions, $\braket{\psi^{\text{post-}Z} | \mathcal{O}_2 | \psi^{\text{post-}Z}} = \braket{\psi^{\text{post-}ZZ} | \mathcal{O}_2 | \psi^{\text{post-}ZZ}}$.
Figure~\ref{fig:Zmeas}(a) verifies for $K=1.5$ that properties (\romannumeral 2) and (\romannumeral 3) indeed hold under projective single-$Z$ measurement---including with the same exponents. 
Finally, Fig.~\ref{fig:Zmeas}(b) confirms property (\romannumeral 4) for the single-$Z$ case, i.e., area-law entanglement of the post-measurement state. We have also verified the same properties for $K=1/2$ and $1$ (not shown). Since  $Z$ measurement induces a relevant perturbation, we expect (\romannumeral 1) through (\romannumeral 4) to persist also in the weak measurement regime. 

\section{Decoding protocol for the gapless parent of the cluster state}\label{sec:decoding}

For the canonical cluster state with partial projective measurements, we reviewed in Sec.~\ref{sec:canonical_cluster} how combining classical information with quantum correlations reveals GHZ correlations without resorting to post-selection; recall in particular Eq.~\eqref{nontrivial_ave}.  In this section we show that a similar protocol applied to the gapless parent of the cluster state reveals nontrivially measurement-altered power-law correlations.

Specifically, we take the gapless parent wavefunction $\ket{\psi_{\Delta}}$ as the initial state and projectively measure $\{X_{j,1}\}$ on the top chain. The measurement outcome  
$\boldsymbol{s}=\{s_j\}$ and  correlations $\braket{{Z}_{j,2} {Z}_{k,2}}_{\boldsymbol{s}}$ on the bottom chain of the post-measurement state are recorded.
Just as for the canonical cluster state in Eq.~\eqref{nontrivial_ave}, we define the \textit{decoded} $ZZ$ correlator as the following average over all measurement outcomes weighted by the Born probability $p_{\boldsymbol{s}}$:
\begin{equation}
    \braket{Z_{j,2}Z_{k,2}}_{\rm d}\equiv\sum_{\bm{s}} p_{\bm{s}} \braket{{Z}_{j,2} {Z}_{k,2}}_{\boldsymbol{s}} s_{j+1} s_{j+2} \cdots s_{k}.
    \label{ZZd}
\end{equation}
Contrary to the cluster state, however, the right-hand side does not simply evaluate to 1. 
 Upon transforming to the $\mathcal{XY}$ basis using Eq.~\eqref{eq:map}, this average instead reduces to the 2-point $\mathcal{XX}$ correlator of the initial \emph{pre-measurement} state,
\begin{equation}\label{eq:decoding3}
    \braket{Z_{j,2}Z_{k,2}}_{\rm d} = \braket{\mathcal{X}_j \mathcal{X}_k}.
\end{equation}
One can efficiently obtain this rewriting by bringing the $s_j$ factors inside of the expectation value and replacing them with $X_{j,1}$ operators. Note that this step relies only an operator identity and is thus independent of the particular initial state under consideration.
For the case of our gapless parent state, Eq.~\eqref{eq:premeas} yields decoded power-law correlations
\begin{equation}\label{eq:ZZdecoded}
    \braket{Z_{j,2}Z_{k,2}}_{\rm d} \sim |j-k|^{-\frac{1}{2K}}.
\end{equation}

As an illuminating reference, $ZZ$ correlations in the pre-measurement gapless parent state take a very different form compared to Eq.~\eqref{eq:decoding3}: 
\begin{equation}\label{eq:ZZinitial}
    \braket{Z_{j,2}Z_{k,2}} = 
    \left[{
    \begin{matrix}
         & \mc{X}_{j+1} & \mc{X}_{j+2}  & \cdots & \mc{X}_{k}\\
        \mc{\tilde X}_j& \mc{\tilde X}_{j+1} &\mc{\tilde X}_{j+2}  & \cdots & \\
    \end{matrix}
    }\right]\sim  |j-k|^{-\frac{1}{2}}.
\end{equation}
The power-law on the right follows from the discussion after Eq.~\eqref{eq:dim_Z} and in Appendix \ref{app:Z1scalingdim}.  At $K = 1$ the decoded and unmeasured correlators coincidentally exhibit the same exponent, though at $K \neq 1$ the former are nontrivially measurement altered as claimed.

\section{Intermediate fixed points from tilted measurements}
\label{sec:tiltmeas}

Section \ref{sub:xbasis} showed that weakly measuring $X_{j,1}$ in the gapless parent of the cluster state and post-selecting the uniform outcome $X_{j,1} = +1$ yields long-range order in the second chain; i.e., the two-point function $\braket{Z_{j,2}Z_{k,2}}_{\rm uni}$ tends to a nonzero constant as $|j-k|\rightarrow \infty$. This result reflects the fact that weak $X_{j,1}$ measurements generate a relevant perturbation to the Luttinger liquid fixed point in this post-selection sector, revealing distinct behavior from the gapped cluster state SPT. 
Indeed in the gapped SPT, long-range $Z$ order arises only when projectively measuring $X_{j,1}$ (recall Appendix~\ref{app:weakandtilted}).  This section investigates whether measurement-induced long-range order in the gapless parent state persists upon rotating the measurement basis away from $X$ toward $Z$.  Here too we will uncover richer behavior than for the gapped SPT, where long-range $Z$ order induced by projective $X$ measurements immediately disappears upon tilting the measuring basis \cite{Lee2022} (see also Appendix~\ref{app:weakandtilted}).  

One can partially address the impact of tilting the measuring basis by studying the stability of fixed points driven by $X$- or $Z$-type weak measurements.  In particular, if both fixed points are stable, then $(i)$ long-range $Z$ order induced by $X$ measurements persists over a finite tilt-angle regime and $(ii)$ an intermediate (unstable) fixed point naturally occurs at some nontrivial critical tilt angle. Therefore, this sets up the stage for exploring possible measurement-induced boundary transitions. 

As a warm-up related to the start of Sec.~\ref{sub:zbasis}, in Sec.~\ref{sub:X_ZZ} we study weak measurements of the $\mathbb{Z}_2$-preserving operators $\cos(\omega) X_{j,1} + \sin(\omega) Z_{j,1} Z_{j+1,1}$, where $\omega \in [0, \pi/2]$ denotes the tilt angle.  In Sec.~\ref{sub:X_Z}, we then consider an on-site tilted measurement basis $\cos{(\omega)} X_{j,1} + \sin{(\omega)} Z_{j,1}$ that generically breaks $\mathbb{Z}_2$ spin-flip symmetry for the measured chain.  Both subsections focus on uniform post-selection sectors.  For both types of tilted measurement bases, we present numerical and analytical evidence that $\omega = 0$ and $\omega = \pi/2$ indeed correspond to stable fixed points separated by an intervening fixed point at a critical $\omega_c$.  Section~\ref{sec:nonlinear} analyzes non-linear averages over measurement outcomes and argues that signatures of such intermediate fixed points persist as measured by such quantities.

\subsection{$\mathbb{Z}_2$-preserving measurement basis}\label{sub:X_ZZ}

Consider a weak measurement enacted by the non-unitary operator 
\begin{equation}\label{eq:measXandZZ}
   M_{\rm sym}(\omega) = \prod_{j \in \rm even} e^{\beta [\cos(\omega) X_{j,1} + \sin(\omega) Z_{j,1} Z_{j+1,1}]},
\end{equation}
which preserves the $\mathbb{Z}_2$ symmetry in the first chain. Each term in the product represents local weak measurement of a two-site operator $\cos(\omega) X_{j,1} + \sin(\omega) Z_{j,1} Z_{j+1,1}$ whose eigenvalues are $\pm1$. Note that the product runs over all \emph{even} sites, such that all the local measurements mutually commute.  The operator $M_{\rm sym}(\omega)$ corresponds to a weak measurement outcome that uniformly amplifies contribution from the $+1$ eigenvalues for each constituent two-site operator. 
Since both $X_{j,1}$ and $Z_{j,1}Z_{j+1,1}$ admit a local representation in terms of the $\mathcal{XXZ}$ spin chains [Eq.~\eqref{eq:2XXZ}], it is easier to understand the transition between different fixed points compared to the case (examined later) where we weakly measure single-spin operators in a tilted basis.

We essentially already studied the extreme cases $\omega = 0$ and $\omega = \pi/2$ in Secs.~\ref{sub:xbasis} and \ref{sub:zbasis}, respectively.  The sole difference is that the weak measurement operator $M_{\rm sym}(\omega)$ covers only half the sites, which does not affect the leading contribution to the measurement-induced boundary term in the action.  For $\omega = 0$ the boundary term again takes the form of Eq.~\eqref{eq:pert}, which generates long-range order in the disorder operator for the upper chain, long-range $Z$ order for the lower chain, and area-law entanglement.  For $\omega = \pi/2$, the boundary term is given in Eq.~\eqref{eq:pertZZ} and generates dual behavior: long-range $Z$ order for the upper chain and long-range order in the disorder operator for the lower chain, again with area-law entanglement.  If the fixed points arising from measurements in these extreme limits are stable, it is natural to expect an intervening unstable fixed point at a critical tilt angle $\omega_c$---which we indeed establish below.  
The duality transformation in Eq.~\eqref{eq:duality} sends $\omega \to \frac{\pi}{2} - \omega$ and hence fixes the critical angle exactly to $\omega_c=\pi/4$.

We can access the intermediate fixed point from a field-theoretic viewpoint as follows. For general tilt angles $\omega$, the measurement encoded in Eq.~\eqref{eq:measXandZZ} induces a boundary term
\begin{equation}\label{eq:pert3}
    \delta S_{\rm meas} \propto \beta \int_{x,\tau} \delta(\tau) [\cos\omega\cos{\theta} \cos{\tilde \theta} +\sin\omega \sin{\theta} \sin{\tilde \theta}],
\end{equation}
which is relevant for $K>1/2$.
Upon passing to symmetric and antisymmetric field combinations $\theta_\pm$ and $\phi_\pm$---defined below Eq.~\eqref{eq:pert}---we equivalently obtain 
\begin{align}
    \delta S_{\rm meas} \propto \beta \int_{x,\tau}\delta(\tau)\bigg{[}&\sin{\left(\frac{\pi}{4} - \omega\right)} \cos\theta_{+} 
    \nonumber \\
    &+ \cos{\left(\frac{\pi}{4} - \omega\right)} \cos\theta_{-}\bigg{]}.
    \label{eq:defect_X_ZZ} 
\end{align}
For $0\leq\omega<\pi/4$, both relevant cosines have nonzero, positive prefactors, thus pinning $\theta_+$ and $\theta_-$ to minima of the boundary perturbation.  In this tilt-angle regime we expect the system to flow to the same boundary fixed point as in the limiting case with $\omega = 0$ (pure $X$ measurement).  For $\pi/4<\omega \leq \pi/2$, both cosines again have nonzero coefficients, though the $\cos\theta_+$ prefactor exhibits the opposite sign, pinning $\theta_+$ to distinct minima compared to the preceding regime.  Here we expect the system to flow to the same boundary fixed point that arises with $\omega = \pi/2$ (pure $ZZ$ measurement). 

We check the stability of these two fixed points by computing the correlation functions of several $\mathbb{Z}_2$-preserving operators 
and extracting their scaling dimensions.  In particular, a symmetry-allowed operator $\mathcal{O}$ with scaling dimension less than 1 would constitute a relevant perturbation and indicate instability of the underlying fixed point; conversely, if the scaling dimensions of all such operators are larger than 1, then the fixed point is stable. 
As reported in Table \ref{tab:exponents_x}, we find that with $\omega = 0$ all scaling dimensions for the $\mathbb{Z}_2$-preserving operators that we extracted indeed exceed 1 for any $K>1/2$---evidencing that this fixed point is stable against symmetry-allowed perturbations. Duality implies that the $\omega = \pi/2$ fixed point has the same stability as the $\omega = 0$ fixed point. 

\begin{figure}[h]
    \centering
    \includegraphics[width=\linewidth]{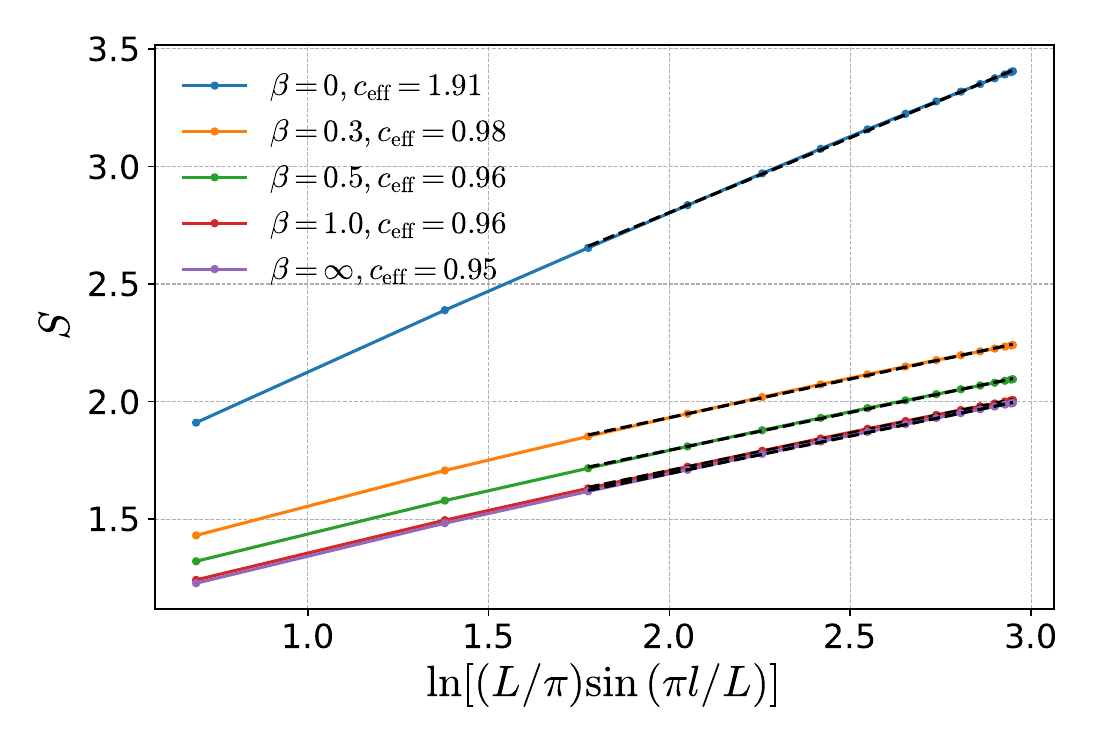}
    \caption{Entanglement entropy scaling at the measurement-induced boundary transition for the $\mathbb{Z}_2$-preserving case [$\omega = \pi/4$ in Eq.~\eqref{eq:measXandZZ}].  Data were obtained from DMRG with bond dimension $\chi = 2000$ assuming $K = 1.5$, $L = 60$, and periodic boundary conditions.  In the horizontal axis $l$ is the subsystem size.    The effective central charge $c_{\rm eff}$ is extracted by fitting to $S = \frac{c_{\rm eff}}{3} \ln[(L/\pi) \sin(\pi l/L)] + {\rm const}$ (see black dashed lines).  For all non-zero measurement strengths shown, we obtain $c_{\rm eff} \approx 1$ in harmony with field-theory predictions.  We verified that the fitted effective central charge is essentially unchanged for $K = 1$.} 
    \label{fig:measXandZZ_EE}
\end{figure}

Exactly at $\omega = \pi/4$, the $\theta_-$ sector remains pinned while the $\theta_+$ sector becomes unpinned (due to vanishing of the relevant cosine term), indicating an 
analytically tractable measurement-induced boundary transition in the $\mathbb{Z}_2$-preserving measurement basis.  One can crudely view the pinned $\theta_-$ sector as contributing area-law entanglement while the unpinned $\theta_+$ sector underlies logarithmic entanglement scaling with an associated effective central charge $c_{\rm eff} = 1$ that is reduced from the value $c = 2$ characteristic of the unmeasured theory.  Formally, $c_{\rm eff}$ follows by fitting the system's overall entanglement entropy to $S=\frac{c_{\rm eff}}{3} \log[\frac{L}{\pi} \sin\frac{\pi l}{L}] + {\rm const}$. 
 We indeed numerically extract an effective central charge $c_{\rm eff} \approx 1$ following this procedure as shown in Fig.~\ref{fig:measXandZZ_EE}. 

The intermediate fixed point is unstable, as one can infer from the scaling dimension $[X_{j,1}] = \frac{1}{2K} < 1$ for $K>\frac{1}{2}$; see Appendix~\ref{app:measX1andX2} for the derivation.  Physically, adding a relevant term composed of $X_{j,1}$ operators to the boundary action has the same effect as changing the tilt angle away from $\omega = \pi/4$---driving the system into one of the two stable fixed points depending on the sign of the coefficient.  

\begin{figure}[ht]
    \centering
    \includegraphics[width=\linewidth]{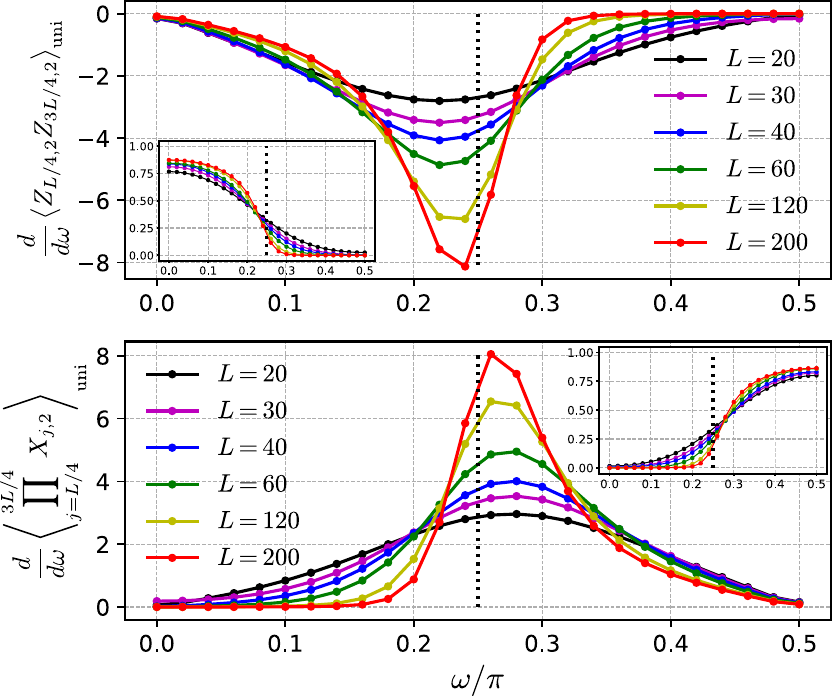}
    \caption{Derivative of order and disorder parameter correlations for the lower chain of the post-measurement gapless parent of the cluster state with $K=1.5$ as a function of the measurement tilt angle $\omega$ in the $\mathbb{Z}_2$-preserving case [Eq.~\eqref{eq:measXandZZ}]. Black dotted lines mark the measurement-induced boundary transition at $\omega_c = \pi/4$. Insets: order parameter  correlations $\braket{Z_{\frac{L}{4},2} Z_{\frac{3L}{4},2}}_{\rm uni}$ (upper panel) and disorder parameter correlations $\left\langle{\prod_{j = L/4}^{3L/4} X_{j,2}}\right\rangle_{\rm uni}$ (bottom panel) as a function of $\omega/\pi$. The gapless parent state is obtained using DMRG with open boundary conditions  and bond dimension $\chi=1200$.}
    \label{fig:measXandZZ_orderparams}
\end{figure}

We numerically examine the measurement-induced boundary transition by probing order and disorder operators as a function of tilt angle. More precisely, for a system of length $L$, we find that in the post-measurement wavefunction, the derivative of order and disorder operator correlations, 
\begin{equation}\label{eq:orderanddisorder}
    \frac{d}{d\omega} \braket{Z_{\frac{L}{4},2} Z_{\frac{3L}{4},2}}_{\rm uni}\quad \text{and} \quad \frac{d}{d\omega} \left\langle{\prod_{j = L/4}^{3L/4} X_{j,2}}\right\rangle_{\rm uni},
\end{equation}
both exhibit a peak near $\omega = \pi/4$ that sharpens as $L$ increases---consistent with onset of a divergence in the thermodynamic limit. Figure~\ref{fig:measXandZZ_orderparams} plots the two quantities above in the projective measurement limit; we have checked that similar results also arise for weak measurements, e.g., with $\beta = 0.5$ (see Appendix~\ref{app:measXandZZ_weak}). Because of duality, the two panels in Fig.~\ref{fig:measXandZZ_orderparams} are related by $\omega \to \frac{\pi}{2} - \omega$, up to a minus sign. 

\begin{table}[h!]
    \centering
    \setlength{\extrarowheight}{2pt}
        \caption{Operators with long-range order in two distinct measurement-tilt-angle regimes, $0 \leq \omega < \frac{\pi}{4}$ and $\frac{\pi}{4} < \omega \leq \frac{\pi}{2}$, for the case with $\mathbb{Z}_2$-preserving tilted measurement operators. At $\omega = \pi/4$ the measurement-induced boundary transition is described by a $c=1$ boson CFT.}
    \begin{tabularx}{\linewidth}{|c|Y|Y|}
    \hline
          & $0 \leq \omega < \frac{\pi}{4}$  & $\frac{\pi}{4} < \omega \leq \frac{\pi}{2}$\\
          \hline
          upper chain & $\braket{\prod_{j=0}^x X_{j,1}}$ & $\braket{Z_{0,1}Z_{x,1}}$ \\
          lower chain & $\braket{Z_{0,2}Z_{x,2}}$ &  $\braket{\prod_{j=0}^x X_{j,2}}$ \\
          \hline
    \end{tabularx}
    \label{tab:measXandZZ_phases}
\end{table}

To summarize, the measurement in Eq.~\eqref{eq:measXandZZ} yields a transition between two stable fixed points when the tilt angle $\omega$ varies from $0$ to $\pi/2$ (see Table~\ref{tab:measXandZZ_phases}).  The transition is described by a $c = 1$ boson CFT that one can view as a two-channel Luttinger liquid in which weak measurements pin a field at $\tau = 0$ for one channel but not the other.   As an aside, the same low-energy theory in Eq.~\eqref{eq:defect_X_ZZ} alternatively arises from measurement of both $X_{j,1}$ and $X_{j,2}$:
\begin{equation}\label{eq:measX1andX2}
    M_{\rm sym}'(\omega) = \prod_j e^{\beta \cos{(\omega)} X_{j,1}} e^{\beta \sin{(\omega)} X_{j,2}}.
\end{equation}
We explore this measurement operator in Appendix~\ref{app:measX1andX2} and find a similar intermediate fixed point with $c_{\rm eff} = 1$.

\subsection{$\mathbb{Z}_2$-breaking measurement basis}\label{sub:X_Z}

The example above was analytically tractable thanks to duality arguments and bosonization.  Next we address the more subtle case of tilted on-site measurements implemented by
\begin{equation}\label{eq:measXandZ}
   M(\omega) =  e^{\beta \sum_j[\cos(\omega) X_{j,1} + \sin(\omega) Z_{j,1}]}. 
\end{equation}
The limits $\omega = 0$ and $\omega = \pi/2$ correspond to two different fixed points that we analyzed in Sections \ref{sub:xbasis} and \ref{sub:zbasis}, respectively.  
For intermediate $\omega$ values, the relevant $X_{j,1}$ and $Z_{j,1}$ measurements compete with each other, and the winner determines the long-distance physics.  We will show that, as in the previous subsection, these fixed points are separated by an intervening unstable fixed point at an intermediate critical tilt angle $\omega_c$ (which here becomes $\beta$ dependent).

\begin{figure}[h]
    \centering
    \includegraphics[width=\linewidth]{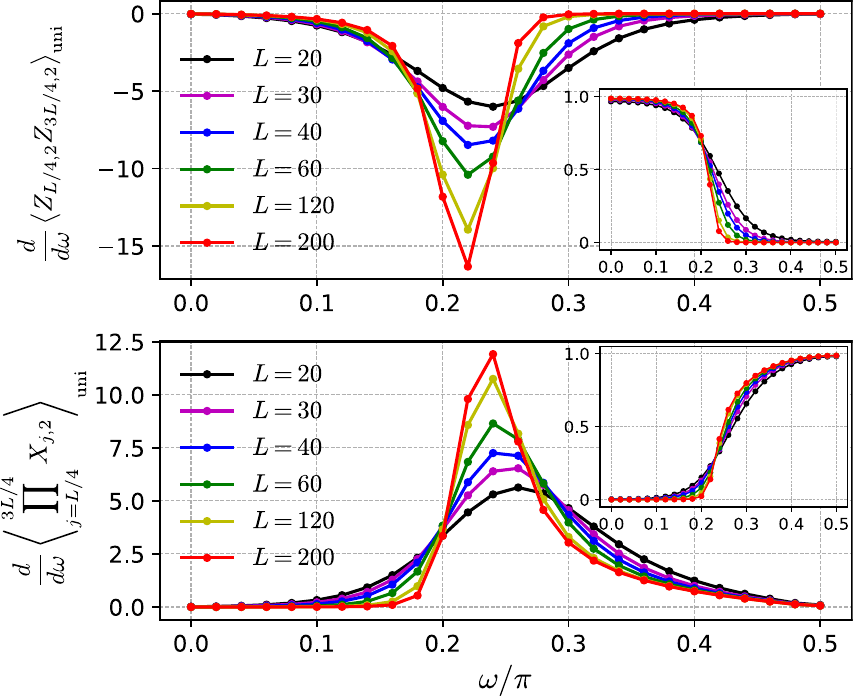}
    \caption{
    Same as Fig.~\ref{fig:measXandZZ_orderparams} but for the $\mathbb{Z}_2$-breaking measurement operator [Eq.~\eqref{eq:measXandZ}].  Here the critical tilt angle for the measurement-induced boundary transition is around $\omega_c \approx 0.22\pi$.    
    }
    \label{fig:orderparams}
\end{figure}

To investigate the stability of the $\omega = 0,\pi/2$ fixed points, we compute the scaling dimension of several different operators. 
Tables \ref{tab:exponents_x} and \ref{tab:exponents_z} report our results for the $\omega=0$ and $\omega=\pi/2$ cases, respectively. 
Unlike the previous subsection, it is now essential to consider operators that break  $\mathbb{Z}_2$ in the first chain, since the tilted measurement operator in Eq.~\eqref{eq:measXandZ} explicitly breaks that symmetry. 
In particular, we find numerically that at $K = 1$, $Z_{j,1}$ is a marginal operator at the $\omega = 0$ fixed point; in this case we can not infer the stability of that fixed point by a simple scaling dimension analysis.  
However, for $K >1$, all the operators we have checked have scaling dimension larger than 1 for both fixed points, indicating that they are likely stable and hence that a transition exists at a nontrivial critical tilt angle $\omega_c$.
Figure~\ref{fig:orderparams} corroborates this picture by plotting correlations of the order and the disorder operators for the lower chain at different system sizes $L$ with $K = 1.5$.  Note the striking similarity to Fig.~\ref{fig:measXandZZ_orderparams} obtained for a fully $\mathbb{Z}_2$ preserving measurement.  The breaking of $\mathbb{Z}_2$ symmetry for the upper chain combined with the loss of duality in the present case pushes the critical tilt angle $\omega_c$ slightly below $\pi/4$; an intermediate fixed point nevertheless clearly still arises---representing a second example of a measurement-induced boundary transition.  

\begin{figure}[ht]
    \centering
    \includegraphics[width=\linewidth]{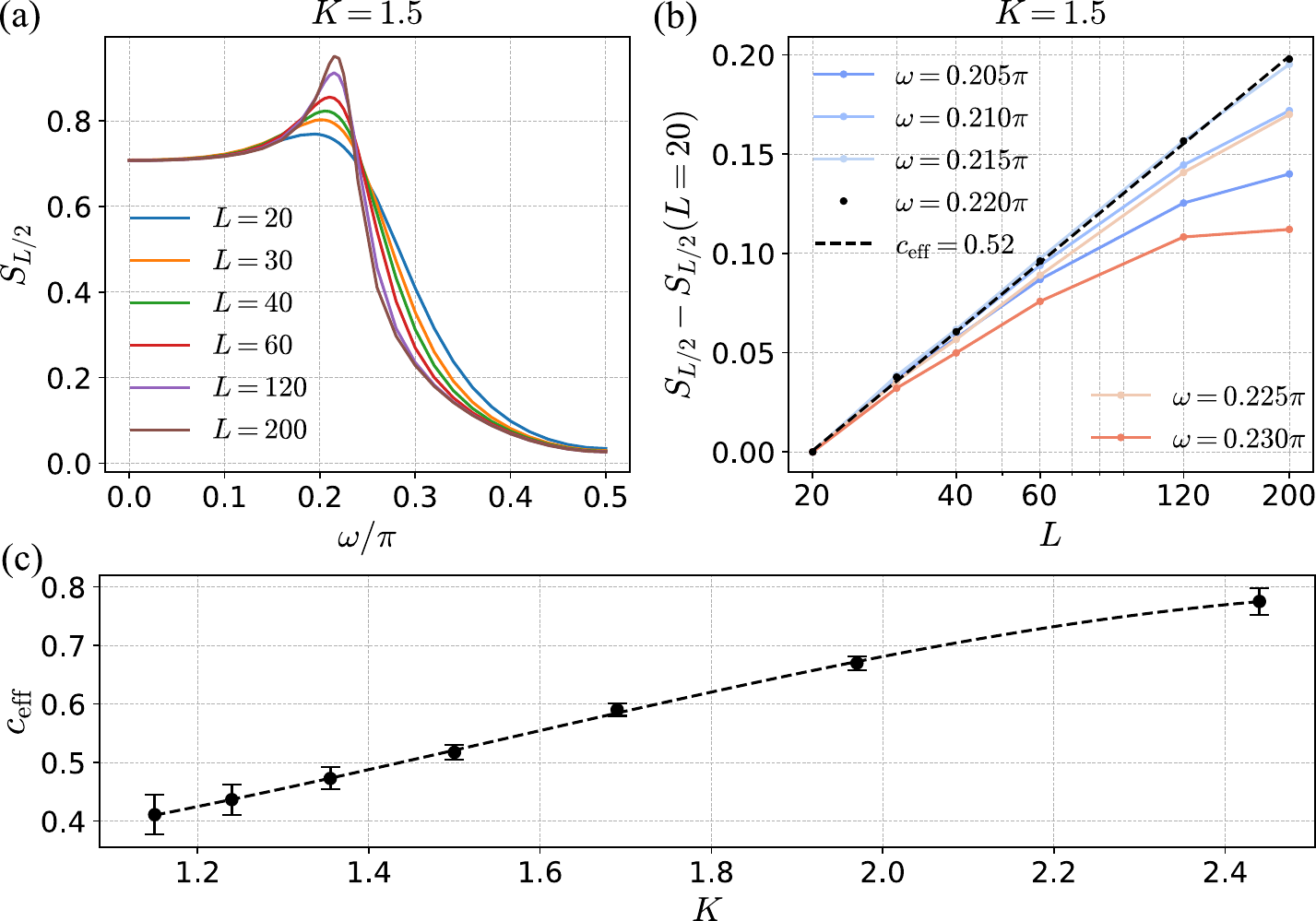}
    \caption{(a,b) Dependence of half-chain entanglement entropy $S_{L/2}$ on tilt angle $\omega$ and system size $L$ for the $\mathbb{Z}_2$-breaking measurement protocol [Eq.~\eqref{eq:measXandZ}], assuming $K = 1.5$. 
    At the intermediate fixed point $\omega = \omega_c \approx 0.22 \pi$ [see black dots in (b)] the data fit well to $S_{L/2} = \frac{c_{\rm eff}}{6} \log(L) + {\rm const}$ with $c_{\rm eff} \approx 0.52$. (c) Effective central charge obtained obtain by the same procedure in (b) but for different Luttinger parameters. 
    The dashed line is simply a guide to the eye. Data were obtained from DMRG with bond dimension $\chi=1200$; see Appendix~\ref{app:dmrgconvergence} for numerical details.}
    \label{fig:EE_measXandZ}
\end{figure}

Unlike the symmetry-preserving measurement in the previous section, it is hard to analytically assess the entanglement scaling at the intermediate fixed point due to the nonlocality of the $\mathbb{Z}_2$-breaking measurement operator in the $\mathcal{XXZ}$ basis. Nonetheless, we numerically compute the effective central charge at the transition for systems with different Luttinger parameters. As an example, in Fig.~\ref{fig:EE_measXandZ}(a) we observe that the entanglement entropy peaks at the intermediate fixed point $\omega_c \approx 0.22 \pi$ for $K=1.5$, and we can extract the effective central charge $c_{\rm eff} \approx 0.52$ using the relation between half-chain entanglement entropy and system size for one interval attached to the boundary, 
$S_{L/2} = \frac{c_{\rm eff}}{6} \log(L) + {\rm const}$ [see Fig.~\ref{fig:EE_measXandZ}(b)]. Using the same procedure for systems with different Luttinger parameters, we also observe that $c_{\rm eff}$ varies as a function of $K$; see Fig.~\ref{fig:EE_measXandZ}(c).

To understand the relation between intermediate fixed points from $\mathbb{Z}_2$-preserving and $\mathbb{Z}_2$-breaking measurements, we note that the former is not stable against symmetry-breaking measurements. More specifically, the $\mathbb{Z}_2$-breaking operator $Z_{j,1}$ 
is relevant at the $\mathbb{Z}_2$-preserving intermediate fixed point, as shown 
in Appendix~\ref{app:measX1andX2}. It is therefore expected that a weak $Z_{j,1}$ measurement drives an RG flow away from the $\mathbb{Z}_2$-preserving intermediate fixed point to the $\omega = \pi/2$ fixed point in Eq.~\eqref{eq:measXandZ}. Meanwhile, adding a small $X_{j,1}$ measurement to the $\mathbb{Z}_2$-preserving intermediate fixed point forces the system to flow to the $\omega = 0$ fixed point. With some fine-tuned combination of $Z_{j,1}$ and $X_{j,1}$ measurements, we expect a flow from the $\mathbb{Z}_2$-preserving intermediate fixed point to the $\mathbb{Z}_2$-breaking one. The origin and significance of the  continuous dependence of $c_{\rm eff}$ on $K$ remains an intriguing question for future work.

\subsection{Intermediate fixed point from an ensemble of measurement outcomes}\label{sec:nonlinear}

\begin{figure}[h]
    \centering
    \includegraphics[width=0.9\linewidth]{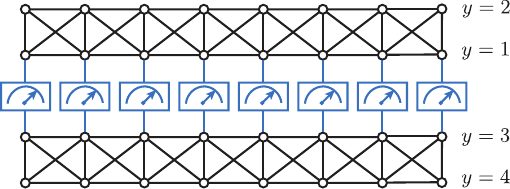}
    \caption{Graphical representation of the setup considered to compute the non-linear averages in Eq.~\eqref{eq:non-linear1}. The replicated system consists of four chains: Chains 1 and 2 are the original gapless parent of the cluster state (swapped relative to Fig.~\ref{fig:cluster}), while chains 3 and 4 form an identical copy of 1 and 2, respectively. To compute the non-linear observables in Eq.~\eqref{eq:non-linear1}, a weak measurement of $O_{j,1} O_{j,3}$ is applied at all sites with post-selection for a uniform outcome.}
    \label{fig:replica}
\end{figure}

So far in this section we focused only on post-selecting for uniform measurement outcomes---which generically require exponentially many trials in system size to capture with reasonable probability.  
A way to bypass post-selection is to average quantities that are nonlinear in the density matrix $\rho_{\bf s}=\ket{\psi_{\bm{s}}} \bra{\psi_{\bm{s}}}$ \cite{Garratt2023,Weinstein2023,Yang2023}.
For instance, 
one can take advantage of ``non-linear observables'' defined as 
\begin{equation} \label{eq:non-linear1}
    \langle\langle \Gamma \rangle\rangle \equiv \frac{\sum_{\bm{s}} p_{\bm{s}}^2 \braket{\Gamma}_{\bm{s}}}{\sum_{\bm{s}} p_{\bm{s}}^2}, 
\end{equation}
where $\Gamma$ is any operator on the second chain of our 
gapless parent of the cluster state. By doing so we effectively weight each set of measurement outcomes $ \bm{s}$ by the probability distribution $p_{\bm{s}}^2$, thereby favoring the most likely outcomes. 
Following Refs.~\onlinecite{Weinstein2023,Yang2023}, we show in Appendix~\ref{app:nonlinear} that these nonlinear, measurement-averaged observables are exactly equivalent to observables evaluated in a replicated theory---i.e., describing four chains instead of two---with uniform post-selection. 
The price we pay in considering a replicated version of the model is thus balanced by the simplicity of focusing on uniform measurement outcomes. 

Suppose that in some original (un-replicated) model we measure operators $O_j$ with strength $\beta$ and perform a non-linear average over measurement outcomes.  Appendix~\ref{app:nonlinear} recasts Eq.~\eqref{eq:non-linear1} as 
\begin{equation}\label{eq:replica}
\begin{aligned}
    \langle \langle \Gamma \rangle \rangle &= \frac{\bra{\psi^{A}_{\Delta}} \bra{\psi^{B}_{\Delta}} \Gamma^{A} e^{2\beta' \sum_j O_j^{A} O_j^{B}}\ket{\psi^{A}_{\Delta}} \ket{\psi^{B}_{\Delta}}}{\bra{\psi^{A}_{\Delta}} \bra{\psi^{B}} e^{2\beta' \sum_j O_j^{A} O_j^{B}}\ket{\psi^{A}_{\Delta}} \ket{\psi^{B}_{\Delta}}},
\end{aligned}
\end{equation}
where the superscripts $(A,B)$ denote the copy of the replicated theory, $\ket{\psi_{\Delta}}$ is the unmeasured Hamiltonian ground state, and $\tanh(2\beta') = \tanh^2 (2\beta)$.  Notice that, as stated above, the right side of Eq.~\eqref{eq:replica} does not involve an explicit sum over measurement outcomes, but rather takes the form of a normalized expectation value involving the joint state $\ket{\psi^A_{\Delta}}\ket{\psi^B_{\Delta}}$ modified by a uniform weak-measurement operator $e^{\beta' \sum_jO_j^A O_j^B}$ acting on both copies.   Numerically, Eq.~\eqref{eq:replica} provides a straightforward way of computing non-linearly averages from a many-body state (obtained from, e.g., tensor networks), in contrast to Eq.~\eqref{eq:non-linear1} which requires summing over exponentially many outcomes.

\begin{figure}[ht]
    \centering
    \includegraphics[width=\linewidth]{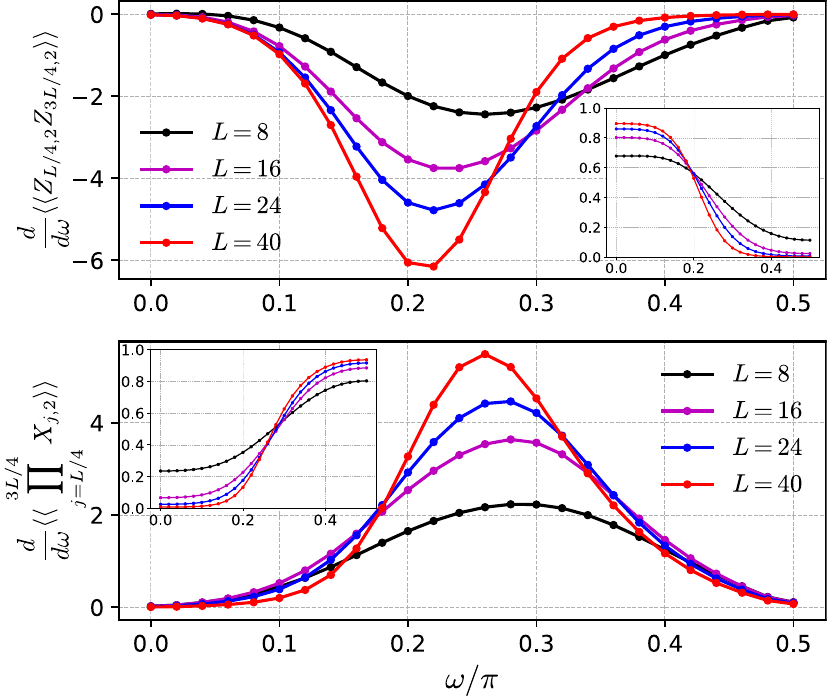}
    \caption{Derivative of order and disorder parameter correlations in the $y=2$ chain of the post-measurement replicated gapless parent of the cluster state as a function of the tilting angle $\omega$. A measurement-induced boundary transition remains visible at a critical value of around $\omega_c\approx 0.2\pi$. Insets: order parameter correlations $\langle\langle Z_{\frac{L}{4},2} Z_{\frac{3L}{4},2}\rangle\rangle$ and disorder parameter correlations $\langle\langle{\prod_{j = L/4}^{3L/4} X_{j,2}}\rangle\rangle$ as a function of $\omega/\pi$. The replicated gapless parent of the cluster state is obtained using DMRG with open boundary conditions and bond dimension $\chi=1600$, assuming $K = 1.5$. Appendix~\ref{app:replica_numerics} analyzes different $K$ values.
    }
    \label{fig:orderparams_replica}
\end{figure}

In our setup, $\ket{\psi}$ is the ground state of the Hamiltonian \eqref{eq:2XXZ} defined on a two-chain system; in the replicated theory we associated copy $A$ with chains $y = 1,2$, and copy $B$ with chains $y = 3,4$ (see Fig.~\ref{fig:replica}).  Measurement of the tilted operator $\cos(\omega)X_{j,1}+\sin(\omega)Z_{j,1}$ in the un-replicated theory translates into a uniform weak measurement operator 
\begin{align}
    e^{\beta' \sum_j[\cos(\omega)X_{j,1}+\sin(\omega)Z_{j,1}][\cos(\omega)X_{j,3}+\sin(\omega)Z_{j,3}]}
\end{align}
in the replica problem specified on the right side of Eq.~\eqref{eq:replica}.  
We are specifically interested in $\omega$ dependence of correlations 
$\langle\langle Z_{j,2} Z_{k,2}\rangle\rangle$ and $\langle\langle \prod_{i=j}^k X_{i,2} \rangle\rangle$ of the order and disorder operators, respectively, in the unmeasured chain 2.  
Computing these quantities analytically is complicated, so we content ourselves with the numerical evaluation presented in Fig.~\ref{fig:orderparams_replica}.   
We observe behavior very similar to that shown in Fig.~\ref{fig:orderparams} for an un-replicated system with uniform post-selection. 
Even though the finite size effects are more pronounced in the replicated theory because of the reduced maximum system size we can achieve using DMRG, an intermediate fixed point at some nontrivial $\omega_c$ clearly persists when non-linearly averaging over measurement outcomes, suggesting that a measurement-induced boundary transition occurs also in this case.

\section{Measurement-induced boundary transitions in minimal models}\label{sec:minimal}

The previous section established that tilted weak measurements on the gapless parent of the cluster state generate a measurement-induced boundary transition.  The key ingredient was the generation of multiple competing relevant perturbations under weak measurement---a property that can certainly arise in broader contexts.  Armed with this insight, we now investigate alternate setups involving minimal models that indeed host similar transitions: namely, ground states of the tricritical Ising model and the three-state Potts model after a single round of measurements that yield a uniform outcome.

\subsection{Tricritical Ising}

The tricritical Ising CFT---realized by various systems including Rydberg atom arrays \cite{FSS,SlagleTCI} and spin chains \cite{Obrien}---provides an experimentally relevant example of a minimal model that exhibits a measurement-induced boundary transition. 
The CFT is characterized by a central charge $c=7/10$ and hosts six primary fields of scaling dimensions $0, 3/40, 1/5, 7/8, 6/5$, and $3$. Particularly relevant here are the spin field $\sigma$ (dimension $3/40$) and the energy field $\varepsilon$ (dimension $1/5$).

We consider O'Brien and Fendley's  microscopic implementation that arises upon adding a self-dual 3-spin interactions to the critical transverse-field Ising model \cite{Obrien}: 
\begin{equation}\label{eq:TCI}
H = - \sum_j [Z_j Z_{j+1} + X_j - \lambda (Z_{j-1} Z_{j} X_{j+1} + X_{j-1} Z_{j} Z_{j+1})]
\end{equation}
with a parameter $\lambda \geq 0$. (Upcoming work will examine similar physics in the Rydberg array context \cite{Naus2025}.) For $\lambda < \lambda_c \approx 0.428$, the low-energy physics is simply governed by an Ising CFT that describes a continuous transition between ferromagnetic and paramagnetic phases. Increasing $\lambda$ beyond $\lambda_c$ generates a spectral gap and renders the transition first order.  At the critical value $\lambda = \lambda_c$, the transition belongs to the tricritical Ising universality class; there we can relate the microscopic spin operators to the CFT fields as $Z \sim \sigma$, $X \sim \varepsilon$.

Suppose that we prepare the ground state of Eq.~\eqref{eq:TCI} at the tricritical Ising point, and then perform a uniform-post-selection weak measurement associated with the non-unitary operator 
\begin{equation}\label{eq:measop_TCI}
    e^{\beta \sum_j [\cos(\omega) X_{j} + \sin(\omega) Z_{j}]}.
\end{equation}
At $\omega = 0$, the measurement operator $X\sim\varepsilon$ induces a relevant defect-line perturbation to the CFT that triggers an RG flow to free boundary conditions.  [Here we assume $\beta >0$, which at $\omega = 0$ amplifies configurations favored by the transverse field in Eq.~\eqref{eq:TCI}.] Post-measurement correlation functions exhibit power-law decay with exponents that can be obtained from BCFT calculations~\cite{Cardy1989,Affleck2000} (see Appendix~\ref{app:BCFT} for details); for example, one finds
\begin{equation}\label{eq:TCI_measX}
    \braket{Z_0 Z_x}_{\rm uni} \sim \braket{Y_0 Y_x}_{\rm uni} \sim x^{-3}.
\end{equation}
The power-law exponent above is sufficiently large that both $Z$ and $Y$ operators comprise irrelevant perturbations to the free boundary condition fixed point, suggesting local stability. 
Similarly, at $\omega = \pi/2$, the measurement operator $Z\sim\sigma$ induces a relevant defect-line perturbation that triggers a flow to fixed boundary conditions with post-measurement correlations
\begin{equation}\label{eq:TCI_measZ}
    \braket{X_0 X_x}_{\rm uni} \sim \braket{Y_0 Y_x}_{\rm uni} \sim x^{-4},
\end{equation}
again suggesting local stability of the corresponding fixed point.  

\begin{figure}[h]
    \centering
    \includegraphics[width=1\linewidth]{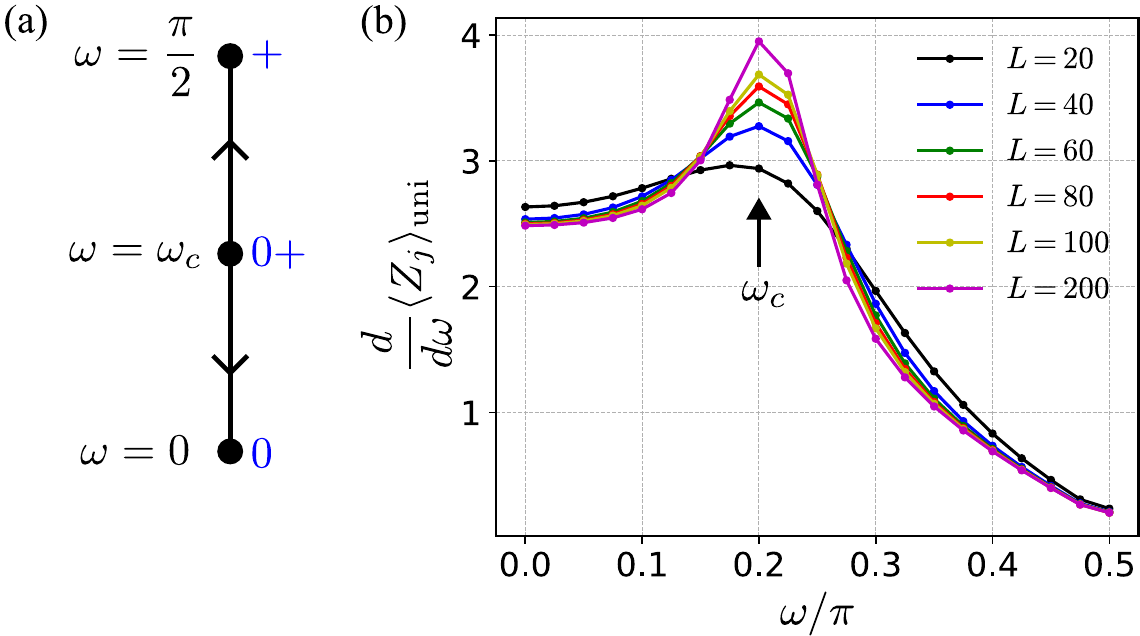}
    \caption{(a) Boundary renormalization group flow involving select fixed points for the tricritical Ising CFT, corresponding to free (0), fixed ($+$), and partially polarized ($0+$) boundary conditions.  The unstable $0+$ fixed point represents a measurement-induced boundary transition obtained at a fine-tuned measurement-basis tilt angle $\omega_c$. 
  (b) Derivative of the post-measurement 1-point function $\braket{Z_j}_{\rm uni}$ versus tilt angle. Data were obtained by DMRG using periodic boundary conditions and bond dimension $\chi = 1000$. The measurement strength is $\beta = 0.5$.}
    \label{fig:TCI}
\end{figure}

In fact, both fixed points are known in the tricritical Ising BCFT and are stable against any local perturbations~\cite{Affleck2000}. Interestingly, in the boundary RG flow---illustrated in Fig.~\ref{fig:TCI}(a)---there exists an unstable intermediate fixed point between the free ($0$) and fixed ($+$) boundary conditions, corresponding to the ``partially polarized'' boundary condition ($0+$) of the tricritical Ising theory. In our weak measurement protocol, this intermediate fixed point arises at a critical tilt angle $\omega_c$, and can be identified by a second-order transition in the order parameter $\braket{Z}_{\rm uni}$; see Fig.~\ref{fig:TCI}(b). This agreement highlights the robustness of our results within the established theoretical framework of BCFT. 
From a perturbative analysis, we note that at the intermediate fixed point, the post-measurement entanglement entropy obeys an area law, in contrast to the case of the gapless parent of the 1D cluster state. For further details, see Appendix~\ref{app:BCFT}.

\subsection{Three-state Potts}

The BCFT analysis capturing possible interesting measurement-induced transitions generalizes to other critical theories as well. 
A further example is the three-state Potts model, described by a CFT with central charge $c=4/5$ and six primary fields with scaling dimensions $0,2/15,4/5,4/3,14/5,6$. 
We are also primarily interested in the spin field $\sigma$ (dimension $2/15$) and energy field $\varepsilon$ (dimension $4/5$).   

For a microscopic realization, let us define on-site operators 
\begin{equation}
    U_j=\begin{pmatrix}
    1 & 0 & 0 \\
    0 & e^{i 2\pi/3} & 0 \\
    0 & 0 & e^{-i 2\pi/3}
    \end{pmatrix}, ~~V_j=\begin{pmatrix}
    0 & 0 & 1 \\
    1 & 0 & 0 \\
    0 & 1 & 0
    \end{pmatrix}.
\end{equation}
The eigenvalues of $U_j$ label three local spin states---which we denote by $A,B,C$---while $V_j$ cycles the spin among those three values.  A lattice Hamiltonian realizing the $c = 4/5$ CFT reads
\begin{equation}
    H=-\sum_j\left(U_j U^{\dagger}_{j+1}+U^{\dagger}_j U_{j+1}+V_j+V_j^{\dagger}\right).
\end{equation}
Notice that $H$ preserves a $\mathbb{Z}_3$ symmetry that sends $U_j \rightarrow e^{i2\pi/3} U_j$.  
The rich operator content of this model gives rise to various possible boundary conditions \cite{CARDY_1989,SALEUR_1989}, some of which we can readily realize via weak measurement.  Exploring the impact of weak measurements is facilitated by the dictionary between lattice operators and CFT fields derived in Ref.~\onlinecite{Mong_2014}: one finds $U_j \sim \sigma$ and $V_j + V_j^\dagger \sim \varepsilon$.  Note that in this context $\sigma$ is not a Hermitian field.  

Consider first the weak measurement operator
\begin{equation}
    e^{\beta \sum_j(V_j+V_j^\dagger)},
\end{equation}
which generates a relevant defect-line action involving the $\varepsilon$ field.  (Here too we assume $\beta>0$ such that the measurement amplifies configurations favored by the $V_j$ terms in $H$.)  The perturbation induces a flow to a BCFT corresponding to free boundary conditions.  There we find power-law correlations (see Appendix \ref{app:BCFT}) 
\begin{equation}
\braket{U_0U^{\dagger}_j}\sim |j|^{-4/3}, \quad \braket{V_0V_j^\dagger}\sim |j|^{-8/3}.
\end{equation}
The fact that the scaling dimension of $U_j$ is smaller than one indicates instability of this fixed point under $\mathbb{Z}_3$-breaking measurement operators.  

\begin{figure}[h]
    \centering
    \includegraphics[width=\linewidth]{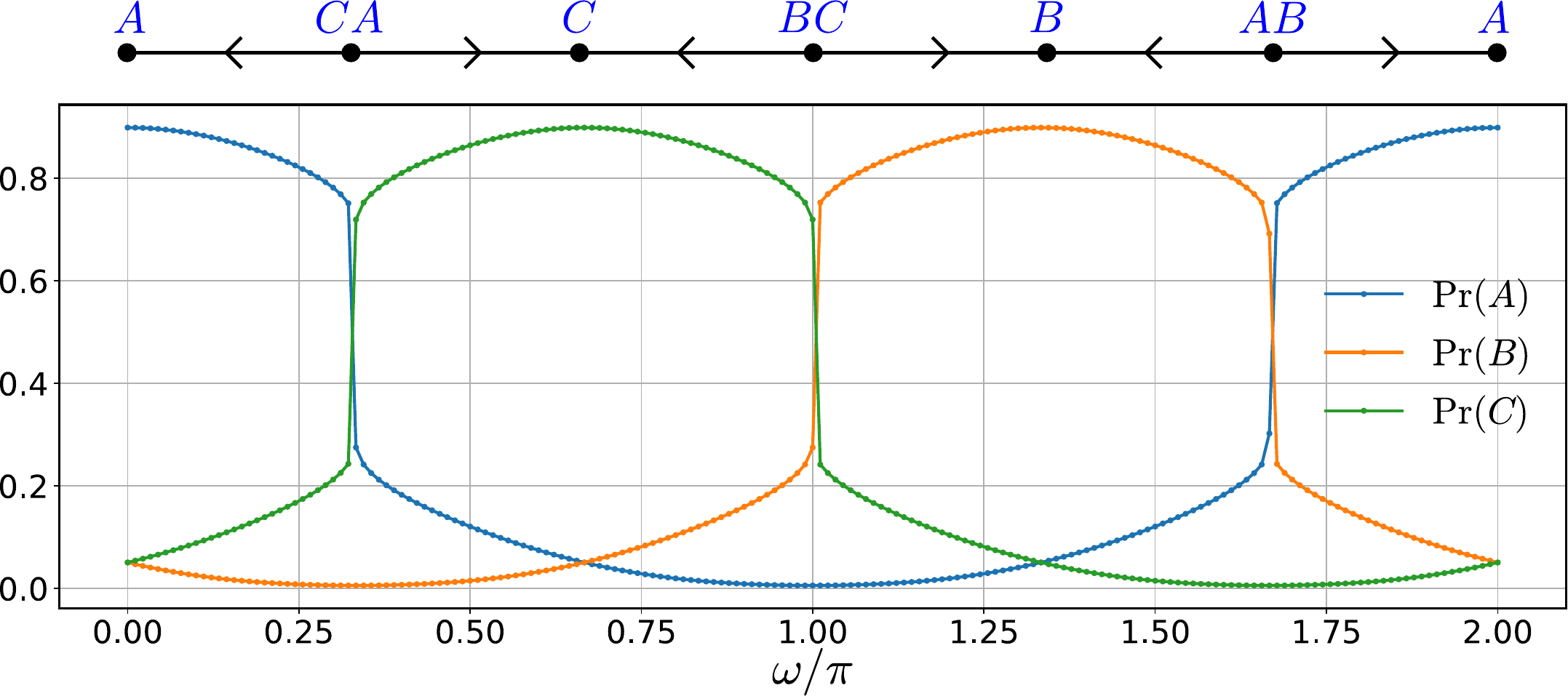}
    \caption{Boundary RG flow between fixed and mixed boundary conditions of the three-state Potts model. The three fixed boundary conditions are labeled by $A, B, C$ and the mixed boundary conditions are labeled by $AB, BC, CA$. We plot the post-measurement probability of observing each Potts state as a function of tilt angle $\omega$ specified in Eq.~\eqref{eq:measPotts}. Data were obtained with iDMRG using bond dimension $\chi = 300$ and measurement strength $\beta = 0.1$.}
    \label{fig:Potts_RGflow}
\end{figure}

Next we consider a family of such $\mathbb{Z}_3$-symmetry breaking operators given by
\begin{equation}\label{eq:measPotts}
    e^{-\beta \sum_j(V_j + V_j^\dagger)(e^{i\omega}U_j + e^{-i \omega} U_j^\dagger)(V_j + V_j^\dagger)} \quad (\beta>0),
\end{equation}  
which generates a defect-line action involving $e^{i \omega} \sigma + e^{-i \omega} \sigma^\dagger$, up to irrelevant $\mathbb{Z}_3$ breaking operators. [The $(V + V^\dagger)$ terms do not change the symmetry of the measurement, but produce the expected boundary fixed points numerically.]
Here $\omega$ 
plays the role of a tilt angle for the measurement basis. When $\omega$ varies, the measurement operator favors different Potts states, or certain superposition of them. At $\omega = 0$, for example, the measurement operator (weakly) projects each site to a state close to $A$, while $\omega = \pi$ equally favors $B$ and $C$ states.  We find that when $\omega \in (-\pi/3, \frac{\pi}{3})$, $(\frac{\pi}{3}, \pi)$ and $(\pi, \frac{5\pi}{3})$, the perturbation induces a flow to $A$, $B$ and $C$ fixed boundary conditions, respectively, with power-law correlations,
\begin{equation}
    \braket{U_0U^{\dagger}_j}\sim |j|^{-4}, \quad \braket{V_0V_j^\dagger}\sim |j|^{-4}.
\end{equation}
At $\omega = \frac{\pi}{3}$, $\pi$ and $\frac{5\pi}{3}$, the measurement induces a flow to mixed boundary conditions labeled by $CA$, $BC$, and $AB$ respectively, featuring power-law correlations,
\begin{equation}
    \braket{U_0U^{\dagger}_j}\sim |j|^{-4/5}, \quad \braket{V_0V_j^\dagger}\sim |j|^{-4/5}.
\end{equation}
These mixed boundary condition fixed points are unstable against fixed boundary conditions and thus can be viewed as measurement-induced boundary transitions between $A,B,C$ fixed points; see Fig.~\ref{fig:Potts_RGflow}.

\section{Discussion and outlook}\label{sec:conclusion}

\begin{figure*}[ht]
    \centering
    \includegraphics[width=0.99\linewidth]{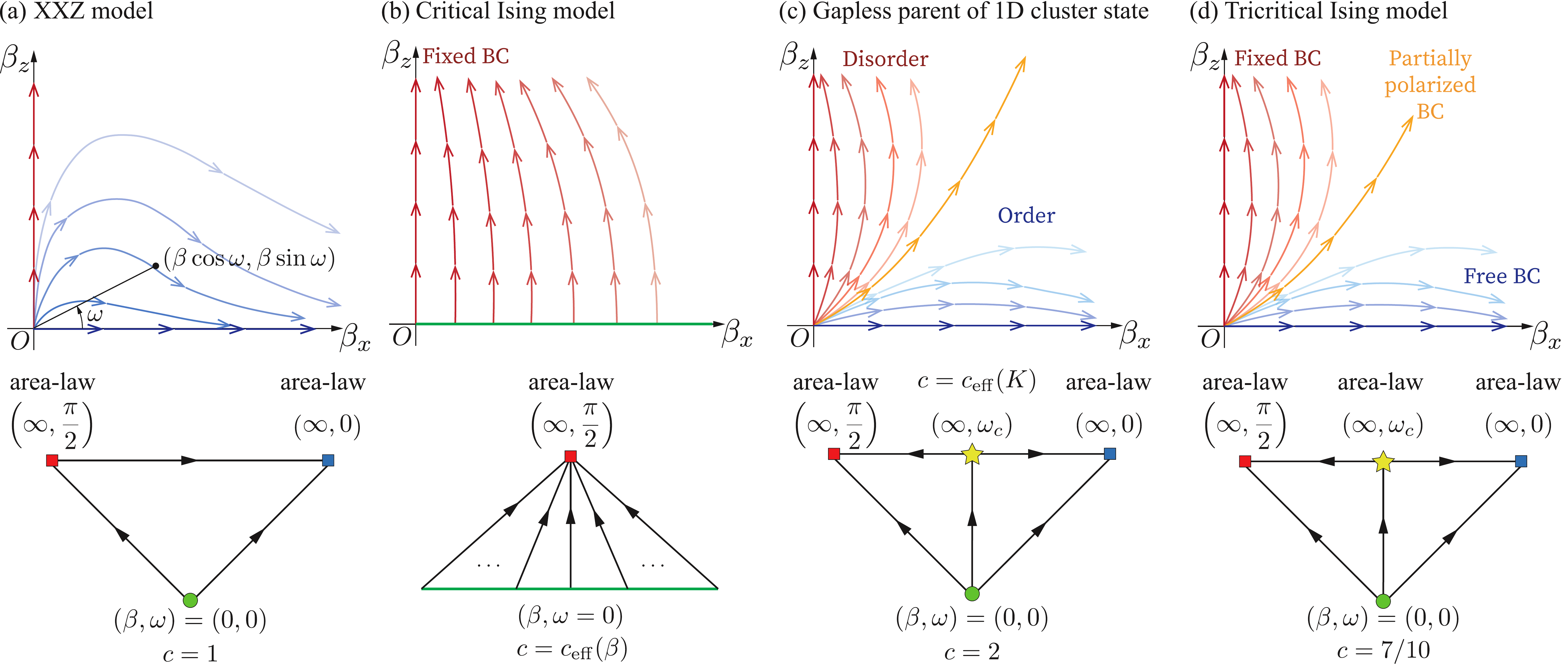}
    \caption{Upper row: Schematic RG flow induced by measurements enacted by $\exp[\beta \sum_j (\cos \omega X_j + \sin \omega Z_j)]$ in different gapless models; $\omega$ specifies the measurement-basis tilt angle, while $\beta_x = \beta \cos\omega$ and $\beta_z = \beta \sin\omega$. 
    Lower row: Phase diagram of different boundary fixed points located at $(\beta, \omega)$ values specified in the parentheses. Green circles represent the pristine unmeasured CFTs; red (blue) squares show the boundary fixed points induced by measurement in the $Z$ ($X$) basis; and yellow stars represent intermediate fixed points realizing measurement-induced boundary transitions [present in (c,d) but not (a,b)]. In (b) the green line is a line of fixed points. All fixed points feature power-law correlations and thus have infinite correlation length.}
    \label{fig:RGflows}
\end{figure*}

We have introduced a gapless parent of the 1D cluster state SPT that exactly maps to two decoupled Luttinger liquids. 
Local measurements together with post-selection of the uniform outcome introduce a defect line in the CFT action that triggers an RG flow to various BCFT fixed points---allowing us to derive universal properties of the post-measurement wavefunction.  We uncovered richer behavior compared to the descendant gapped cluster state SPT, including nontrivial correlations persisting under weak measurements and measurement-induced boundary transitions accessed by tilting the measurement basis.  

It is useful to additionally contrast with several previous works exploring  measurement-altered quantum critical states---which can also be understood within BCFT. For instance, Ref.~\onlinecite{Garratt2023} investigated weak measurement of the density operator $\cos(2\phi)$ in a single-channel Luttinger liquid [in spin language, $Z$-basis measurement in an $\mathcal{XXZ}$ spin chain governed by the first line of Eq.~\eqref{eq:2XXZ}]. 
By post-selecting the outcome $\boldsymbol{s}=\{(-1)^j\}$, the weak measurement induces a boundary condition $\phi = n\pi, n\in\mathbb{Z}$.
References~\onlinecite{usmeasurementaltered,Weinstein2023,Yang2023,Sala2024} studied weak measurements in the critical Ising chain described by the Hamiltonian $H = -\sum_j Z_j Z_{j+1} - \sum_j X_j$; here different measurement bases yield significantly different behaviors. The underlying low-energy theory is a $c = 1/2$ Ising CFT with three primary fields, $1, \sigma, \varepsilon$ with respective scaling dimensions $0, 1/8$, $1$.
After post-selection of the uniform outcome, $X$-basis measurements induce a marginal perturbation $\sim \varepsilon$ that yields a series of fixed points whose features depend on the measurement strength; $Z$-basis measurements, however, generate a relevant perturbation $\sim \sigma$ and a flow to the fixed boundary condition of the Ising CFT. 
In all these previous setups exhibiting measurement-altered criticality, tilting the measurement basis away from the 
$Z$-direction 
does not lead to a boundary transition, contrary to the gapless parent of the cluster state SPT.

Indeed, we can generalize the protocol of Ref.~\onlinecite{Garratt2023} by applying a tilted-basis weak measurement operator, $e^{\beta \sum_j [\cos\omega  X_j  +  \sin(\omega) (-1)^j Z_j]}$, similarly to what we did for the gapless parent state in Section \ref{sec:tiltmeas}. We focus on the $K<1$ case where the measurement of $(-1)^j Z_j$ induces a relevant perturbation to the pristine Luttinger liquid. By computing the scaling dimension of $X_j, Y_j, (-1)^j Z_j$ operators, we can study the stability of the two fixed points corresponding to $(\beta, \omega) = (\infty, 0)$ and $(\beta, \omega) = (\infty, \pi/2)$. The former is stable against any perturbation, while the latter is unstable against uniform $X$ perturbations (or $Y$ by symmetry). Relevance of weak $X$ measurement in the second case follows from the fact that, after $Z$-basis measurements, the correlator $\braket{X_0X_x}$ decays as $\sim x^{-1/K}$.
\sfigure{fig:RGflows}{a} illustrates the RG flow: by tuning $\omega$ from $0$ to $\pi/2$, the system flows to $(\beta, \omega) = (\infty, 0)$, without any intermediate fixed point.

Similar logic applies to the critical Ising chain under tilted weak measurements enacted by 
$e^{\sum_j \beta [\cos(\omega) X_{j} + \sin(\omega) Z_{j}]}$.  The series of fixed points induced by $X$ measurement at $\omega = 0$ is always unstable to a $\sigma$ perturbation induced by finite $\omega$, while the fixed point arising at $\omega = \pi/2$ is stable to arbitrary perturbations. 
Therefore, for any $\omega > 0$, the long-range physics is determined by the latter. 
In BCFT language, the boundary fixed points of the Ising model (that preserve the conformal symmetry and the RG flow between them are known \cite{Cardy1984,Cardy1989} (see also \sfigref{fig:RGflows}{b}). 
Interested readers may wonder what happens when tilting the measurement basis in the ($Y,Z$) instead of the $(X,Z$) plane, since symmetry does not relate the two problems in this case.  Reference~\onlinecite{Sala2024} shows that, at the CFT level, weak measurements in the $Y$ basis generate a marginal $\varepsilon$ perturbation---similar to $X$ but with reduced coupling strength. Consequently, weak measurements in a basis chosen arbitrarily in the $(X,Y)$ plane are always unstable to rotating the basis towards $Z$.

One goal of this work was to identify protocols that, after a single round of measurements on critical systems, induce a non-trivial transition between two fixed points. 
Such measurement-induced boundary transitions can arise when different types of measurements yield at least two \emph{stable}  boundary fixed points, which in the gapless parent of the cluster state correspond to $\omega = 0$ and $\omega = \pi/2$ in Eqs.~\eqref{eq:measXandZZ} and \eqref{eq:measXandZ}. These two fixed points are disconnected in the sense that any path in the parameter space connecting them goes across an intermediate fixed point. If the measurement preserves the $\mathbb{Z}_2$ symmetry of the gapless parent state, the intermediate fixed point is a critical theory with central charge $c=1$. On the other hand, if we consider a $\mathbb{Z}_2$ symmetry-breaking measurement operator, surprisingly the intermediate fixed point appears to show an effective central charge that continuously varies with the Luttinger parameter $K$. \sfigure{fig:RGflows}{c} illustrates the resulting schematic RG flow diagram indicating the competition between the two relevant measurement-induced perturbations.
Other measurement-induced boundary transitions can be found by studying tilted measurements on the ground state of minimal models such as tricritical Ising (see \sfigref{fig:RGflows}{d})  and the three-state Potts model, whose rich operator content also enables non-trivial transitions between different fixed points.

We mainly focused our analysis of measurement-induced boundary transitions by post-selecting for uniform measurement outcomes 
(but see Sec.~\ref{sec:nonlinear}). Can analogous transitions persist when post-selecting for \emph{typical} outcomes, i.e., those
obtained by sampling according to the 
Born probability distribution $p_{\bm s}$? Following Kadanoff’s decimation argument, Refs.~\cite{Garratt2023,Sala2024} found that if the scaling dimension of the leading perturbation generated by the measurement operator is less than $1/2$, then the \emph{disordered} defect-line perturbation emerging with typical outcomes remains relevant. 
Critical theories supporting multiple measurement operators with dimensions below $1/2$ might then support novel measurement-induced boundary transitions for typical outcomes---provided the fixed points generated by each operator are also stable.  The minimal models studied in Sec.~\ref{sec:minimal} provide potentially fruitful test cases.  For tricritical Ising, tilting between $X$ (dimension $1/5$) and $Z$ (dimension $3/40$) measurements would potentially yield interesting typical-outcome boundary transitions.  In the three-state Potts case, the Hermitian operators $U + U^\dagger$ and $i(U-U^\dagger)$ both carry dimension $2/15$, potentially fulfilling the above criterion as well. 
Further investigation along these lines would be worthwhile.

Our work also highlights various other interesting future directions.   
For instance, can one connect the $c=1$ intermediate fixed point studied in Sec.~\ref{sub:X_ZZ} to that of Sec.~\ref{sub:X_Z} by adding a $\mathbb{Z}_2$-breaking perturbation, and gain an analytical understanding of the $K$-dependent effective central charge found in the latter case? Is it possible to induce additional boundary conditions via measurements beyond those studied here, and possibly new measurement-induced boundary transitions, in minimal models such as the tricritical Ising and three-state Potts examples? 
Are there metrological applications of the measurement-altered critical states studied in this work?  And finally, can one devise generalized decoding protocols (extending that of Sec.~\ref{sec:decoding}) to a more general class of measurement bases, ideally to probe intermediate fixed points without post-selection?

\section*{Acknowledgments}

We are grateful to Ehud Altman, Nandu Manoj, Stephen Naus, Lorenzo Piroli, Pablo Sala, Kevin Slagle, and Nathanan Tantivasadakarn for insightful discussions. 
This work was primarily supported by the U.S.~Department of Energy, Office of Science, National Quantum Information Science Research Centers, Quantum Science Center. 
 Additional support was provided by the Caltech Institute for Quantum Information and Matter, an NSF Physics Frontiers Center with support of the Gordon and Betty Moore Foundation through Grant GBMF1250, the Walter Burke Institute for Theoretical Physics at Caltech, and the Israel Science Foundation (ISF) under grant 2572/21.

\newpage
\bibliographystyle{apsrev4-2}
\bibliography{ref}

\begin{thebibliography}{63}%
\makeatletter
\providecommand \@ifxundefined [1]{%
 \@ifx{#1\undefined}
}%
\providecommand \@ifnum [1]{%
 \ifnum #1\expandafter \@firstoftwo
 \else \expandafter \@secondoftwo
 \fi
}%
\providecommand \@ifx [1]{%
 \ifx #1\expandafter \@firstoftwo
 \else \expandafter \@secondoftwo
 \fi
}%
\providecommand \natexlab [1]{#1}%
\providecommand \enquote  [1]{``#1''}%
\providecommand \bibnamefont  [1]{#1}%
\providecommand \bibfnamefont [1]{#1}%
\providecommand \citenamefont [1]{#1}%
\providecommand \href@noop [0]{\@secondoftwo}%
\providecommand \href [0]{\begingroup \@sanitize@url \@href}%
\providecommand \@href[1]{\@@startlink{#1}\@@href}%
\providecommand \@@href[1]{\endgroup#1\@@endlink}%
\providecommand \@sanitize@url [0]{\catcode `\\12\catcode `\$12\catcode
  `\&12\catcode `\#12\catcode `\^12\catcode `\_12\catcode `\%12\relax}%
\providecommand \@@startlink[1]{}%
\providecommand \@@endlink[0]{}%
\providecommand \url  [0]{\begingroup\@sanitize@url \@url }%
\providecommand \@url [1]{\endgroup\@href {#1}{\urlprefix }}%
\providecommand \urlprefix  [0]{URL }%
\providecommand \Eprint [0]{\href }%
\providecommand \doibase [0]{https://doi.org/}%
\providecommand \selectlanguage [0]{\@gobble}%
\providecommand \bibinfo  [0]{\@secondoftwo}%
\providecommand \bibfield  [0]{\@secondoftwo}%
\providecommand \translation [1]{[#1]}%
\providecommand \BibitemOpen [0]{}%
\providecommand \bibitemStop [0]{}%
\providecommand \bibitemNoStop [0]{.\EOS\space}%
\providecommand \EOS [0]{\spacefactor3000\relax}%
\providecommand \BibitemShut  [1]{\csname bibitem#1\endcsname}%
\let\auto@bib@innerbib\@empty
\bibitem [{\citenamefont {Li}\ \emph {et~al.}(2018)\citenamefont {Li},
  \citenamefont {Chen},\ and\ \citenamefont {Fisher}}]{Li2018}%
  \BibitemOpen
  \bibfield  {author} {\bibinfo {author} {\bibfnamefont {Y.}~\bibnamefont
  {Li}}, \bibinfo {author} {\bibfnamefont {X.}~\bibnamefont {Chen}},\ and\
  \bibinfo {author} {\bibfnamefont {M.~P.~A.}\ \bibnamefont {Fisher}},\ }\href
  {https://doi.org/10.1103/PhysRevB.98.205136} {\bibfield  {journal} {\bibinfo
  {journal} {Phys. Rev. B}\ }\textbf {\bibinfo {volume} {98}},\ \bibinfo
  {pages} {205136} (\bibinfo {year} {2018})}\BibitemShut {NoStop}%
\bibitem [{\citenamefont {Skinner}\ \emph {et~al.}(2019)\citenamefont
  {Skinner}, \citenamefont {Ruhman},\ and\ \citenamefont
  {Nahum}}]{Skinner2019}%
  \BibitemOpen
  \bibfield  {author} {\bibinfo {author} {\bibfnamefont {B.}~\bibnamefont
  {Skinner}}, \bibinfo {author} {\bibfnamefont {J.}~\bibnamefont {Ruhman}},\
  and\ \bibinfo {author} {\bibfnamefont {A.}~\bibnamefont {Nahum}},\ }\href
  {https://doi.org/10.1103/PhysRevX.9.031009} {\bibfield  {journal} {\bibinfo
  {journal} {Phys. Rev. X}\ }\textbf {\bibinfo {volume} {9}},\ \bibinfo {pages}
  {031009} (\bibinfo {year} {2019})}\BibitemShut {NoStop}%
\bibitem [{\citenamefont {Fisher}\ \emph {et~al.}(2023)\citenamefont {Fisher},
  \citenamefont {Khemani}, \citenamefont {Nahum},\ and\ \citenamefont
  {Vijay}}]{Fisher2023}%
  \BibitemOpen
  \bibfield  {author} {\bibinfo {author} {\bibfnamefont {M.~P.~A.}\
  \bibnamefont {Fisher}}, \bibinfo {author} {\bibfnamefont {V.}~\bibnamefont
  {Khemani}}, \bibinfo {author} {\bibfnamefont {A.}~\bibnamefont {Nahum}},\
  and\ \bibinfo {author} {\bibfnamefont {S.}~\bibnamefont {Vijay}},\ }\href
  {https://doi.org/10.1146/annurev-conmatphys-031720-030658} {\bibfield
  {journal} {\bibinfo  {journal} {Annu. Rev. Condens. Matter Phys.}\ ,\
  \bibinfo {pages} {335}} (\bibinfo {year} {2023})}\BibitemShut {NoStop}%
\bibitem [{\citenamefont {Raussendorf}\ and\ \citenamefont
  {Briegel}(2001)}]{Raussendorf01}%
  \BibitemOpen
  \bibfield  {author} {\bibinfo {author} {\bibfnamefont {R.}~\bibnamefont
  {Raussendorf}}\ and\ \bibinfo {author} {\bibfnamefont {H.~J.}\ \bibnamefont
  {Briegel}},\ }\href {https://doi.org/10.1103/PhysRevLett.86.5188} {\bibfield
  {journal} {\bibinfo  {journal} {Phys. Rev. Lett.}\ }\textbf {\bibinfo
  {volume} {86}},\ \bibinfo {pages} {5188} (\bibinfo {year}
  {2001})}\BibitemShut {NoStop}%
\bibitem [{\citenamefont {Lee}\ \emph {et~al.}(2022)\citenamefont {Lee},
  \citenamefont {Ji}, \citenamefont {Bi},\ and\ \citenamefont
  {Fisher}}]{Lee2022}%
  \BibitemOpen
  \bibfield  {author} {\bibinfo {author} {\bibfnamefont {J.~Y.}\ \bibnamefont
  {Lee}}, \bibinfo {author} {\bibfnamefont {W.}~\bibnamefont {Ji}}, \bibinfo
  {author} {\bibfnamefont {Z.}~\bibnamefont {Bi}},\ and\ \bibinfo {author}
  {\bibfnamefont {M.~P.~A.}\ \bibnamefont {Fisher}},\ }\bibfield  {journal}
  {\bibinfo  {journal} {arXiv}\ }\href
  {https://doi.org/10.48550/arXiv.2208.11699} {10.48550/arXiv.2208.11699}
  (\bibinfo {year} {2022})\BibitemShut {NoStop}%
\bibitem [{\citenamefont {Garratt}\ \emph {et~al.}(2023)\citenamefont
  {Garratt}, \citenamefont {Weinstein},\ and\ \citenamefont
  {Altman}}]{Garratt2023}%
  \BibitemOpen
  \bibfield  {author} {\bibinfo {author} {\bibfnamefont {S.~J.}\ \bibnamefont
  {Garratt}}, \bibinfo {author} {\bibfnamefont {Z.}~\bibnamefont {Weinstein}},\
  and\ \bibinfo {author} {\bibfnamefont {E.}~\bibnamefont {Altman}},\ }\href
  {https://doi.org/10.1103/PhysRevX.13.021026} {\bibfield  {journal} {\bibinfo
  {journal} {Phys. Rev. X}\ }\textbf {\bibinfo {volume} {13}},\ \bibinfo
  {pages} {021026} (\bibinfo {year} {2023})}\BibitemShut {NoStop}%
\bibitem [{\citenamefont {Sun}\ \emph {et~al.}(2023)\citenamefont {Sun},
  \citenamefont {Yao},\ and\ \citenamefont {Jian}}]{Sun2023}%
  \BibitemOpen
  \bibfield  {author} {\bibinfo {author} {\bibfnamefont {X.}~\bibnamefont
  {Sun}}, \bibinfo {author} {\bibfnamefont {H.}~\bibnamefont {Yao}},\ and\
  \bibinfo {author} {\bibfnamefont {S.-K.}\ \bibnamefont {Jian}},\ }\href
  {https://arxiv.org/abs/2301.11337} {\bibinfo {title} {New critical states
  induced by measurement}} (\bibinfo {year} {2023}),\ \Eprint
  {https://arxiv.org/abs/2301.11337} {arXiv:2301.11337 [quant-ph]} \BibitemShut
  {NoStop}%
\bibitem [{\citenamefont {Ashida}\ \emph {et~al.}(2024)\citenamefont {Ashida},
  \citenamefont {Furukawa},\ and\ \citenamefont {Oshikawa}}]{Ashida2023}%
  \BibitemOpen
  \bibfield  {author} {\bibinfo {author} {\bibfnamefont {Y.}~\bibnamefont
  {Ashida}}, \bibinfo {author} {\bibfnamefont {S.}~\bibnamefont {Furukawa}},\
  and\ \bibinfo {author} {\bibfnamefont {M.}~\bibnamefont {Oshikawa}},\ }\href
  {https://doi.org/10.1103/PhysRevB.110.094404} {\bibfield  {journal} {\bibinfo
   {journal} {Phys. Rev. B}\ }\textbf {\bibinfo {volume} {110}},\ \bibinfo
  {pages} {094404} (\bibinfo {year} {2024})}\BibitemShut {NoStop}%
\bibitem [{\citenamefont {Tang}\ and\ \citenamefont {Wen}(2024)}]{tang2024}%
  \BibitemOpen
  \bibfield  {author} {\bibinfo {author} {\bibfnamefont {Q.}~\bibnamefont
  {Tang}}\ and\ \bibinfo {author} {\bibfnamefont {X.}~\bibnamefont {Wen}},\
  }\href {https://arxiv.org/abs/2411.13705} {\bibinfo {title} {A critical state
  under weak measurement is not critical}} (\bibinfo {year} {2024})\BibitemShut
  {NoStop}%
\bibitem [{\citenamefont {Yang}\ \emph {et~al.}(2023)\citenamefont {Yang},
  \citenamefont {Mao},\ and\ \citenamefont {Jian}}]{Yang2023}%
  \BibitemOpen
  \bibfield  {author} {\bibinfo {author} {\bibfnamefont {Z.}~\bibnamefont
  {Yang}}, \bibinfo {author} {\bibfnamefont {D.}~\bibnamefont {Mao}},\ and\
  \bibinfo {author} {\bibfnamefont {C.-M.}\ \bibnamefont {Jian}},\ }\href
  {https://doi.org/10.1103/PhysRevB.108.165120} {\bibfield  {journal} {\bibinfo
   {journal} {Phys. Rev. B}\ }\textbf {\bibinfo {volume} {108}},\ \bibinfo
  {pages} {165120} (\bibinfo {year} {2023})}\BibitemShut {NoStop}%
\bibitem [{\citenamefont {Weinstein}\ \emph {et~al.}(2023)\citenamefont
  {Weinstein}, \citenamefont {Sajith}, \citenamefont {Altman},\ and\
  \citenamefont {Garratt}}]{Weinstein2023}%
  \BibitemOpen
  \bibfield  {author} {\bibinfo {author} {\bibfnamefont {Z.}~\bibnamefont
  {Weinstein}}, \bibinfo {author} {\bibfnamefont {R.}~\bibnamefont {Sajith}},
  \bibinfo {author} {\bibfnamefont {E.}~\bibnamefont {Altman}},\ and\ \bibinfo
  {author} {\bibfnamefont {S.~J.}\ \bibnamefont {Garratt}},\ }\href
  {https://doi.org/10.1103/PhysRevB.107.245132} {\bibfield  {journal} {\bibinfo
   {journal} {Phys. Rev. B}\ }\textbf {\bibinfo {volume} {107}},\ \bibinfo
  {pages} {245132} (\bibinfo {year} {2023})}\BibitemShut {NoStop}%
\bibitem [{\citenamefont {Murciano}\ \emph {et~al.}(2023)\citenamefont
  {Murciano}, \citenamefont {Sala}, \citenamefont {Liu}, \citenamefont {Mong},\
  and\ \citenamefont {Alicea}}]{usmeasurementaltered}%
  \BibitemOpen
  \bibfield  {author} {\bibinfo {author} {\bibfnamefont {S.}~\bibnamefont
  {Murciano}}, \bibinfo {author} {\bibfnamefont {P.}~\bibnamefont {Sala}},
  \bibinfo {author} {\bibfnamefont {Y.}~\bibnamefont {Liu}}, \bibinfo {author}
  {\bibfnamefont {R.~S.~K.}\ \bibnamefont {Mong}},\ and\ \bibinfo {author}
  {\bibfnamefont {J.}~\bibnamefont {Alicea}},\ }\href
  {https://doi.org/10.1103/PhysRevX.13.041042} {\bibfield  {journal} {\bibinfo
  {journal} {Phys. Rev. X}\ }\textbf {\bibinfo {volume} {13}},\ \bibinfo
  {pages} {041042} (\bibinfo {year} {2023})}\BibitemShut {NoStop}%
\bibitem [{\citenamefont {Sala}\ \emph {et~al.}(2024)\citenamefont {Sala},
  \citenamefont {Murciano}, \citenamefont {Liu},\ and\ \citenamefont
  {Alicea}}]{Sala2024}%
  \BibitemOpen
  \bibfield  {author} {\bibinfo {author} {\bibfnamefont {P.}~\bibnamefont
  {Sala}}, \bibinfo {author} {\bibfnamefont {S.}~\bibnamefont {Murciano}},
  \bibinfo {author} {\bibfnamefont {Y.}~\bibnamefont {Liu}},\ and\ \bibinfo
  {author} {\bibfnamefont {J.}~\bibnamefont {Alicea}},\ }\href
  {https://doi.org/10.1103/PRXQuantum.5.030307} {\bibfield  {journal} {\bibinfo
   {journal} {PRX Quantum}\ }\textbf {\bibinfo {volume} {5}},\ \bibinfo {pages}
  {030307} (\bibinfo {year} {2024})}\BibitemShut {NoStop}%
\bibitem [{\citenamefont {Paviglianiti}\ \emph {et~al.}(2023)\citenamefont
  {Paviglianiti}, \citenamefont {Turkeshi}, \citenamefont {Schir\`o},\ and\
  \citenamefont {Silva}}]{Paviglianiti2023}%
  \BibitemOpen
  \bibfield  {author} {\bibinfo {author} {\bibfnamefont {A.}~\bibnamefont
  {Paviglianiti}}, \bibinfo {author} {\bibfnamefont {X.}~\bibnamefont
  {Turkeshi}}, \bibinfo {author} {\bibfnamefont {M.}~\bibnamefont {Schir\`o}},\
  and\ \bibinfo {author} {\bibfnamefont {A.}~\bibnamefont {Silva}},\
  }\href@noop {} {\bibfield  {journal} {\bibinfo  {journal} {arXiv}\ }
  (\bibinfo {year} {2023})},\ \Eprint {https://arxiv.org/abs/2310.02686}
  {arXiv:2310.02686 [quant-ph]} \BibitemShut {NoStop}%
\bibitem [{\citenamefont {Patil}\ and\ \citenamefont
  {Ludwig}(2024)}]{patilLud2024}%
  \BibitemOpen
  \bibfield  {author} {\bibinfo {author} {\bibfnamefont {R.~A.}\ \bibnamefont
  {Patil}}\ and\ \bibinfo {author} {\bibfnamefont {A.~W.~W.}\ \bibnamefont
  {Ludwig}},\ }\href@noop {} {\bibfield  {journal} {\bibinfo  {journal}
  {arXiv}\ } (\bibinfo {year} {2024})},\ \Eprint
  {https://arxiv.org/abs/2409.02107} {arXiv:2409.02107 [cond-mat.stat-mech]}
  \BibitemShut {NoStop}%
\bibitem [{\citenamefont {Lee}\ \emph {et~al.}(2023{\natexlab{a}})\citenamefont
  {Lee}, \citenamefont {Jian},\ and\ \citenamefont {Xu}}]{Lee2023}%
  \BibitemOpen
  \bibfield  {author} {\bibinfo {author} {\bibfnamefont {J.~Y.}\ \bibnamefont
  {Lee}}, \bibinfo {author} {\bibfnamefont {C.-M.}\ \bibnamefont {Jian}},\ and\
  \bibinfo {author} {\bibfnamefont {C.}~\bibnamefont {Xu}},\ }\href
  {https://doi.org/10.1103/PRXQuantum.4.030317} {\bibfield  {journal} {\bibinfo
   {journal} {PRX Quantum}\ }\textbf {\bibinfo {volume} {4}},\ \bibinfo {pages}
  {030317} (\bibinfo {year} {2023}{\natexlab{a}})}\BibitemShut {NoStop}%
\bibitem [{\citenamefont {Zou}\ \emph {et~al.}(2023)\citenamefont {Zou},
  \citenamefont {Sang},\ and\ \citenamefont {Hsieh}}]{Zou2023}%
  \BibitemOpen
  \bibfield  {author} {\bibinfo {author} {\bibfnamefont {Y.}~\bibnamefont
  {Zou}}, \bibinfo {author} {\bibfnamefont {S.}~\bibnamefont {Sang}},\ and\
  \bibinfo {author} {\bibfnamefont {T.~H.}\ \bibnamefont {Hsieh}},\ }\href
  {https://doi.org/10.1103/PhysRevLett.130.250403} {\bibfield  {journal}
  {\bibinfo  {journal} {Phys. Rev. Lett.}\ }\textbf {\bibinfo {volume} {130}},\
  \bibinfo {pages} {250403} (\bibinfo {year} {2023})}\BibitemShut {NoStop}%
\bibitem [{\citenamefont {Lee}\ \emph {et~al.}(2023{\natexlab{b}})\citenamefont
  {Lee}, \citenamefont {Jian},\ and\ \citenamefont {Xu}}]{leequantum2023}%
  \BibitemOpen
  \bibfield  {author} {\bibinfo {author} {\bibfnamefont {J.~Y.}\ \bibnamefont
  {Lee}}, \bibinfo {author} {\bibfnamefont {C.-M.}\ \bibnamefont {Jian}},\ and\
  \bibinfo {author} {\bibfnamefont {C.}~\bibnamefont {Xu}},\ }\href
  {https://doi.org/10.1103/PRXQuantum.4.030317} {\bibfield  {journal} {\bibinfo
   {journal} {PRX Quantum}\ }\textbf {\bibinfo {volume} {4}},\ \bibinfo {pages}
  {030317} (\bibinfo {year} {2023}{\natexlab{b}})}\BibitemShut {NoStop}%
\bibitem [{\citenamefont {Ma}(2023)}]{ma2023tmy}%
  \BibitemOpen
  \bibfield  {author} {\bibinfo {author} {\bibfnamefont {R.}~\bibnamefont
  {Ma}},\ }\href@noop {} {\bibfield  {journal} {\bibinfo  {journal} {arXiv}\ }
  (\bibinfo {year} {2023})},\ \Eprint {https://arxiv.org/abs/2304.08277}
  {arXiv:2304.08277 [quant-ph]} \BibitemShut {NoStop}%
\bibitem [{\citenamefont {Garratt}\ and\ \citenamefont
  {Altman}(2024)}]{garratt2024probe}%
  \BibitemOpen
  \bibfield  {author} {\bibinfo {author} {\bibfnamefont {S.~J.}\ \bibnamefont
  {Garratt}}\ and\ \bibinfo {author} {\bibfnamefont {E.}~\bibnamefont
  {Altman}},\ }\href {https://doi.org/10.1103/PRXQuantum.5.030311} {\bibfield
  {journal} {\bibinfo  {journal} {PRX Quantum}\ }\textbf {\bibinfo {volume}
  {5}},\ \bibinfo {pages} {030311} (\bibinfo {year} {2024})}\BibitemShut
  {NoStop}%
\bibitem [{\citenamefont {McGinley}(2024)}]{McGinley2024}%
  \BibitemOpen
  \bibfield  {author} {\bibinfo {author} {\bibfnamefont {M.}~\bibnamefont
  {McGinley}},\ }\href {https://doi.org/10.1103/PRXQuantum.5.020347} {\bibfield
   {journal} {\bibinfo  {journal} {PRX Quantum}\ }\textbf {\bibinfo {volume}
  {5}},\ \bibinfo {pages} {020347} (\bibinfo {year} {2024})}\BibitemShut
  {NoStop}%
\bibitem [{\citenamefont {McGinley}\ and\ \citenamefont
  {Fava}(2023)}]{McGinleyLett2023}%
  \BibitemOpen
  \bibfield  {author} {\bibinfo {author} {\bibfnamefont {M.}~\bibnamefont
  {McGinley}}\ and\ \bibinfo {author} {\bibfnamefont {M.}~\bibnamefont
  {Fava}},\ }\href {https://doi.org/10.1103/PhysRevLett.131.160601} {\bibfield
  {journal} {\bibinfo  {journal} {Phys. Rev. Lett.}\ }\textbf {\bibinfo
  {volume} {131}},\ \bibinfo {pages} {160601} (\bibinfo {year}
  {2023})}\BibitemShut {NoStop}%
\bibitem [{\citenamefont {Rossini}\ and\ \citenamefont
  {Vicari}(2020)}]{vicari}%
  \BibitemOpen
  \bibfield  {author} {\bibinfo {author} {\bibfnamefont {D.}~\bibnamefont
  {Rossini}}\ and\ \bibinfo {author} {\bibfnamefont {E.}~\bibnamefont
  {Vicari}},\ }\href {https://doi.org/10.1103/PhysRevB.102.035119} {\bibfield
  {journal} {\bibinfo  {journal} {Phys. Rev. B}\ }\textbf {\bibinfo {volume}
  {102}},\ \bibinfo {pages} {035119} (\bibinfo {year} {2020})}\BibitemShut
  {NoStop}%
\bibitem [{\citenamefont {Lu}\ \emph {et~al.}(2022)\citenamefont {Lu},
  \citenamefont {Lessa}, \citenamefont {Kim},\ and\ \citenamefont
  {Hsieh}}]{Lu22}%
  \BibitemOpen
  \bibfield  {author} {\bibinfo {author} {\bibfnamefont {T.-C.}\ \bibnamefont
  {Lu}}, \bibinfo {author} {\bibfnamefont {L.~A.}\ \bibnamefont {Lessa}},
  \bibinfo {author} {\bibfnamefont {I.~H.}\ \bibnamefont {Kim}},\ and\ \bibinfo
  {author} {\bibfnamefont {T.~H.}\ \bibnamefont {Hsieh}},\ }\href
  {https://doi.org/10.1103/PRXQuantum.3.040337} {\bibfield  {journal} {\bibinfo
   {journal} {PRX Quantum}\ }\textbf {\bibinfo {volume} {3}},\ \bibinfo {pages}
  {040337} (\bibinfo {year} {2022})}\BibitemShut {NoStop}%
\bibitem [{\citenamefont {Chan}\ \emph {et~al.}(2019)\citenamefont {Chan},
  \citenamefont {Nandkishore}, \citenamefont {Pretko},\ and\ \citenamefont
  {Smith}}]{Chan2019}%
  \BibitemOpen
  \bibfield  {author} {\bibinfo {author} {\bibfnamefont {A.}~\bibnamefont
  {Chan}}, \bibinfo {author} {\bibfnamefont {R.~M.}\ \bibnamefont
  {Nandkishore}}, \bibinfo {author} {\bibfnamefont {M.}~\bibnamefont
  {Pretko}},\ and\ \bibinfo {author} {\bibfnamefont {G.}~\bibnamefont
  {Smith}},\ }\href {https://doi.org/10.1103/PhysRevB.99.224307} {\bibfield
  {journal} {\bibinfo  {journal} {Phys. Rev. B}\ }\textbf {\bibinfo {volume}
  {99}},\ \bibinfo {pages} {224307} (\bibinfo {year} {2019})}\BibitemShut
  {NoStop}%
\bibitem [{\citenamefont {Cao}\ \emph {et~al.}(2019)\citenamefont {Cao},
  \citenamefont {Tilloy},\ and\ \citenamefont {Luca}}]{deluca}%
  \BibitemOpen
  \bibfield  {author} {\bibinfo {author} {\bibfnamefont {X.}~\bibnamefont
  {Cao}}, \bibinfo {author} {\bibfnamefont {A.}~\bibnamefont {Tilloy}},\ and\
  \bibinfo {author} {\bibfnamefont {A.~D.}\ \bibnamefont {Luca}},\ }\href
  {https://doi.org/10.21468/SciPostPhys.7.2.024} {\bibfield  {journal}
  {\bibinfo  {journal} {SciPost Phys.}\ }\textbf {\bibinfo {volume} {7}},\
  \bibinfo {pages} {024} (\bibinfo {year} {2019})}\BibitemShut {NoStop}%
\bibitem [{\citenamefont {Li}\ \emph {et~al.}(2021)\citenamefont {Li},
  \citenamefont {Chen}, \citenamefont {Ludwig},\ and\ \citenamefont
  {Fisher}}]{Li2021}%
  \BibitemOpen
  \bibfield  {author} {\bibinfo {author} {\bibfnamefont {Y.}~\bibnamefont
  {Li}}, \bibinfo {author} {\bibfnamefont {X.}~\bibnamefont {Chen}}, \bibinfo
  {author} {\bibfnamefont {A.~W.~W.}\ \bibnamefont {Ludwig}},\ and\ \bibinfo
  {author} {\bibfnamefont {M.~P.~A.}\ \bibnamefont {Fisher}},\ }\href
  {https://doi.org/10.1103/PhysRevB.104.104305} {\bibfield  {journal} {\bibinfo
   {journal} {Phys. Rev. B}\ }\textbf {\bibinfo {volume} {104}},\ \bibinfo
  {pages} {104305} (\bibinfo {year} {2021})}\BibitemShut {NoStop}%
\bibitem [{\citenamefont {Friedman}\ \emph {et~al.}(2022)\citenamefont
  {Friedman}, \citenamefont {Yin}, \citenamefont {Hong},\ and\ \citenamefont
  {Lucas}}]{Friedman}%
  \BibitemOpen
  \bibfield  {author} {\bibinfo {author} {\bibfnamefont {A.~J.}\ \bibnamefont
  {Friedman}}, \bibinfo {author} {\bibfnamefont {C.}~\bibnamefont {Yin}},
  \bibinfo {author} {\bibfnamefont {Y.}~\bibnamefont {Hong}},\ and\ \bibinfo
  {author} {\bibfnamefont {A.}~\bibnamefont {Lucas}},\ }\href@noop {}
  {\bibfield  {journal} {\bibinfo  {journal} {arXiv}\ } (\bibinfo {year}
  {2022})},\ \Eprint {https://arxiv.org/abs/2206.09929} {arXiv:2206.09929
  [quantum-ph]} \BibitemShut {NoStop}%
\bibitem [{\citenamefont {Lin}\ \emph {et~al.}(2023)\citenamefont {Lin},
  \citenamefont {Ye}, \citenamefont {Zou}, \citenamefont {Sang},\ and\
  \citenamefont {Hsieh}}]{Hsieh1}%
  \BibitemOpen
  \bibfield  {author} {\bibinfo {author} {\bibfnamefont {C.-J.}\ \bibnamefont
  {Lin}}, \bibinfo {author} {\bibfnamefont {W.}~\bibnamefont {Ye}}, \bibinfo
  {author} {\bibfnamefont {Y.}~\bibnamefont {Zou}}, \bibinfo {author}
  {\bibfnamefont {S.}~\bibnamefont {Sang}},\ and\ \bibinfo {author}
  {\bibfnamefont {T.~H.}\ \bibnamefont {Hsieh}},\ }\href
  {https://doi.org/10.22331/q-2023-02-02-910} {\bibfield  {journal} {\bibinfo
  {journal} {Quantum}\ }\textbf {\bibinfo {volume} {7}},\ \bibinfo {pages}
  {910} (\bibinfo {year} {2023})}\BibitemShut {NoStop}%
\bibitem [{\citenamefont {Rajabpour}(2016)}]{Rajabpour_2016}%
  \BibitemOpen
  \bibfield  {author} {\bibinfo {author} {\bibfnamefont {M.~A.}\ \bibnamefont
  {Rajabpour}},\ }\href {https://doi.org/10.1088/1742-5468/2016/06/063109}
  {\bibfield  {journal} {\bibinfo  {journal} {J. Stat. Mech.}\ }\textbf
  {\bibinfo {volume} {2016}},\ \bibinfo {pages} {063109} (\bibinfo {year}
  {2016})}\BibitemShut {NoStop}%
\bibitem [{\citenamefont {Rajabpour}(2015)}]{Rajabpour_2015}%
  \BibitemOpen
  \bibfield  {author} {\bibinfo {author} {\bibfnamefont {M.~A.}\ \bibnamefont
  {Rajabpour}},\ }\href {https://doi.org/10.1103/PhysRevB.92.075108} {\bibfield
   {journal} {\bibinfo  {journal} {Phys. Rev. B}\ }\textbf {\bibinfo {volume}
  {92}},\ \bibinfo {pages} {075108} (\bibinfo {year} {2015})}\BibitemShut
  {NoStop}%
\bibitem [{\citenamefont {Cardy}(1984)}]{Cardy1984}%
  \BibitemOpen
  \bibfield  {author} {\bibinfo {author} {\bibfnamefont {J.~L.}\ \bibnamefont
  {Cardy}},\ }\href
  {https://doi.org/https://doi.org/10.1016/0550-3213(84)90241-4} {\bibfield
  {journal} {\bibinfo  {journal} {Nuclear Physics B}\ }\textbf {\bibinfo
  {volume} {240}},\ \bibinfo {pages} {514} (\bibinfo {year}
  {1984})}\BibitemShut {NoStop}%
\bibitem [{\citenamefont {Qian}\ and\ \citenamefont
  {Wang}(2024)}]{Dongheng2024}%
  \BibitemOpen
  \bibfield  {author} {\bibinfo {author} {\bibfnamefont {D.}~\bibnamefont
  {Qian}}\ and\ \bibinfo {author} {\bibfnamefont {J.}~\bibnamefont {Wang}},\
  }\href {https://doi.org/10.1103/PhysRevB.109.024301} {\bibfield  {journal}
  {\bibinfo  {journal} {Phys. Rev. B}\ }\textbf {\bibinfo {volume} {109}},\
  \bibinfo {pages} {024301} (\bibinfo {year} {2024})}\BibitemShut {NoStop}%
\bibitem [{\citenamefont {Sang}\ \emph {et~al.}(2023)\citenamefont {Sang},
  \citenamefont {Li}, \citenamefont {Hsieh},\ and\ \citenamefont
  {Yoshida}}]{yoshida2023}%
  \BibitemOpen
  \bibfield  {author} {\bibinfo {author} {\bibfnamefont {S.}~\bibnamefont
  {Sang}}, \bibinfo {author} {\bibfnamefont {Z.}~\bibnamefont {Li}}, \bibinfo
  {author} {\bibfnamefont {T.~H.}\ \bibnamefont {Hsieh}},\ and\ \bibinfo
  {author} {\bibfnamefont {B.}~\bibnamefont {Yoshida}},\ }\href
  {https://doi.org/10.1103/PRXQuantum.4.040332} {\bibfield  {journal} {\bibinfo
   {journal} {PRX Quantum}\ }\textbf {\bibinfo {volume} {4}},\ \bibinfo {pages}
  {040332} (\bibinfo {year} {2023})}\BibitemShut {NoStop}%
\bibitem [{\citenamefont {Feng}\ \emph {et~al.}(2023)\citenamefont {Feng},
  \citenamefont {Skinner},\ and\ \citenamefont {Nahum}}]{nahum2023}%
  \BibitemOpen
  \bibfield  {author} {\bibinfo {author} {\bibfnamefont {X.}~\bibnamefont
  {Feng}}, \bibinfo {author} {\bibfnamefont {B.}~\bibnamefont {Skinner}},\ and\
  \bibinfo {author} {\bibfnamefont {A.}~\bibnamefont {Nahum}},\ }\href
  {https://doi.org/10.1103/PRXQuantum.4.030333} {\bibfield  {journal} {\bibinfo
   {journal} {PRX Quantum}\ }\textbf {\bibinfo {volume} {4}},\ \bibinfo {pages}
  {030333} (\bibinfo {year} {2023})}\BibitemShut {NoStop}%
\bibitem [{\citenamefont {Bao}\ \emph {et~al.}(2020)\citenamefont {Bao},
  \citenamefont {Choi},\ and\ \citenamefont {Altman}}]{bao2020}%
  \BibitemOpen
  \bibfield  {author} {\bibinfo {author} {\bibfnamefont {Y.}~\bibnamefont
  {Bao}}, \bibinfo {author} {\bibfnamefont {S.}~\bibnamefont {Choi}},\ and\
  \bibinfo {author} {\bibfnamefont {E.}~\bibnamefont {Altman}},\ }\href
  {https://doi.org/10.1103/PhysRevB.101.104301} {\bibfield  {journal} {\bibinfo
   {journal} {Phys. Rev. B}\ }\textbf {\bibinfo {volume} {101}},\ \bibinfo
  {pages} {104301} (\bibinfo {year} {2020})}\BibitemShut {NoStop}%
\bibitem [{\citenamefont {Zabalo}\ \emph {et~al.}(2020)\citenamefont {Zabalo},
  \citenamefont {Gullans}, \citenamefont {Wilson}, \citenamefont
  {Gopalakrishnan}, \citenamefont {Huse},\ and\ \citenamefont
  {Pixley}}]{Zabalo2020}%
  \BibitemOpen
  \bibfield  {author} {\bibinfo {author} {\bibfnamefont {A.}~\bibnamefont
  {Zabalo}}, \bibinfo {author} {\bibfnamefont {M.~J.}\ \bibnamefont {Gullans}},
  \bibinfo {author} {\bibfnamefont {J.~H.}\ \bibnamefont {Wilson}}, \bibinfo
  {author} {\bibfnamefont {S.}~\bibnamefont {Gopalakrishnan}}, \bibinfo
  {author} {\bibfnamefont {D.~A.}\ \bibnamefont {Huse}},\ and\ \bibinfo
  {author} {\bibfnamefont {J.~H.}\ \bibnamefont {Pixley}},\ }\href
  {https://doi.org/10.1103/PhysRevB.101.060301} {\bibfield  {journal} {\bibinfo
   {journal} {Phys. Rev. B}\ }\textbf {\bibinfo {volume} {101}},\ \bibinfo
  {pages} {060301} (\bibinfo {year} {2020})}\BibitemShut {NoStop}%
\bibitem [{\citenamefont {Jian}\ \emph {et~al.}(2020)\citenamefont {Jian},
  \citenamefont {You}, \citenamefont {Vasseur},\ and\ \citenamefont
  {Ludwig}}]{chaoming2020}%
  \BibitemOpen
  \bibfield  {author} {\bibinfo {author} {\bibfnamefont {C.-M.}\ \bibnamefont
  {Jian}}, \bibinfo {author} {\bibfnamefont {Y.-Z.}\ \bibnamefont {You}},
  \bibinfo {author} {\bibfnamefont {R.}~\bibnamefont {Vasseur}},\ and\ \bibinfo
  {author} {\bibfnamefont {A.~W.~W.}\ \bibnamefont {Ludwig}},\ }\href
  {https://doi.org/10.1103/PhysRevB.101.104302} {\bibfield  {journal} {\bibinfo
   {journal} {Phys. Rev. B}\ }\textbf {\bibinfo {volume} {101}},\ \bibinfo
  {pages} {104302} (\bibinfo {year} {2020})}\BibitemShut {NoStop}%
\bibitem [{\citenamefont {Lavasani}\ \emph {et~al.}(2021)\citenamefont
  {Lavasani}, \citenamefont {Alavirad},\ and\ \citenamefont
  {Barkeshli}}]{Lavasani2020xea}%
  \BibitemOpen
  \bibfield  {author} {\bibinfo {author} {\bibfnamefont {A.}~\bibnamefont
  {Lavasani}}, \bibinfo {author} {\bibfnamefont {Y.}~\bibnamefont {Alavirad}},\
  and\ \bibinfo {author} {\bibfnamefont {M.}~\bibnamefont {Barkeshli}},\ }\href
  {https://doi.org/10.1038/s41567-020-01112-z} {\bibfield  {journal} {\bibinfo
  {journal} {Nature Phys.}\ }\textbf {\bibinfo {volume} {17}},\ \bibinfo
  {pages} {342} (\bibinfo {year} {2021})},\ \Eprint
  {https://arxiv.org/abs/2004.07243} {arXiv:2004.07243 [quant-ph]} \BibitemShut
  {NoStop}%
\bibitem [{\citenamefont {Sierant}\ and\ \citenamefont
  {Turkeshi}(2023)}]{turkeshi2023}%
  \BibitemOpen
  \bibfield  {author} {\bibinfo {author} {\bibfnamefont {P.}~\bibnamefont
  {Sierant}}\ and\ \bibinfo {author} {\bibfnamefont {X.}~\bibnamefont
  {Turkeshi}},\ }\href {https://doi.org/10.1103/PhysRevLett.130.120402}
  {\bibfield  {journal} {\bibinfo  {journal} {Phys. Rev. Lett.}\ }\textbf
  {\bibinfo {volume} {130}},\ \bibinfo {pages} {120402} (\bibinfo {year}
  {2023})}\BibitemShut {NoStop}%
\bibitem [{\citenamefont {Turkeshi}\ \emph {et~al.}(2020)\citenamefont
  {Turkeshi}, \citenamefont {Fazio},\ and\ \citenamefont
  {Dalmonte}}]{Turkeshi2020}%
  \BibitemOpen
  \bibfield  {author} {\bibinfo {author} {\bibfnamefont {X.}~\bibnamefont
  {Turkeshi}}, \bibinfo {author} {\bibfnamefont {R.}~\bibnamefont {Fazio}},\
  and\ \bibinfo {author} {\bibfnamefont {M.}~\bibnamefont {Dalmonte}},\ }\href
  {https://doi.org/10.1103/PhysRevB.102.014315} {\bibfield  {journal} {\bibinfo
   {journal} {Phys. Rev. B}\ }\textbf {\bibinfo {volume} {102}},\ \bibinfo
  {pages} {014315} (\bibinfo {year} {2020})}\BibitemShut {NoStop}%
\bibitem [{\citenamefont {Kennedy}\ and\ \citenamefont
  {Tasaki}(1992{\natexlab{a}})}]{Kennedy1992}%
  \BibitemOpen
  \bibfield  {author} {\bibinfo {author} {\bibfnamefont {T.}~\bibnamefont
  {Kennedy}}\ and\ \bibinfo {author} {\bibfnamefont {H.}~\bibnamefont
  {Tasaki}},\ }\href {https://doi.org/10.1007/BF02097239} {\bibfield  {journal}
  {\bibinfo  {journal} {Commun. Math. Phys.}\ }\textbf {\bibinfo {volume}
  {147}},\ \bibinfo {pages} {431} (\bibinfo {year}
  {1992}{\natexlab{a}})}\BibitemShut {NoStop}%
\bibitem [{\citenamefont {Kennedy}\ and\ \citenamefont
  {Tasaki}(1992{\natexlab{b}})}]{Kennedy1992_2}%
  \BibitemOpen
  \bibfield  {author} {\bibinfo {author} {\bibfnamefont {T.}~\bibnamefont
  {Kennedy}}\ and\ \bibinfo {author} {\bibfnamefont {H.}~\bibnamefont
  {Tasaki}},\ }\href {https://doi.org/10.1103/PhysRevB.45.304} {\bibfield
  {journal} {\bibinfo  {journal} {Phys. Rev. B}\ }\textbf {\bibinfo {volume}
  {45}},\ \bibinfo {pages} {304} (\bibinfo {year}
  {1992}{\natexlab{b}})}\BibitemShut {NoStop}%
\bibitem [{\citenamefont {Li}\ \emph {et~al.}(2023{\natexlab{a}})\citenamefont
  {Li}, \citenamefont {Oshikawa},\ and\ \citenamefont {Zheng}}]{Li23}%
  \BibitemOpen
  \bibfield  {author} {\bibinfo {author} {\bibfnamefont {L.}~\bibnamefont
  {Li}}, \bibinfo {author} {\bibfnamefont {M.}~\bibnamefont {Oshikawa}},\ and\
  \bibinfo {author} {\bibfnamefont {Y.}~\bibnamefont {Zheng}},\ }\href
  {https://doi.org/10.1103/PhysRevB.108.214429} {\bibfield  {journal} {\bibinfo
   {journal} {Phys. Rev. B}\ }\textbf {\bibinfo {volume} {108}},\ \bibinfo
  {pages} {214429} (\bibinfo {year} {2023}{\natexlab{a}})}\BibitemShut
  {NoStop}%
\bibitem [{\citenamefont {Li}\ \emph {et~al.}(2023{\natexlab{b}})\citenamefont
  {Li}, \citenamefont {Oshikawa},\ and\ \citenamefont {Zheng}}]{Li23_2}%
  \BibitemOpen
  \bibfield  {author} {\bibinfo {author} {\bibfnamefont {L.}~\bibnamefont
  {Li}}, \bibinfo {author} {\bibfnamefont {M.}~\bibnamefont {Oshikawa}},\ and\
  \bibinfo {author} {\bibfnamefont {Y.}~\bibnamefont {Zheng}},\ }\href
  {https://arxiv.org/abs/2307.04788} {\bibfield  {journal} {\bibinfo  {journal}
  {arXiv}\ } (\bibinfo {year} {2023}{\natexlab{b}})},\ \Eprint
  {https://arxiv.org/abs/2307.04788} {arXiv:2307.04788 [cond-mat.str-el]}
  \BibitemShut {NoStop}%
\bibitem [{\citenamefont {White}(1992)}]{white1992}%
  \BibitemOpen
  \bibfield  {author} {\bibinfo {author} {\bibfnamefont {S.~R.}\ \bibnamefont
  {White}},\ }\href {https://doi.org/10.1103/PhysRevLett.69.2863} {\bibfield
  {journal} {\bibinfo  {journal} {Phys. Rev. Lett.}\ }\textbf {\bibinfo
  {volume} {69}},\ \bibinfo {pages} {2863} (\bibinfo {year}
  {1992})}\BibitemShut {NoStop}%
\bibitem [{\citenamefont {Lu}\ \emph {et~al.}(2023)\citenamefont {Lu},
  \citenamefont {Zhang}, \citenamefont {Vijay},\ and\ \citenamefont
  {Hsieh}}]{Lu2023}%
  \BibitemOpen
  \bibfield  {author} {\bibinfo {author} {\bibfnamefont {T.-C.}\ \bibnamefont
  {Lu}}, \bibinfo {author} {\bibfnamefont {Z.}~\bibnamefont {Zhang}}, \bibinfo
  {author} {\bibfnamefont {S.}~\bibnamefont {Vijay}},\ and\ \bibinfo {author}
  {\bibfnamefont {T.~H.}\ \bibnamefont {Hsieh}},\ }\href
  {https://doi.org/10.1103/PRXQuantum.4.030318} {\bibfield  {journal} {\bibinfo
   {journal} {PRX Quantum}\ }\textbf {\bibinfo {volume} {4}},\ \bibinfo {pages}
  {030318} (\bibinfo {year} {2023})}\BibitemShut {NoStop}%
\bibitem [{\citenamefont {Verresen}\ \emph {et~al.}(2017)\citenamefont
  {Verresen}, \citenamefont {Moessner},\ and\ \citenamefont
  {Pollmann}}]{Verresen17}%
  \BibitemOpen
  \bibfield  {author} {\bibinfo {author} {\bibfnamefont {R.}~\bibnamefont
  {Verresen}}, \bibinfo {author} {\bibfnamefont {R.}~\bibnamefont {Moessner}},\
  and\ \bibinfo {author} {\bibfnamefont {F.}~\bibnamefont {Pollmann}},\ }\href
  {https://doi.org/10.1103/PhysRevB.96.165124} {\bibfield  {journal} {\bibinfo
  {journal} {Phys. Rev. B}\ }\textbf {\bibinfo {volume} {96}},\ \bibinfo
  {pages} {165124} (\bibinfo {year} {2017})}\BibitemShut {NoStop}%
\bibitem [{\citenamefont {Liu}\ \emph {et~al.}(2023)\citenamefont {Liu},
  \citenamefont {Tantivasadakarn}, \citenamefont {Slagle}, \citenamefont
  {Mross},\ and\ \citenamefont {Alicea}}]{Liu2023}%
  \BibitemOpen
  \bibfield  {author} {\bibinfo {author} {\bibfnamefont {Y.}~\bibnamefont
  {Liu}}, \bibinfo {author} {\bibfnamefont {N.}~\bibnamefont
  {Tantivasadakarn}}, \bibinfo {author} {\bibfnamefont {K.}~\bibnamefont
  {Slagle}}, \bibinfo {author} {\bibfnamefont {D.~F.}\ \bibnamefont {Mross}},\
  and\ \bibinfo {author} {\bibfnamefont {J.}~\bibnamefont {Alicea}},\ }\href
  {https://doi.org/10.1103/PhysRevB.108.184406} {\bibfield  {journal} {\bibinfo
   {journal} {Phys. Rev. B}\ }\textbf {\bibinfo {volume} {108}},\ \bibinfo
  {pages} {184406} (\bibinfo {year} {2023})}\BibitemShut {NoStop}%
\bibitem [{\citenamefont {Giamarchi}(2003)}]{Giamarchi2003}%
  \BibitemOpen
  \bibfield  {author} {\bibinfo {author} {\bibfnamefont {T.}~\bibnamefont
  {Giamarchi}},\ }\href
  {https://doi.org/10.1093/acprof:oso/9780198525004.001.0001} {\emph {\bibinfo
  {title} {{Quantum Physics in One Dimension}}}}\ (\bibinfo  {publisher}
  {Oxford University Press},\ \bibinfo {year} {2003})\BibitemShut {NoStop}%
\bibitem [{\citenamefont {Kane}\ and\ \citenamefont {Fisher}(1992)}]{Kane1992}%
  \BibitemOpen
  \bibfield  {author} {\bibinfo {author} {\bibfnamefont {C.~L.}\ \bibnamefont
  {Kane}}\ and\ \bibinfo {author} {\bibfnamefont {M.~P.~A.}\ \bibnamefont
  {Fisher}},\ }\href {https://doi.org/10.1103/PhysRevB.46.15233} {\bibfield
  {journal} {\bibinfo  {journal} {Phys. Rev. B}\ }\textbf {\bibinfo {volume}
  {46}},\ \bibinfo {pages} {15233} (\bibinfo {year} {1992})}\BibitemShut
  {NoStop}%
\bibitem [{\citenamefont {Fendley}\ \emph {et~al.}(2004)\citenamefont
  {Fendley}, \citenamefont {Sengupta},\ and\ \citenamefont {Sachdev}}]{FSS}%
  \BibitemOpen
  \bibfield  {author} {\bibinfo {author} {\bibfnamefont {P.}~\bibnamefont
  {Fendley}}, \bibinfo {author} {\bibfnamefont {K.}~\bibnamefont {Sengupta}},\
  and\ \bibinfo {author} {\bibfnamefont {S.}~\bibnamefont {Sachdev}},\ }\href
  {https://doi.org/10.1103/PhysRevB.69.075106} {\bibfield  {journal} {\bibinfo
  {journal} {Phys. Rev. B}\ }\textbf {\bibinfo {volume} {69}},\ \bibinfo
  {pages} {075106} (\bibinfo {year} {2004})}\BibitemShut {NoStop}%
\bibitem [{\citenamefont {Slagle}\ \emph {et~al.}(2021)\citenamefont {Slagle},
  \citenamefont {Aasen}, \citenamefont {Pichler}, \citenamefont {Mong},
  \citenamefont {Fendley}, \citenamefont {Chen}, \citenamefont {Endres},\ and\
  \citenamefont {Alicea}}]{SlagleTCI}%
  \BibitemOpen
  \bibfield  {author} {\bibinfo {author} {\bibfnamefont {K.}~\bibnamefont
  {Slagle}}, \bibinfo {author} {\bibfnamefont {D.}~\bibnamefont {Aasen}},
  \bibinfo {author} {\bibfnamefont {H.}~\bibnamefont {Pichler}}, \bibinfo
  {author} {\bibfnamefont {R.~S.~K.}\ \bibnamefont {Mong}}, \bibinfo {author}
  {\bibfnamefont {P.}~\bibnamefont {Fendley}}, \bibinfo {author} {\bibfnamefont
  {X.}~\bibnamefont {Chen}}, \bibinfo {author} {\bibfnamefont {M.}~\bibnamefont
  {Endres}},\ and\ \bibinfo {author} {\bibfnamefont {J.}~\bibnamefont
  {Alicea}},\ }\href {https://doi.org/10.1103/PhysRevB.104.235109} {\bibfield
  {journal} {\bibinfo  {journal} {Phys. Rev. B}\ }\textbf {\bibinfo {volume}
  {104}},\ \bibinfo {pages} {235109} (\bibinfo {year} {2021})}\BibitemShut
  {NoStop}%
\bibitem [{\citenamefont {O'Brien}\ and\ \citenamefont
  {Fendley}(2018)}]{Obrien}%
  \BibitemOpen
  \bibfield  {author} {\bibinfo {author} {\bibfnamefont {E.}~\bibnamefont
  {O'Brien}}\ and\ \bibinfo {author} {\bibfnamefont {P.}~\bibnamefont
  {Fendley}},\ }\href {https://doi.org/10.1103/PhysRevLett.120.206403}
  {\bibfield  {journal} {\bibinfo  {journal} {Phys. Rev. Lett.}\ }\textbf
  {\bibinfo {volume} {120}},\ \bibinfo {pages} {206403} (\bibinfo {year}
  {2018})}\BibitemShut {NoStop}%
\bibitem [{\citenamefont {Naus}\ \emph {et~al.}(tion)\citenamefont {Naus},
  \citenamefont {Liu}, \citenamefont {Murciano}, \citenamefont {Sala},
  \citenamefont {Endres},\ and\ \citenamefont {Alicea}}]{Naus2025}%
  \BibitemOpen
  \bibfield  {author} {\bibinfo {author} {\bibfnamefont {S.}~\bibnamefont
  {Naus}}, \bibinfo {author} {\bibfnamefont {Y.}~\bibnamefont {Liu}}, \bibinfo
  {author} {\bibfnamefont {S.}~\bibnamefont {Murciano}}, \bibinfo {author}
  {\bibfnamefont {P.}~\bibnamefont {Sala}}, \bibinfo {author} {\bibfnamefont
  {M.}~\bibnamefont {Endres}},\ and\ \bibinfo {author} {\bibfnamefont
  {J.}~\bibnamefont {Alicea}},\ }\href@noop {} {} (\bibinfo {year} {in
  preparation})\BibitemShut {NoStop}%
\bibitem [{\citenamefont {Cardy}(1989{\natexlab{a}})}]{Cardy1989}%
  \BibitemOpen
  \bibfield  {author} {\bibinfo {author} {\bibfnamefont {J.~L.}\ \bibnamefont
  {Cardy}},\ }\href {https://doi.org/10.1016/0550-3213(89)90521-X} {\bibfield
  {journal} {\bibinfo  {journal} {Nucl. Phys. B}\ }\textbf {\bibinfo {volume}
  {324}},\ \bibinfo {pages} {581} (\bibinfo {year}
  {1989}{\natexlab{a}})}\BibitemShut {NoStop}%
\bibitem [{\citenamefont {Affleck}(2000)}]{Affleck2000}%
  \BibitemOpen
  \bibfield  {author} {\bibinfo {author} {\bibfnamefont {I.}~\bibnamefont
  {Affleck}},\ }\href {https://doi.org/10.1088/0305-4470/33/37/301} {\bibfield
  {journal} {\bibinfo  {journal} {J. Phys. A: Math. Gen.}\ }\textbf {\bibinfo
  {volume} {33}},\ \bibinfo {pages} {6473} (\bibinfo {year}
  {2000})}\BibitemShut {NoStop}%
\bibitem [{\citenamefont {Cardy}(1989{\natexlab{b}})}]{CARDY_1989}%
  \BibitemOpen
  \bibfield  {author} {\bibinfo {author} {\bibfnamefont {J.~L.}\ \bibnamefont
  {Cardy}},\ }\href
  {https://doi.org/https://doi.org/10.1016/0550-3213(89)90521-X} {\bibfield
  {journal} {\bibinfo  {journal} {Nuclear Physics B}\ }\textbf {\bibinfo
  {volume} {324}},\ \bibinfo {pages} {581} (\bibinfo {year}
  {1989}{\natexlab{b}})}\BibitemShut {NoStop}%
\bibitem [{\citenamefont {Saleur}\ and\ \citenamefont
  {Bauer}(1989)}]{SALEUR_1989}%
  \BibitemOpen
  \bibfield  {author} {\bibinfo {author} {\bibfnamefont {H.}~\bibnamefont
  {Saleur}}\ and\ \bibinfo {author} {\bibfnamefont {M.}~\bibnamefont {Bauer}},\
  }\href {https://doi.org/https://doi.org/10.1016/0550-3213(89)90014-X}
  {\bibfield  {journal} {\bibinfo  {journal} {Nuclear Physics B}\ }\textbf
  {\bibinfo {volume} {320}},\ \bibinfo {pages} {591} (\bibinfo {year}
  {1989})}\BibitemShut {NoStop}%
\bibitem [{\citenamefont {Mong}\ \emph {et~al.}(2014)\citenamefont {Mong},
  \citenamefont {Clarke}, \citenamefont {Alicea}, \citenamefont {Lindner},\
  and\ \citenamefont {Fendley}}]{Mong_2014}%
  \BibitemOpen
  \bibfield  {author} {\bibinfo {author} {\bibfnamefont {R.~S.~K.}\
  \bibnamefont {Mong}}, \bibinfo {author} {\bibfnamefont {D.~J.}\ \bibnamefont
  {Clarke}}, \bibinfo {author} {\bibfnamefont {J.}~\bibnamefont {Alicea}},
  \bibinfo {author} {\bibfnamefont {N.~H.}\ \bibnamefont {Lindner}},\ and\
  \bibinfo {author} {\bibfnamefont {P.}~\bibnamefont {Fendley}},\ }\href
  {https://doi.org/10.1088/1751-8113/47/45/452001} {\bibfield  {journal}
  {\bibinfo  {journal} {Journal of Physics A: Mathematical and Theoretical}\
  }\textbf {\bibinfo {volume} {47}},\ \bibinfo {pages} {452001} (\bibinfo
  {year} {2014})}\BibitemShut {NoStop}%
\bibitem [{\citenamefont {You}\ \emph {et~al.}(2014)\citenamefont {You},
  \citenamefont {Bi}, \citenamefont {Rasmussen}, \citenamefont {Slagle},\ and\
  \citenamefont {Xu}}]{You14}%
  \BibitemOpen
  \bibfield  {author} {\bibinfo {author} {\bibfnamefont {Y.-Z.}\ \bibnamefont
  {You}}, \bibinfo {author} {\bibfnamefont {Z.}~\bibnamefont {Bi}}, \bibinfo
  {author} {\bibfnamefont {A.}~\bibnamefont {Rasmussen}}, \bibinfo {author}
  {\bibfnamefont {K.}~\bibnamefont {Slagle}},\ and\ \bibinfo {author}
  {\bibfnamefont {C.}~\bibnamefont {Xu}},\ }\href
  {https://doi.org/10.1103/PhysRevLett.112.247202} {\bibfield  {journal}
  {\bibinfo  {journal} {Phys. Rev. Lett.}\ }\textbf {\bibinfo {volume} {112}},\
  \bibinfo {pages} {247202} (\bibinfo {year} {2014})}\BibitemShut {NoStop}%
\bibitem [{\citenamefont {Di~Francesco}\ \emph {et~al.}(1997)\citenamefont
  {Di~Francesco}, \citenamefont {Mathieu},\ and\ \citenamefont
  {Senechal}}]{DiFrancesco}%
  \BibitemOpen
  \bibfield  {author} {\bibinfo {author} {\bibfnamefont {P.}~\bibnamefont
  {Di~Francesco}}, \bibinfo {author} {\bibfnamefont {P.}~\bibnamefont
  {Mathieu}},\ and\ \bibinfo {author} {\bibfnamefont {D.}~\bibnamefont
  {Senechal}},\ }\href {https://doi.org/10.1007/978-1-4612-2256-9} {\emph
  {\bibinfo {title} {{Conformal Field Theory}}}},\ Graduate Texts in
  Contemporary Physics\ (\bibinfo  {publisher} {Springer-Verlag},\ \bibinfo
  {address} {New York},\ \bibinfo {year} {1997})\BibitemShut {NoStop}%
\bibitem [{\citenamefont {Chim}(1996)}]{Chim1996}%
  \BibitemOpen
  \bibfield  {author} {\bibinfo {author} {\bibfnamefont {L.}~\bibnamefont
  {Chim}},\ }\href {https://doi.org/10.1142/S0217751X9600208X} {\bibfield
  {journal} {\bibinfo  {journal} {Int. J. Mod. Phys. A}\ }\textbf {\bibinfo
  {volume} {11}},\ \bibinfo {pages} {4491} (\bibinfo {year}
  {1996})}\BibitemShut {NoStop}%
\end{thebibliography}%
\newpage
\appendix
\onecolumngrid
\section{More properties of the canonical cluster state SPT}\label{app:moreproperties}

\subsection{Measurement on generic $\mathbb{Z}_2 \times \mathbb{Z}_2$ SPTs}

In Sec.~\ref{sec:canonical_cluster} we primarily reviewed properties of the $\mathbb{Z}_2 \times \mathbb{Z}_2$ cluster state SPT by specializing to the zero-correlation limit.  There we recalled how projective measurement of $X_{j,1}$ for all $j$ in the upper chain yields long-range ordered correlations with $|\braket{Z_{j,1} Z_{k,1}}| = 1$ in the lower chain.  Here we show that such measurement also induces long-range order in generic $\mathbb{Z}_2 \times \mathbb{Z}_2$ SPTs with finite correlation length.

Let $\ket{\psi_{\rm SPT}}$ denote a generic ground state in the $\mathbb{Z}_2 \times \mathbb{Z}_2$ SPT phase.  We continue to view $\ket{\psi_{\rm SPT}}$ as living on the ladder from Fig.~\ref{fig:cluster}(a); moreover, we assume that the two $\mathbb{Z}_2$ symmetries protecting the order are generated by the spin-flip operators in Eq.~\eqref{Z2generators}.  
By assumption, one of the SPT string order parameters evaluated with respect to $\ket{\psi_{\rm SPT}}$ gives
\begin{equation}\label{appeq:LRO}
\left\langle \left[{
    \begin{matrix}
        & X_{j+1} &\cdots& X_{k-1}&X_{k} &\\
        Z_j&  & & & Z_{k} \\
    \end{matrix}
    }\right] \right\rangle \neq 0~~{\rm as}~~|j-k| \to \infty.
\end{equation}
Applying the string order parameter directly to $\ket{\psi_{\rm SPT}}$, we can write the resulting wavefunction as
\begin{equation}\label{appeq:sopspt}
     \left[{
    \begin{matrix}
        & X_{j+1} &\cdots& X_{k-1}&X_{k} &\\
        Z_j&  & & & Z_{k} \\
    \end{matrix}
    }\right] \ket{\psi_{\rm SPT}} = c \ket{\psi_{\rm SPT}} + c_{\perp} \ket{\psi_{\rm SPT}^\perp},
\end{equation}
where $\ket{\psi_{\rm SPT}^\perp}$ lives in the orthogonal complement space of 
$\ket{\psi_{\rm SPT}}$, i.e., $\braket{\psi_{\rm SPT}^\perp|\psi_{\rm SPT}}=0$. 
The coefficients $c$ and $c_\perp$ generally depend on $j$ and $k$.  Compatibility with Eq.~\eqref{appeq:LRO}, however, requires that $c \neq 0$ for $|j-k| \to \infty$. 

Now consider measuring $\{X_{j,1}\}$ with outcome $ \bm{s} = \{s_j\}$.  Applying the associated projector
\begin{equation}
    \mathcal{P}_{\bm{s}} = \prod_{j=1}^{N} \left( \frac{1+s_j X_{j,1}}{2}\right)
\end{equation}
to the left of Eq.~\eqref{appeq:sopspt}, one obtains
\begin{equation}
    \begin{aligned}
    &\mathcal{P}_{\bm{s}}  \left[{
    \begin{matrix}
        & X_{j+1} &\cdots& X_{k-1}&X_{k} &\\
        Z_j&  & & & Z_{k} \\
    \end{matrix}
    }\right] \ket{\psi_{\rm SPT}} \\
    & = s_{j+1} \cdots s_{k} Z_{j,2} Z_{k,2} \mathcal{P}_{\bm{s}} \ket{\psi_{\rm SPT}} 
    \nonumber \\
    &= c \mathcal{P}_{\bm{s}} \ket{\psi_{\rm SPT}} + c_\perp \mathcal{P}_{\bm{s}} \ket{\psi_{\rm{SPT}}^\perp}.
    \end{aligned}
\end{equation}
The $Z_{j,2}Z_{k,2}$ correlator of the post-measurement state is therefore
\begin{equation}\label{eq:spt_result}
    \begin{aligned}
        \braket{Z_{j,2} Z_{k,2}}_{\bm{s}} &\equiv \frac{\braket{\psi_{\rm SPT} |\mathcal{P}_{\bm{s}} Z_{j,2} Z_{k,2} \mathcal{P}_{\bm{s}} |\psi_{\rm SPT}}}{\braket{\psi_{\rm SPT}|\mathcal{P}_{\bm{s}}|\psi_{\rm SPT}}} \\
        &= s_{j+1} \dots s_{k} \left(c + c_\perp \frac{\braket{\psi_{\rm SPT}|\mathcal{P}_{\bm{s}}|\psi_{\rm SPT}^\perp}}{\braket{\psi_{\rm SPT}|\mathcal{P}_{\bm{s}}|\psi_{\rm SPT}}}\right).\\
    \end{aligned}
\end{equation}
For the canonical cluster state, $c = 1$ and $c_\perp = 0$, so one immediately gets $\braket{Z_{j,2} Z_{k,2}}_{\bm{s}} = s_{j+1} \dots s_{k}$ in agreement with Eq.~\eqref{ZZcorr}. For generic SPT states, since $c$ remains non-zero even at $|j-k|\rightarrow \infty$, $|\langle Z_{j,2} Z_{k,2} \rangle_{\bm{s}}|$, or alternatively $s_{j+1} \dots s_{k} \braket{Z_{j,2} Z_{k,2}}_{\bm{s}}$, generically exhibits long-range order in the post-measurement state, regardless of the particular measurement outcome ${\bm s}$. 

The decoding protocol reviewed in Sec.~\ref{sec:canonical_cluster} also extends straightforwardly to the finite-correlation-length regime.  Consider again the generic SPT string order parameter expectation value from Eq.~\eqref{appeq:LRO}.  Upon inserting $X$-basis resolutions of the identity for chain 1, one can exactly write 
\begin{equation}
\begin{aligned}\label{appeq:decoding}
     & \left\langle \left[{
    \begin{matrix}
        & X_{j+1} &\cdots& X_{k-1}&X_{k} &\\
        Z_j&  & & & Z_{k} \\
    \end{matrix}
    }\right] \right\rangle = \sum_{\bm{s}} p_{\bm{s}} \braket{Z_{j,2} Z_{k,2}}_{\bm{s}} s_{j+1} \cdots s_{k}. \\
\end{aligned}
\end{equation}
Since the left side exhibits long-range order (again by assumption), so too must the right side, i.e., 
\begin{equation}
    \sum_{\bm{s}} p_{\bm{s}} \braket{Z_{j,2} Z_{k,2}}_{\bm{s}}s_{j+1}\cdots s_{k} \neq 0~~{\rm as}~~|j-k| \to \infty.
\end{equation}
The finite correlation length merely reduces the nonzero constant from its maximal value of 1; cf.~Eq.~\eqref{nontrivial_ave}.  

\subsection{Weak measurements in a tilted basis}\label{app:weakandtilted}

For the remainder of this Appendix we return to the zero-correlation-length SPT ground state, $\ket{\psi_0}$, defined with periodic boundary conditions. Suppose now that we \emph{weakly} measure the `tilted' operator $\cos{\omega} X_{j,1} + \sin{\omega} Z_{j,1}$ for all sites $j$ in the top chain, post-selecting for simplicity on a spatially uniform measurement outcome.  We are interested in the fate of the two-point $ZZ$ correlator for the lower chain in this generalized scenario.  The correlator explicitly reads 
\begin{equation}   \braket{Z_{j,2}Z_{k,2}}_{\rm uni}=\frac{\braket{\psi_0|Z_{j,2} Z_{k,2}e^{2\beta\sum_j (\cos{\omega} X_{j,1} + \sin{\omega} Z_{j,1}) }|\psi_0}}{\braket{\psi_0|e^{2\beta \sum_j (\cos{\omega} X_{j,1} + \sin{\omega} Z_{j,1})} |\psi_0}},
\label{tiltedZZ}
\end{equation}
where $\beta$ parameterizes the weak measurement strength.  Using the fact that 
\begin{equation}
    Z_{j,2} X_{j+1,1} \dots X_{k,1} Z_{k,2} = 1
\end{equation}
when acting on the zero-correlation-length ground state, we can rewrite Eq.~\eqref{tiltedZZ} in terms of expectation values of operators exclusively acting on the top chain:
\begin{equation}   \braket{Z_{j,2}Z_{k,2}}_{\rm uni}=\frac{\braket{\psi_0|\left(\prod_{i = j+1}^{k} X_{i,1}\right)e^{2\beta\sum_j (\cos{\omega} X_{j,1} + \sin{\omega} Z_{j,1}) }|\psi_0}}{\braket{\psi_0|e^{2\beta \sum_j (\cos{\omega} X_{j,1} + \sin{\omega} Z_{j,1})} |\psi_0}},
\label{tiltedZZ2}
\end{equation}

Reference~\cite{Lee2022} showed that such expectation values in the $\ket{\psi_0}$ state are nonvanishing only if the operator under consideration commutes with all $ZXZ$ terms in the Hamiltonian; for an operator living exclusively on one chain, as in the case above, it must be either the identity or a symmetry generator. 
We then obtain
\begin{equation}
\begin{aligned}
    \braket{Z_{j,2}Z_{k,2}}_{\rm uni} &= \frac{(\tanh 2\beta\cos{\omega})^{|j-k|} + (\tanh 2\beta\cos{\omega})^{N-|j-k|}}{1 + (\tanh{2\beta} \cos{\omega})^N},
\end{aligned}
\label{ZZuni}
\end{equation}
where $N$ is the number of sites in each chain and we used the fact that $\langle \prod_{i = 1}^N X_{i,1}\rangle = +1$. 
With $\omega = 0$ and in the projective-measurement limit $\beta \rightarrow \infty$, the right side of Eq.~\eqref{ZZuni} becomes unity, recovering the standard long-range-order result.  Suppose instead that $\beta$ is finite and/or $\omega \neq 0$.  Taking $|j-k| \ll N$ and then $N \rightarrow \infty$, here we find
\begin{equation}
\begin{aligned}
    \braket{Z_{j,2}Z_{k,2}}_{\rm uni} &\approx  (\tanh 2\beta \cos{\omega})^{|j-k|} \equiv e^{-|j-k|/\xi}.
\end{aligned}
\end{equation}
Above we defined a finite correlation length
\begin{equation}
    \xi = \frac{1}{\ln\left(\frac{1}{\tanh2\beta \cos\omega}\right)}
\end{equation}
that signals the exponential decay of the $ZZ$ correlator.  
Therefore, long-range order only appears when the measurement is projective \emph{and} when the tilt angle is $\omega = 0$~\cite{Lee2022}.

\section{Perturbative approach in the projective measurement limit}\label{app:pert}
\begin{figure}[h]
    \centering
    \includegraphics[width=10cm]{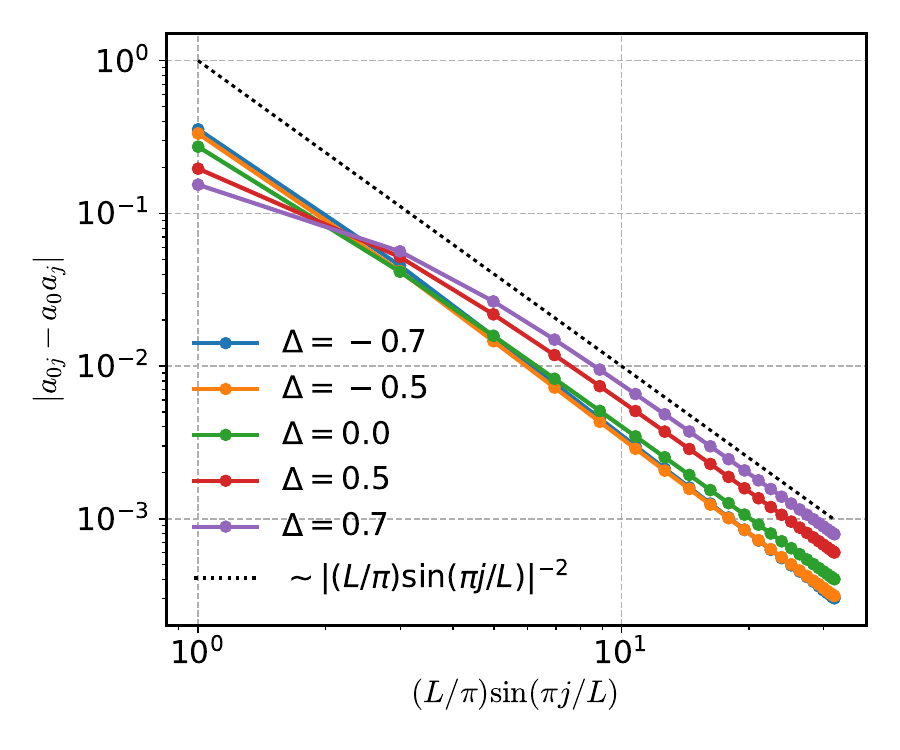}
    \caption{The strange correlator $a_{0j} - a_{0}a_{j}$ of an $\mathcal{XXZ}$ chain as a function of chord length $(L/\pi) \sin(\pi j/L)$. The data for different interaction strength $\Delta$ is obtained by DMRG simulations with periodic boundary conditions and system size $L = 50$, using bond dimension $\chi = 300$. The BCFT analysis of Sec.~\ref{sec:LL} predicts that $a_{0j} - a_{0}a_{j}$ decays as a power law with exponent $-2$ for any $\Delta \in (-1,1)$, which is consistent with the numerical data.}
    \label{appfig:Vjk}
\end{figure}

In this Appendix, we focus on the $\mathcal{X}$ measurement,  $e^{\sum_j \beta \mathcal{X}_j}$, in a single $\mathcal{XXZ}$ chain
\begin{equation}\label{eq:HamXXZapp}
    H = -\sum_{j} (\mathcal{X}_j \mathcal{X}_{j+1} + \mathcal{Y}_{j} \mathcal{Y}_{j+1} + \Delta \mathcal{Z}_{j} \mathcal{Z}_{j+1}).
\end{equation}
Since the $\mathcal{X}$ measurement and post-selection of the uniform measurement outcome induce a relevant perturbation for every $\Delta \in (-1,1)$, any weak measurement with finite measurement strength $\beta$ flows to the projective-measurement fixed point with $\beta \to \infty$. Therefore, we can expand the wavefunction around the projective limit \cite{Sala2024}
\begin{equation}\label{eq:perturbation}
\begin{aligned}
    \ket{\psi}_{\rm uni} \propto e^{\sum_j \beta \mathcal{X}_j}\ket{\psi_c} \propto \ket{\Omega} + u \sum_j a_j \ket{j} + u^2 \sum_{j,k} a_{jk} \ket{j,k} + O(u^3)\\
\end{aligned}
\end{equation}
where $u=\exp(-2\beta)$, $\ket{\psi_c}$ is the ground state of the Hamiltonian \eqref{eq:HamXXZapp}, $\ket{\psi}_{\rm uni}$ is the post-measurement wavefunction, $\ket{\Omega}$ is the product state $\bigotimes_j\ket{\mathcal{X}_j=+1}$ corresponding to the uniform outcome, $\ket{j,k,l,\dots} = \mathcal{Z}_j \mathcal{Z}_k \mathcal{Z}_l \dots \ket{\Omega}$, and $a_{j,k,l,\dots}$ is the \textit{strange correlator}
\begin{equation}
    a_{j,k,l,\dots} \equiv \frac{\braket{\Omega|\mathcal{Z}_j \mathcal{Z}_k \mathcal{Z}_l \dots|\psi_c}}{\braket{\Omega|\psi_c}}.
\end{equation}
The strange correlator has been studied for SPT states~\cite{You14}, but here it is computed for a critical state. In general $\ket{\psi_0}$ can be any critical state, and one can replace $\mathcal{X}$ by any local operator $\mathcal{O}$ and $\mathcal{Z}$ by $\mathcal{O^\perp}$ such that $\{\mathcal{O},\mathcal{O^\perp} \} = 0$. In the large $\beta$ (i.e. $u\ll 1$) limit, the $O(u^3)$ terms in Eq.~\eqref{eq:perturbation} can be neglected and the correlation functions can be computed from the truncated wavefunction
\begin{equation}\label{eq:truncatedWF}
    \ket{\psi_{\rm trunc}} = \ket{\Omega} + u \sum_j a_j \ket{j} + u^2 \sum_{j,k} a_{jk} \ket{j,k}.
\end{equation}
In the cases of interest, we observe that $a_j$ and $a_{jk}$ take the following form,
\begin{equation}\label{eq:ajk}
    a_j = m, \quad a_{jk}-a_j a_k \sim \frac{const.}{|j-k|^\eta}.
\end{equation}
We remark that the expansion \eqref{eq:perturbation} is well-behaved only if $\eta>1$: for the examples analyzed in this work, this is always true when the measurement corresponds to a relevant operator.
The correlators of the truncated wavefunction can thus be computed as, 
\begin{equation}\label{eq:corr_truncatdWF}
\begin{aligned}
\braket{ \mathcal{X}_j}_{\rm trunc} &\sim 1 - c_1 u^2 m^2 \quad   (\text{if}~ m \neq 0)\\
& \sim 1-c_2 u^4 \qquad~   (\text{if}~ m = 0) \\
    \braket{\mathcal{X}_0 \mathcal{X}_j}_{c, \rm trunc} &\sim u^4 m^4 |j|^{-\eta} \quad~  (\text{if}~m \neq 0)\\
    & \sim u^4 |j|^{-2\eta}  \qquad~   (\text{if}~ m = 0) \\
    \braket{\mathcal{Y}_0 \mathcal{Y}_j}_{\rm trunc} &\sim \braket{\mathcal{Z}_0 \mathcal{Z}_j}_{\rm trunc} \sim u^2 |j|^{-\eta}\\
\end{aligned}
\end{equation}
where $c_{1,2}$ are non-universal constants.
For the ground state of the $\mathcal{XXZ}$ model, $m=0$. The exponents above do not depend on the measurement strength and should be consistent with the power-law decay of correlators found from BCFT. Comparing Eq.~\eqref{eq:corr_truncatdWF} with Eq.~\eqref{eq:YY2}, we conclude that the exponent of $a_{jk}-a_j a_k$ should be exactly $\eta = 2$ in the $\mathcal{XXZ}$ model. 
In Fig.~\ref{appfig:Vjk} we numerically compute $a_{jk}-a_j a_k$ using DMRG and indeed observe $\eta = 2$ for different $\Delta$.

In fact, this perturbative approach is generic and $\ket{\psi_c}$ can be any critical state. One can also compute the second R\'enyi entropy of the truncated wavefunction in Eq.~\eqref{eq:corr_truncatdWF},
\begin{equation}
    S_A^{(2)} \simeq 2u^4 \sum_{j \in A} \sum_{k \in \bar{A}} (a_{jk} - a_{j} a_{k})^2 + O(u^6),
\end{equation}
where $A$ is a subsystem consisting of $l$ consecutive sites and $\bar{A}$ is the complement. It is shown in Ref.~\cite{Sala2024} that the scaling of $S_A^{(2)}$ depends on the exponent of the strange correlator, 
\begin{equation}
    S_A^{(2)} \sim 
    \begin{cases}
    \text{area law}. & \text{if } \eta > 1,\\
    \ln(l (L-l)/l) & \text{if } \eta = 1.
  \end{cases}
\end{equation}
The perturbative approach breaks down when $\eta < 1$.

\section{Non-linear observables}\label{app:nonlinear} 

In this Appendix, we derive the relation between non-linear observables (Eq.~\eqref{eq:replica}) and expectation values of the replica system. The weak measurement on site $j$ is described by the Kraus operators,
\begin{equation}
    K_{s_j = \pm} = \frac{e^{s_j \beta O_j}}{\sqrt{2 \cosh{2\beta}}}
\end{equation}
where $O_j$ is a linear combination of Pauli matrices s.t. $O_j^2 = \mathbb{I}$ and $\beta$ is the measurement strength. Let's consider weak measurements on $N$ sites of the initial state $\ket{\psi_{\Delta}}$ with outcome $\bm{s} = \{s_1, \dots, s_N \}$. The post-measurement state is
\begin{equation}
    \ket{\psi_{\bm{s}}} = \frac{K_{\bm{s}} \ket{\psi_{\Delta}}}{\sqrt{\braket{K^2_{\bm{s}} }_0 } }
\end{equation}
where $K_{\bm{s}} = \prod_j K_{s_j}$ and $\braket{\cdot}_0 \equiv \braket{\psi_{\Delta} | \cdot |\psi_{\Delta}}$. We are interested in the non-linear observable 
\begin{equation}
    \frac{\sum_{\bm{s}} p_{\bm{s}}^n \braket{\Gamma}_{\bm{s}}}{\sum_{\bm{s}} p_{\bm{s}}^n} = \frac{\sum_{\bm{s}} \braket{K_{\bm{s}}^2}_0^{n-1} \braket{K_{\bm{s}}^2 \Gamma}_0 }{\sum_{\bm{s}} \braket{K_{\bm{s}}^2}_0^n}
\end{equation}
where $p_{\bm{s}} = \braket{K^2_{\bm{s}}}_0$ is the probability of the outcome $\bm{s}$ and $\Gamma$ is the observable of interest such that $[\Gamma, K_{\bm{s}}] = 0$.

On the other hand, let us consider the $n$-replica version of the initial wavefunction, $\ket{\psi_{\Delta}^n} \equiv \ket{\psi_{\Delta}^{(1)}} \otimes \cdots \otimes \ket{\psi_{\Delta}^{(n)}}$. The superscript refers to the index of replicas. The terms in the numerator and denominator can be expressed in terms of expectation values of the replica wavefunction. For example,
\begin{equation}
    \begin{aligned}
    \sum_{\bm{s}} \braket{K_{\bm{s}}^2}_0^n &= \sum_{\bm{s}} \braket{\psi_{\Delta}^n|\left(K_{\bm{s}}^2\right)^{(1)}\otimes\cdots\otimes \left(K_{\bm{s}}^2\right)^{(n)}|\psi_{\Delta}^n} \\
    &=  \braket{\psi_{\Delta}^n| \sum_{\bm{s}} \left(K_{\bm{s}}^2\right)^{(1)}\otimes\cdots\otimes \left(K_{\bm{s}}^2\right)^{(n)}|\psi_{\Delta}^n}\\
    &= \braket{\psi_{\Delta}^n| \prod_j \left[\sum_{s_j = \pm} \left(K_{s_j}^2\right)^{(1)}\otimes\cdots\otimes \left(K_{s_j}^2\right)^{(n)} \right]|\psi_{\Delta}^n} \\
\end{aligned}
\end{equation}
and also,
\begin{equation}
    \begin{aligned}
    &\sum_{\vec{s}} \braket{K_{\vec{s}}^2}_0^{n-1} \braket{K_{\vec{s}}^2 \Gamma}_0 = \braket{\psi_{\Delta}^n| \Gamma^{(1)} \prod_j \left[\sum_{s_j = \pm} \left(K_{s_j}^2\right)^{(1)}\otimes\cdots\otimes \left(K_{s_j}^2\right)^{(n)} \right]|\psi_{\Delta}^n}. \\
\end{aligned}
\end{equation}
In the $n=2$ case, we have
\begin{equation}
    \sum_{s_j = \pm} \left(K_{s_j}^2\right)^{(1)}\otimes\left(K_{s_j}^2\right)^{(2)} \propto e^{2\beta' O_j^{(1)} O_j^{(2)}},
\end{equation}
where $\tanh(2\beta') = \tanh^2(2\beta)$.
Therefore we have proven Eq.~\eqref{eq:replica},
\begin{equation}
    \frac{\sum_{\bm{s}} p_{\bm{s}}^2 \braket{\Gamma}_{\bm{s}}}{\sum_{\bm{s}} p_{\bm{s}}^2} = \frac{\braket{\psi_{\Delta}^2| \Gamma^{(1)} \prod_j e^{2\beta' O_j^{(1)} O_j^{(2)}}  |\psi_{\Delta}^2}}{\braket{\psi_{\Delta}^2|\prod_j e^{2\beta' O_j^{(1)} O_j^{(2)}}  |\psi_{\Delta}^2}}.
\end{equation}

\section{Correlation functions of the post-measurement state from BCFT}\label{app:correlators}
The low-energy and long-distance theory of an $\mathcal{XXZ}$ chain is described by the TLL theory whose action is reported in Eq. \eqref{eq:TLLaction}. This model is also known as a compact boson, and the compactification radius is related to the Luttinger parameter $K$. The action can be expressed in terms in terms of the field $\theta$, or its dual, $\phi$. Let us start by analyzing some relevant features of the $\theta$ field, which can be helpful to derive part of the results we have presented in the main text.

An important class of primary fields for this model are the vertex operators, $e^{i a \theta}$ (here we implicitly assume that all the fields are normal ordered), 
whose expectation value is given by
\begin{equation} \label{appeq:vertex}
    \braket{\prod_{j} e^{i a_j \theta(z_j, \Bar{z}_j)}} = \prod_{j<k} |z_j - z_k|^{a_j a_k/(2K)} ~ (\text{if}~ \sum_{j} a_j = 0)
\end{equation}
where we have used the complex coordinates $z = x + i \tau$, $\bar{z} = x -i \tau$.

As we discussed in Sec. \ref{sec:LL}, the weak measurement of $X\sim \cos(\theta)$ induces a defect-line perturbation at $\tau = 0$ which is relevant in the RG analysis for $K>1/4$. It effectively cuts the full 1+1D plane into two halves and pins $\theta(\tau^*) = 0$ at all $x$ and at a $\beta$-dependent $\tau^*$. To simplify the following computations, in this Appendix we assume that the field will be pinned at $\tau^*=0$, having in mind that the final result could be always translated by $\tau\to \tau-\tau^*$.

Thus, the correlation functions of the post-measurement wavefunction in the long-distance limit can be computed by studying the Luttinger liquid on the upper half-plane (UHP), $\tau>0$, with Dirichlet boundary condition, $\theta(x,\tau = 0) = 0$. Under the parity transformation $\tau \to -\tau$, in the presence of DBC, the $\theta$ field transforms as
\begin{equation}
    \theta (z, \Bar{z}) \to - \theta(\Bar{z}, z).
\end{equation}
It was shown by Cardy~\cite{Cardy1984} that an $n$-point function on the UHP satisfies the same differential equation as the (holomorphic) $2n$-point function on the entire complex plane $\mathbb{C}$.
By exploiting this doubling trick, we can find that the vertex operators at $z_1 = x_1 + i\tau_1$ and $z_2 = x_2 + i\tau_2$ satisfy~\cite{DiFrancesco}
\begin{equation}\label{appeq:theta}
\begin{aligned}
    &\braket{e^{i a \theta(z_1, \Bar{z}_1)} e^{-i b \theta(z_2, \Bar{z}_2)}}_{\rm UHP} = \braket{e^{ia \theta(z_1)} e^{-ia \theta(\bar{z}_1)} e^{-ib \theta(z_2)} e^{ib \theta(\bar{z}_2)}}_\mathbb{C}\\
    &=\frac{1}{(\textrm{Im} z_1)^{a^2/(4K)} (\textrm{Im} z_2)^{b^2/(4K)}} \left|\frac{z_1-z_2}{z_1-\bar{z}_2}\right|^{-a b/(2K)}\\
    &= \tau_1^{-a^2/(4K)} \tau_2^{-b^2/(4K)} \left[ 1 - \frac{4\tau_1 \tau_2}{(\tau_1 + \tau_2)^2 + (x_1 - x_2)^2} \right]^{-ab/(4K)}.\\
\end{aligned}
\end{equation}
In the second line, we have used the holomorphic part of Eq.~\eqref{appeq:vertex}. From the result above, we can also easily derive the one-point correlator
\begin{equation}   \braket{e^{ia\theta(0,\tau)}}_{\rm DBC}= \tau^{-a^2/(4K)}. 
\end{equation}

As we mentioned earlier, the action Eq.~\eqref{eq:TLLaction} is also written in a dual representation, in terms of the field $\phi$.
If $\theta(z,\bar{z})$ satisfies DBC, the dual field $\phi(z,\bar{z})$ obeys Neumann boundary condition and, under the parity transformation $\tau \to -\tau$, behaves as
\begin{equation}
    \phi (z, \Bar{z}) \to  \phi(\Bar{z}, z).
\end{equation}
Therefore, the correlation function of vertex operators $e^{ia\phi}$ in the presence of NBC reads~\cite{DiFrancesco}
\begin{equation}\label{appeq:phi}
\begin{aligned}
    &\braket{e^{i a \phi(z_1, \Bar{z}_1)} e^{-i b \phi(z_2, \Bar{z}_2)}}_{\rm UHP} = \braket{e^{ia \phi(z_1)} e^{ia \phi(\bar{z}_1)} e^{-ib \phi(z_2)} e^{-ib \phi(\bar{z}_2)}}_\mathbb{C}\\
    &=\delta_{ab} \left( \frac{\textrm{Im} z_1 ~\textrm{Im} z_2}{|z_1-z_2|^2 |z_1-\bar{z}_2|^2} \right) ^{a^2 K/4}\\
    &= \delta_{ab} \left\{ \frac{\tau_1 \tau_2}{[(\tau_1 - \tau_2)^2 + (x_1 - x_2)^2] [(\tau_1 + \tau_2)^2 + (x_1 - x_2)^2]} \right\}^{\frac{a^2 K}{4}},\\
\end{aligned}
\end{equation}
where again we have used the holomorphic part of Eq.~\eqref{appeq:vertex}.
The Eq. \eqref{appeq:theta} has been used in Sec. \ref{sec:LL} to derive the post-measuremnet correlation function \eqref{eq:bcft_tt} and they were also useful to extrapolate a relationship between the measurement strength $\beta$ and the imaginary-time evolution $\tau$. In particular, we can perform an expansion around $\tau=0$, then shift the final result as $\tau\to\tau-\tau^*$, and finally set $\tau=0$ to recover Eq. \eqref{eq:bcft_tt}.

\begin{figure}[h]
    \centering
    \includegraphics[width=0.8\linewidth]{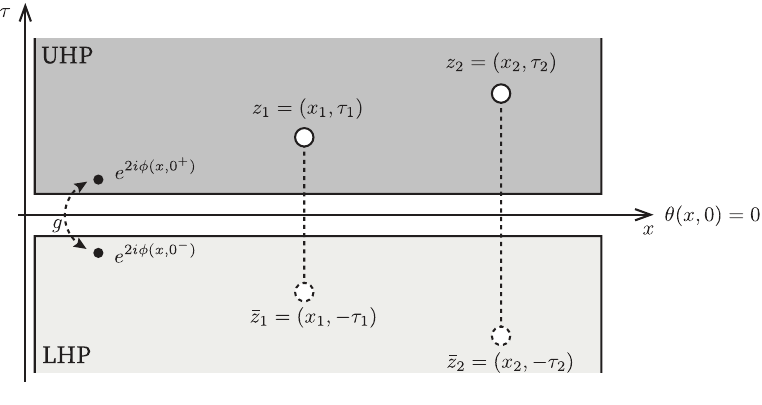}
    \caption{BCFT and method of images. The measurement pins $\theta(0,\tau=0)=0$, and the remnant interaction between UHP and LHP is modeled by $g \cos[2\phi(x, 0^+) - 2 \phi(x, 0^-)]$, as depicted by the black dots on the left. We are interested in the two-point correlators at $z_1$ and $z_2$ in the UHP, which can be computed as holomorphic four-point correlators involving $z_1, z_2, \bar{z}_1, \bar{z}_2$ as shown by the hollow circles~\cite{Cardy1984}. The two circles $\bar{z}_1$ and $\bar{z}_2$ in the LHP represent the mirror images of $z_1$ and $z_2$.}
    \label{fig:bcft}
\end{figure}
Despite BCFT provides a powerful tool to compute the correlation functions in the presence of relevant perturbations, we need to be very careful about the periodicity of the weak measurement in Eq. \eqref{eq:action_imp}. In the large measurement strength limit, the field $\theta $ can tunnel from one minimum of the cosine to the other \cite{Giamarchi2003}, so the field $\theta$ is not really blocked. A similar problem is already discussed in \cite{Garratt2023} computing the correlator $\langle\partial_x\phi(x)\partial_{x'}\phi(x')\rangle$ in the presence of the weak measurement $\int_{x} \cos[2\phi(x)]$. Also in that case, a na\"ive analysis based on BCFT would predict that $\langle\partial_x\phi(x)\partial_{x'}\phi(x')\rangle\sim |x-x'|^{-4}$, while Ref. \cite{Garratt2023} has shown that $\langle\partial_x\phi(x)\partial_{x'}\phi(x')\rangle\sim |x-x'|^{-2/K}$. This result can be obtained by taking into account that $\cos[2\phi(x)]$ is maximized for $\phi(x)=p \pi$, with $p$ integer and the tunneling among the different saddle points of the action determines the behavior of the correlator $\langle\partial_x\phi(x)\partial_{x'}\phi(x')\rangle$.

By taking into account the necessity of a careful treatment of the cosine perturbation in Eq. \eqref{eq:action_imp}, in the second part of this section, we compute the two-point correlation function $\braket{\mathcal{Y}_0\mathcal{Y}_x}\sim \braket{\sin(\theta(0))\sin(\theta(x))}$. 
If we map the measurement problem into the Kane and Fisher study of an
isolated defect in a TLL~\cite{Kane1992}, we can use their analysis of the strong coupling regime in order to get the correct behavior of our correlation functions in the presence of the relevant perturbation \eqref{eq:action_imp}. In the defect problem, the impurity is spatially local, i.e. it acts at all (imaginary) time steps but $x=0$. 
If the field $\theta(x=0)$ was totally blocked, the chain would be cut into two disconnected pieces. As we mentioned above, this cannot be true if the coupling of the impurity is not infinite, so to analyze the strong coupling  regime, one has to take into account the possibility of adding a tunneling term jumping across the impurity. Therefore, in our measurement problem, the effective action that we should consider when $\theta$ is pinned around $\theta(x,\tau=0)=0$ is  
\begin{equation}\label{eq:tunnel}
S' = \frac{K}{2\pi} \int_{x} \int_0^{\infty}d\tau \left[ (\partial_\tau \theta_{1})^2 + (\partial_x \theta_{1})^2 \right]+\frac{K}{2\pi} \int_{x} \int_{-\infty}^0 d\tau \left[ (\partial_\tau \theta_{2})^2 + (\partial_x \theta_{2})^2 \right]
 + g\int_{x} O(x),
\end{equation}
where $O(x)=\cos[2\phi_1(x)-2\phi_2(x)]$, $\phi_1(x)=\phi(x , 0^+)$
and $\phi_2(x)=\phi(x,   0^-)$ 
are copies of the fields on the upper and lower half plane, respectively, and similarly for $\theta_j$, $j=1,2$. We stress that the choice of the Hermitian perturbation $O(x)$ is compatible with the symmetries of our problem and with the fact that it is the first irrelevant operator for $K>1/4$, which is the regime where the weak measurement \eqref{eq:action_imp} is relevant.

Let us analyse how the tunneling term present in Eq. \eqref{eq:tunnel} changes the behavior of the correlation functions.

By expanding the action, we get that 
\begin{equation}\label{eq:tunneltheta}
 \begin{split}
     \braket{\sin[\theta_1(z)] \sin[\theta_1(z')]}= \braket{\sin[\theta_1(z)] \sin[\theta_1(z')]}_{g=0}+g^2\int_{x_1,x_2} \braket{\sin[\theta_1(z)] \sin[\theta_1(z')]O(x_1) O(x_2)}_{g=0}+o(g^4).
 \end{split}   
\end{equation}
From the computation of the correlation functions close to a boundary in Eq.~\eqref{eq:bcft_tt}, we learn that $\braket{\sin[\theta_1(x)] \sin[\theta_1(x')]}_{g=0}\sim |x'-x|^{-2}$ and the scaling dimension of $O(x)$ is $4K$. 
Computing the integral in Eq. \eqref{eq:tunneltheta} is far from being trivial. We observe that the integrand is proportional to 
\begin{equation}
    \braket{e^{i[\theta_1(z)-\theta_1(z')]}e^{2i[\phi_1(x_1)-\phi_1(x_2)]}e^{-2i[\phi_2(x_1)-\phi_2(x_2)]}}= \braket{e^{i[\theta_1(z)-\theta_1(z')]}e^{2i[\phi_1(x_1)-\phi_1(x_2)]}}\braket{e^{-2i[\phi_2(x_1)-\phi_2(x_2)]}}.
\end{equation}
The correlator $\braket{e^{-2i[\phi_2(x_1)-\phi_2(x_2)]}}$ can be found in Eq.~\ref{appeq:phi} and it is $|x_1-x_2|^{-4K}$. However, the first correlator is less trivial to be computed. We can ask what is the operator product expansion between $e^{i\theta}$ and $e^{i\phi}$ and, to address this question, it can be useful to write down both $\phi$ and $\theta$ in terms of the right and left chiral components. In particular, $\phi=\varphi_R+\varphi_L$ and $\theta=\varphi_R-\varphi_L$. 
We can use that (we are implicitly assuming that all the operators of the form $e^{i\alpha \varphi_{L/R}}$ are normal ordered \cite{DiFrancesco})
\begin{equation}\label{eq:ope}
    e^{i\alpha\varphi_L(x)} e^{i\beta\varphi_L(x')}= e^{i\alpha \varphi_L(x)+i\beta \varphi_L(x')} (x-x')^{\alpha \beta K/2},
\end{equation}
such that, by multiplying the chiral and antichiral parts again, we get 
\begin{equation}
    e^{i[\theta_1(z)-\theta_1(z')]}e^{2i[\phi_1(x_1)-\phi_1(x_2)]}\sim \frac{1}{(z-x_1)^K(z-x_2)^K(z'-x_1)^K(z'-x_2)^K}e^{i[\varphi_R(x_2)-\varphi_L(x_2)]}e^{-i[\varphi_R(x_1)-\varphi_L(x_1)]}.
\end{equation}
By recalling that $\varphi_R-\varphi_L=\theta$, and $\braket{e^{i\theta(x_1)}e^{-i\theta(x_2)}}\sim |x_1-x_2|^{-2}$ we have to evaluate the following integral 
\begin{multline}
    \braket{\sin[\theta_1(z)] \sin[\theta_1(z')]}= \braket{\sin[\theta_1(z)] \sin[\theta_1(z')]}_{g=0}\\+g^2\int_{x_1,x_2}\frac{|x_1-x_2|^{-2-4K}}{(z-x_1)^K(z-x_2)^K(z'-x_1)^K(z'-x_2)^K}+o(g^4)
\end{multline}    
Without explicitly solving the integral above, we observe that the scaling dimension of the second term is $|x-x'|^{-8K}$ and we can conclude that 
\begin{equation}\label{eq:onechain}
   \braket{\sin[\theta_1(z)] \sin[\theta_1(z')]}\sim (z-z')^{-\mathrm{min}(8K,2)}.
\end{equation}

Eq. \eqref{eq:onechain} holds as far as the perturbation \eqref{eq:action_imp} is relevant, i.e. $K>1/4$. We can therefore conclude that $\braket{\mathcal{Y}_j\mathcal{Y}_k}\sim |j-k|^{-2}$.

Given this analysis, we can now also understand the result in Eq.~\eqref{eq:2XYcorrelations_1}. In this case, we are interested in the correlation function $\braket{\mathcal{Y}_j\tilde{\mathcal{Y}}_j \mathcal{Y}_k \tilde{\mathcal{Y}}_k}$ when we measure $\mathcal{X}_{j}\tilde{\mathcal{X}}_{j}$, which in the bosonic language amounts to add the weak measurement $\int_{x}\cos[\theta(x)]\cos[\tilde{\theta}(x)]$ (Eq. \eqref{eq:pert}). In order to evaluate the correlation functions above, it is useful to consider the $+,-$ sectors, with $\theta_{\pm}=\theta\pm\tilde{\theta}$. In this way, the two Luttinger theories coupled through $\int_{x}\cos[\theta(x)]\cos[\tilde{\theta}(x)]$ can be written as in Eq. \eqref{eq:pert2},
i.e. a sum of two TLL with $\theta_{\pm}$ pinned around 0. The scaling dimension of $\cos(\theta_{\pm})$ is $1/(2K)$.

\begin{figure}[h]
    \centering
    \includegraphics[width=0.65\linewidth]{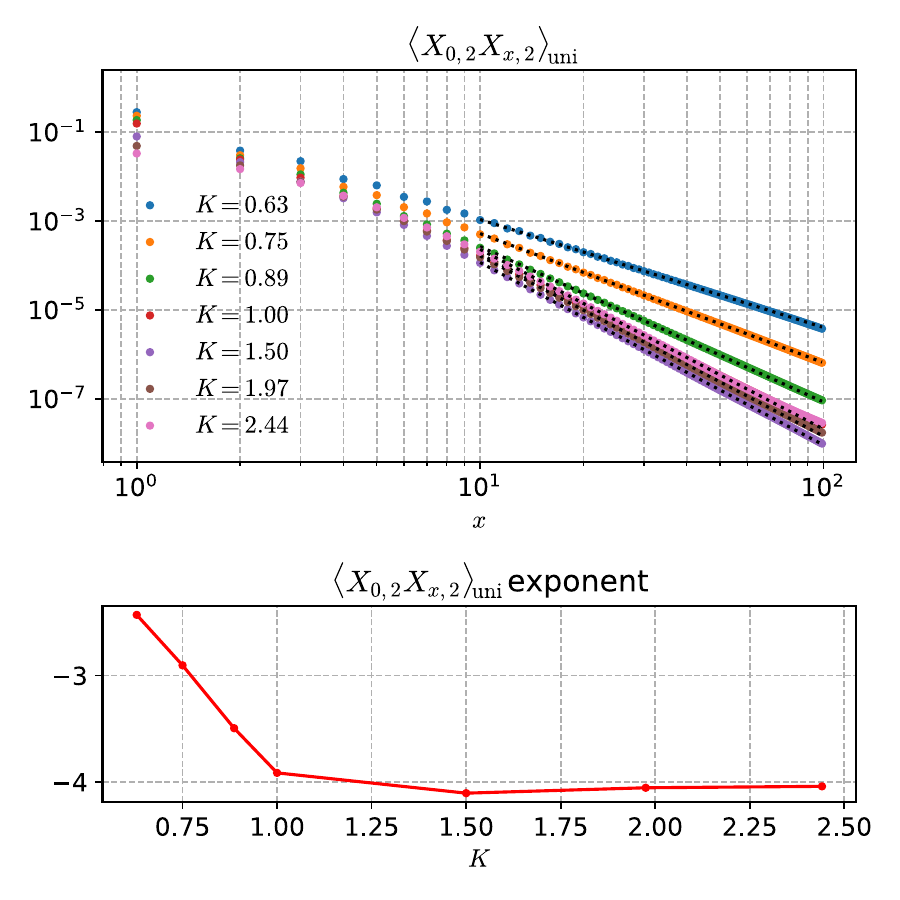}
    \caption{Top panel: the correlation function $\braket{\mathcal{Y}_0 \mathcal{\tilde Y}_0 \mathcal{Y}_x \mathcal{\tilde Y}_x }_{\rm uni} \equiv \braket{X_{0,2} X_{x,2}}_{\rm uni}$ of the gapless parent of the cluster state with different Luttinger parameter $K$ after $X_{j, 1}=1, \forall j$ measurement. The power-law exponents are extracted by fitting the data (see the black dotted lines). Bottom panel: the power-law exponent as a function of $K$, which is consistent with $-\min(4K, 4)$. Data obtained by iDMRG with bond dimension $\chi=2000$.}
    \label{fig:YYYY}
\end{figure}

In order to evaluate the correlation function we are interested in, we must add a term 
\begin{equation}\label{eq:tunnel2}
   \delta S=g\displaystyle\int_{x} O(x)
\end{equation}
where $O(x)=\cos[2\phi^+_1(x)-2\phi^+_2(x)]$. Here we focus on only one single TLL, for instance in the + sector.
We remind that 
\begin{equation}
    \begin{aligned}
        &\braket{\mathcal{Y}_j \mathcal{\tilde Y}_j \mathcal{Y}_k \mathcal{\tilde Y}_k } \sim \braket{\sin[\theta(j)] \sin[\tilde \theta(j)] \sin[\theta(k)] \sin[\tilde \theta(k)]}\\
        &~~~\sim \braket{\cos[\theta_+(j)] \cos(\theta_+(k))} + \braket{\cos[\theta_-(j)] \cos[\theta_-(k)]}\\
        &~~~~~~ - \braket{\cos[\theta_+(j)]}\braket{\cos[\theta_-(k)]}- \braket{\cos[\theta_-(j)]}\braket{\cos[\theta_+(k)]},
    \end{aligned}
\end{equation}
and taking into account the presence the tunneling term
\eqref{eq:tunnel2}, we need to evaluate 
\begin{equation}\label{eq:coscos}
\begin{split}
   &\braket{\cos[\theta_{+,1}(z)] \cos[\theta_{+,1}(z')]}=\braket{\cos[\theta_{+,1}(z)] \cos[\theta_{+,1}(z')]}_{g=0}\\
    &+g^2\int_{x_1,x_2} \braket{\cos[\theta_{+,1}(z)] \cos[\theta_{+,1}(z')]O(x_1) O^{\dagger}(x_2)}+o(g^4).
    \end{split}
\end{equation}
We can expand the second term, such that we get 
\begin{multline}\label{eq:tunnel4}
   \int_{x_1,x_2} \braket{\cos[\theta_{+,1}(z)] \cos[\theta_{+,1}(z')]O(x_1) O^{\dagger}(x_2)}=\\
   \frac{1}{4} \int_{x_1,x_2}[\braket{e^{i\theta_{+,1}(z)}e^{-i\theta_{+,1}(z')}O(x_1) O^{\dagger}(x_2)}+\braket{e^{-i\theta_{+,1}(z)}e^{i\theta_{+,1}(z')}O(x_1) O^{\dagger}(x_2)}].
\end{multline} 
At the strong coupling fixed point, the dimension of $O(x)$ is $2K$.
In order to compute the large distance behavior of the correlator \eqref{eq:tunnel4}, we can again use a decomposition similar to Eq. \eqref{eq:ope}, together with a scaling dimension analysis. Therefore, we can deduce that each integral above behaves as $|x-x'|^{-4K}$. Given that $\braket{\cos[\theta_{+,1}(x)] \cos[\theta_{+,1}(x')]}_{g=0}\sim const+|x-x'|^{-4}$ (see Eq.~\eqref{eq:bcft_tt}), we find that 
\begin{equation}
    \braket{\mathcal{Y}_j \mathcal{\tilde Y}_j \mathcal{Y}_k \mathcal{\tilde Y}_k }\sim |j-k|^{-\mathrm{min}(4K,4)}.
\end{equation}
We cross-check the result above against numerics in Fig. \ref{fig:YYYY}.

\section{Numerical and analytical details on the gapless parent of 1D cluster state}
In this Appendix, we provide some additional numerical evidence of the findings we have presented in Sections \ref{sec:uniform}-\ref{sec:tiltmeas} of the main text. 
\subsection{Power-law exponent of $\braket{\prod_{j=0}^x X_{j,2}}_{\rm uni}$ after uniform projective measurement of $X_{j,1}$}\label{app:2K}
\begin{figure}[h!]
    \centering
    \includegraphics[width=0.65\linewidth]{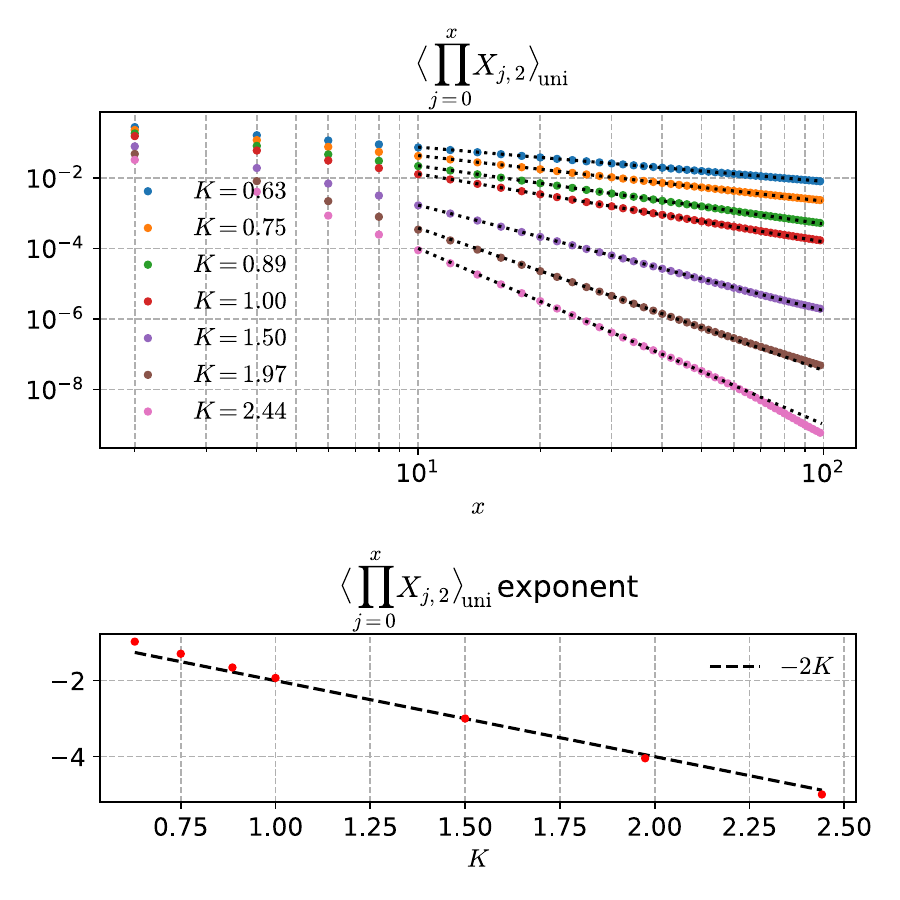}
    \caption{Top panel: the expectation value $\langle \prod_{j=0}^x X_{j,2} \rangle_{\rm uni}$ of the gapless parent of the cluster state with different Luttinger parameter $K$ after $X_{j, 1}=1, \forall j$ measurement. The power-law exponents are extracted by fitting the data (see the black dotted lines). Bottom panel: the power-law exponent as a function of $K$. Data obtained by iDMRG with bond dimension $\chi=2000$.}
    \label{fig:exponent_2K}
\end{figure}

In the gapless parent of the cluster state, if one performs $X_{j,1}$ measurement and post-select the uniform outcome $X_{j,1}=1, \forall j$, it is shown in Sec.~\ref{sub:xbasis} that we expect $\langle \prod_{j=0}^x X_{j,2} \rangle_{\rm uni} \sim x^{-2K}$. We examine this in Fig.~\ref{fig:exponent_2K} using infinite DMRG (iDMRG) simulations for systems with different Luttinger parameters.

\subsection{Scaling dimension of $Z_{j,1}$} \label{app:Z1scalingdim}
\begin{figure}[h!]
    \centering
    \includegraphics[width=0.6\linewidth]{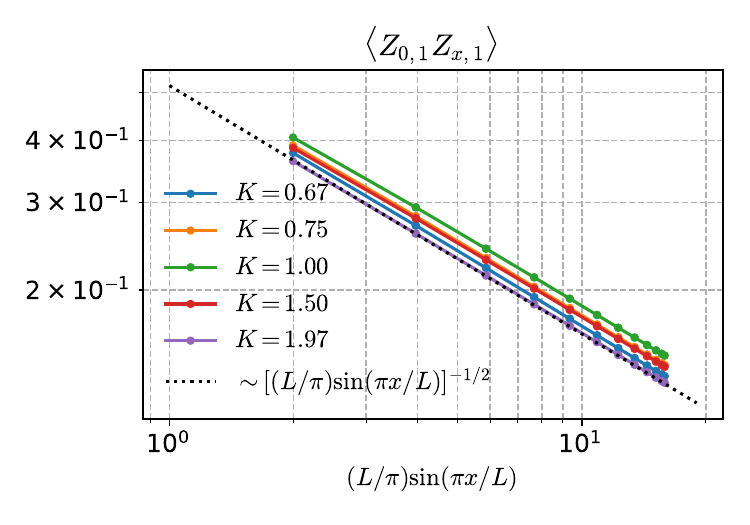}
    \caption{Two-point correlation function of $Z_{x,1}$ as a function of the chord length. Data have been obtained using the decoupled $\mathcal{XXZ}$ representation. We simulate the $\mathcal{XXZ}$ chain with $L = 50$ and periodic boundary conditions using DMRG with bond dimension $\chi = 300$. The data for different values of $K$ show that the scaling dimension $[Z_{x,1}]$ is always $1/4$.}
    \label{appfig:Z1scaling}
\end{figure}

In the decoupled-$\mathcal{XXZ}$-chain representation, we have
\begin{equation}\label{appeq:Z1corr}
    Z_{j,1} Z_{k,1} = \left[{
    \begin{matrix}
         & \mc{Y}_{j+1} &\cdots & \mc{Y}_k & \mc{Y}_{k+1} \\
        \mc{\tilde Y}_j& \mc{\tilde Y}_{j+1} &\cdots & \mc{\tilde Y}_k & \\
    \end{matrix}
    }\right],\qquad
    \braket{Z_{j,1} Z_{k,1}} = \left(\braket{\prod_{i=j}^{k} \mathcal{Y}_i}\right)^2.
\end{equation}
Thus we can compute the correlation function from $\langle \prod_{i=j}^{k} \mathcal{Y}_i \rangle$ of a single $\mathcal{XXZ}$ chain. Numerically, we use DMRG to obtain the ground state of an $\mathcal{XXZ}$ chain with $L = 50$ and periodic boundary conditions, and then compute $\braket{Z_{j,1} Z_{k,1}}$ using Eq.~\ref{appeq:Z1corr}. The numerical data for different Luttinger parameters $K$ is shown in Fig.~\ref{appfig:Z1scaling}. The results suggest that $[Z_{j,1}] = 1/4$ for all values of $K$. Even though, in general, the lack of a bosonic representation for Eq. \eqref{appeq:Z1corr} prevents us from predicting this scaling dimension,
we are able to analytically derive this result for two special cases, $K=1/2$ and $K=1$. For $K=1/2$, the symmetry of the $\mathcal{XXZ}$ spin chain is enhanced to $SU(2)$, and therefore we can use the following equality among correlators $\braket{\prod_{x=j}^{k} \mathcal{Y}_x} = \braket{\prod_{x=j}^{k} \mathcal{Z}_x}$. By using the bosonization dictionary $\mathcal{Z} \sim -\frac{2}{\pi} \partial_x \phi$ in Eq.~\eqref{eq:dictionary}, we observe that $\braket{\prod_{x=j}^{k} \mathcal{Z}_x} \sim \braket{e^{i \frac{\pi}{2} \sum_{x = j}^{k} \mathcal{Z}_x} }\sim \braket{e^{-i \int_{j}^{k} dx \partial \phi}} = \braket{e^{-i \phi(k)} e^{i \phi(j)}} \sim |j-k|^{-K/2}$. Therefore, when $K=1/2$, $\braket{\prod_{x=j}^{k} \mathcal{Y}_x} \sim |j-k|^{-1/4}$ and, after taking into account also the contribution of $\mathcal{\tilde Y}$, we can conclude that $[Z_{j,1}] = \frac{1}{4}$.

When $K=1$, the coupling $\Delta$ in Eq.~\eqref{eq:2XXZ} vanishes, so we get two copies of the $\mathcal{X}\mathcal{Y}$ spin chain. Each of them can be rewritten as
\begin{equation}\label{eq:Ham1}
    H=-\sum_j [(\mathcal{X}_{2j-1}\mathcal{X}_{2j}+\mathcal{Y}_{2j-1}\mathcal{Y}_{2j})+(\mathcal{X}_{2j}\mathcal{X}_{2j+1}+\mathcal{Y}_{2j}\mathcal{Y}_{2j+1})],
\end{equation}
where we have split odd-even and even-odd couplings. We can now define 
\begin{equation}
    \mathcal{X}_j=\sigma^x_j\sigma^x_{j+1}, \quad \mathcal{Y}_j\mathcal{Y}_{j+1}=\sigma^y_j,
\end{equation}
such that Eq.~\eqref{eq:Ham1} becomes 
\begin{equation}
    H=-\sum_j [(\sigma^x_{2j-1}\sigma^x_{2j+1}+\sigma^y_{2j-1})+(\sigma^x_{2j}\sigma^x_{2j+2}+\sigma^y_{2j})],
\end{equation}
i.e. it becomes the sum of two transverse-field Ising Hamiltonians defined on odd and even lattice sites. Notice that, in principle, we could have applied a rotation $e^{i\pi/4 \mathcal{X}_j}$ in Eq. \eqref{eq:Ham1} to send $\mathcal{Y}_j\to \mathcal{Z}_j$ and recover the more canonical form of the Ising Hamiltonian, but it is more convenient for us to work with a transverse field along the $y$-direction. 
Indeed, we are interested in the operator $\prod_{k=j}\mathcal{Y}_k$, and in particular 
\begin{equation}
   \prod_{k=2j}\mathcal{Y}_k= \prod_{k=2j,\rm even}\sigma^y_k=\mu_{2j}P^{\rm even},\qquad \prod_{k=2j-1}\mathcal{Y}_k= \prod_{k=2j-1,\rm odd}\sigma^y_k=\mu_{2j-1}P^{\rm odd}
\end{equation}
where $P^{\rm even}=\prod_{2j}\sigma_{2j}^y$ is the parity operator of the Ising chain defined on the even sites (and similarly for $P^{\rm odd}$), while $\mu$
is the disorder operator with scaling dimension $1/8$ (note that the parity operator is dimensionless).
If we now take into account also the  $\mathcal{\tilde{X}}\mathcal{\tilde{Y}}$ spin chain, then we get that the total scaling dimension for $\prod_{k=2j}\mathcal{Y}_k\mathcal{\tilde{Y}}_{k-1}$ or $\prod_{k=2j+1}\mathcal{Y}_k\mathcal{\tilde{Y}}_{k-1}$ is $1/8+1/8=1/4$, which is consistent with the scaling dimension of the operator $Z_{j,1}$.

\subsection{More numerical results on the weak $\mathbb{Z}_2$-preserving measurements }\label{app:measXandZZ_weak}
\begin{figure}[ht]
    \centering
    \includegraphics[width=\linewidth]{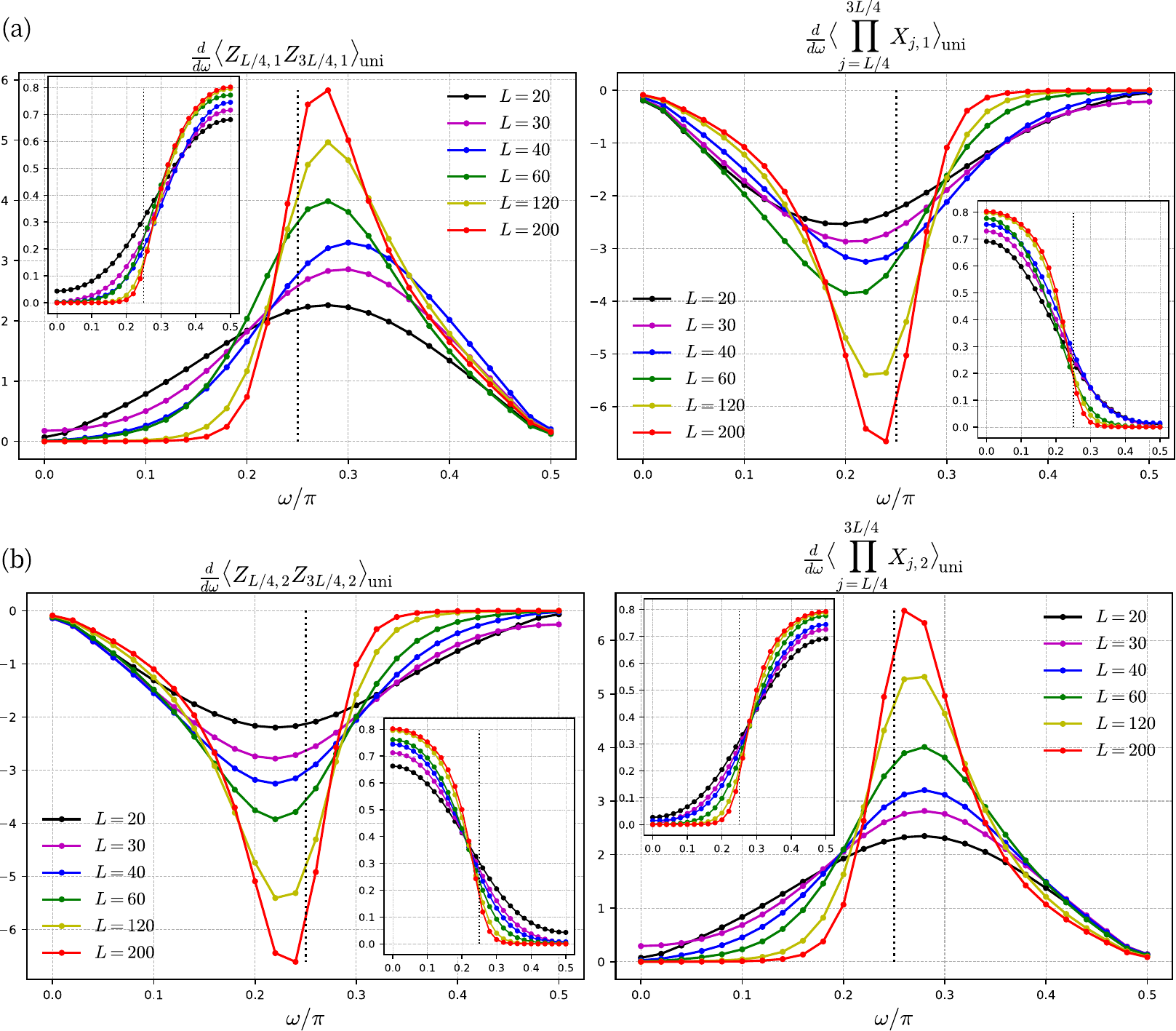}
    \caption{The derivative of order and disorder parameters in the (a) upper and (b) lower chain of the post-measurement gapless parent of the cluster state with $K=1.5$ as a function of the parameter $\omega$. The measurement strength is $\beta = 0.5$. The black dotted lines mark the transition at $\omega_c = \pi/4$. Insets: order and disorder parameters without derivative. The gapless parent state in obtained using DMRG with open boundary conditions and bond dimension $\chi=1200$.}
    \label{fig:measXandZZ_beta0.5_orderparams}
\end{figure}

In Sec.~\ref{sub:X_ZZ}, Fig.~\ref{fig:measXandZZ_orderparams} shows numerically the behavior of the order parameter in the lower chain of the post-measurement wavefunction as a function of the parameter $\omega$ in the projective measurement limit. In Fig.~\ref{fig:measXandZZ_beta0.5_orderparams} we check that for the weak measurement with $\beta = 0.5$, the long-range order and disorder parameters remain the same as those presented in Table~\ref{tab:measXandZZ_phases}: when $0\leq \omega < \frac{\pi}{4}$, $\braket{\prod_{j=0}^x X_{j, 1} }_{\rm uni}$ and  $\braket{Z_{0, 2} Z_{2}}_{\rm uni}$ have long-range order; when $\omega = \frac{\pi}{4}$, the system is gapless; when $\frac{\pi}{4} < \omega \leq \frac{\pi}{2}$, $\braket{Z_{0, 1} Z_{x, 1}}_{\rm uni}$ and $\braket{\prod_{j=0}^x X_{j, 2}}_{\rm uni} $ have long-range order.

\subsection{Uniform weak measurement of $X_{j,1}$ and $X_{j,2}$}\label{app:measX1andX2}

\begin{figure}[ht]
    \centering
    \includegraphics[width=\linewidth]{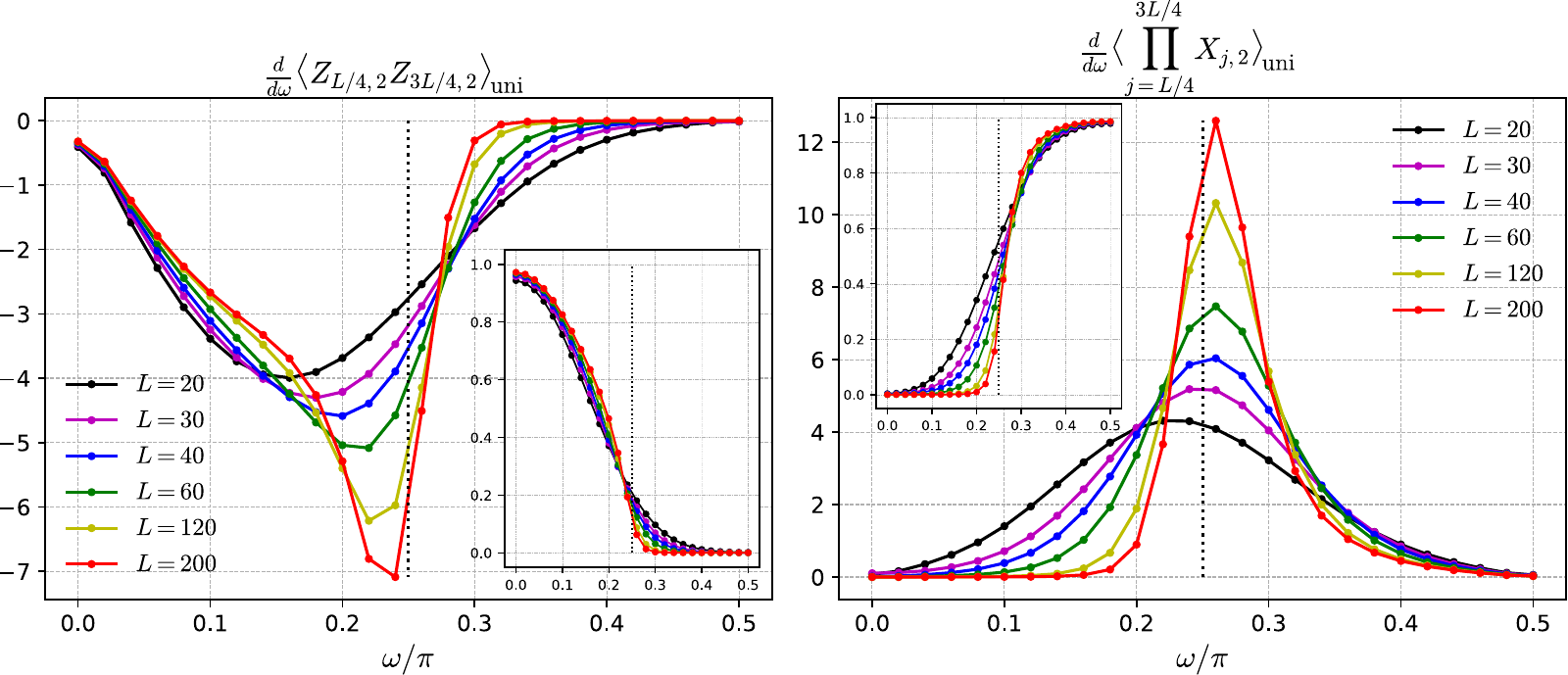}
    \caption{The derivative of order and disorder parameters in the lower chain of the post-measurement gapless parent of the cluster state with $K=1.5$ as a function of the parameter $\omega$. The black dotted lines mark the transition at $\omega_c = \pi/4$. Insets: the order parameter , $\braket{Z_{\frac{L}{4},2} Z_{\frac{3L}{4},2}}_{\rm uni}$, and disorder parameter, $\left\langle{\prod_{j = L/4}^{3L/4} X_{j,2}}\right\rangle_{\rm uni}$, as a function of $\omega$. The gapless parent of the cluster state in obtained using DMRG with open boundary conditions and bond dimension $\chi=1200$.}
    \label{fig:measX1andX2_orderparams}
\end{figure}

As another example of $\mathbb{Z}_2$-preserving measurements, we consider weak measurement of both $X_{j,1}$ and $X_{j,2}$ as described in Eq.~\eqref{eq:measX1andX2}.  Since the low-energy theory is still described by Eq.~\eqref{eq:defect_X_ZZ}, we expect that the results to be similar to those in Sec.~\ref{sub:X_ZZ}. In Fig.~\ref{fig:measX1andX2_orderparams} we apply the weak measurement in Eq.~\eqref{eq:measX1andX2} with $\beta = 0.5$ to the gapless parent state, and probe the order and disorder parameters of the post-measurement wavefunction, $\braket{Z_{\frac{L}{4},2} Z_{\frac{3L}{4},2}}_{\rm uni}$  and $\left\langle{\prod_{j = L/4}^{3L/4} X_{j,2}}\right\rangle_{\rm uni}$, as a function of $\omega$ for different system sizes $L$. The divergence of the derivative of the order/disorder parameter with respect to $\omega$ suggests a phase transition at $\omega = \pi/4$, as expected from Sec.~\ref{sub:X_ZZ}.

\begin{figure}[h]
    \centering
    \includegraphics[width=\linewidth]{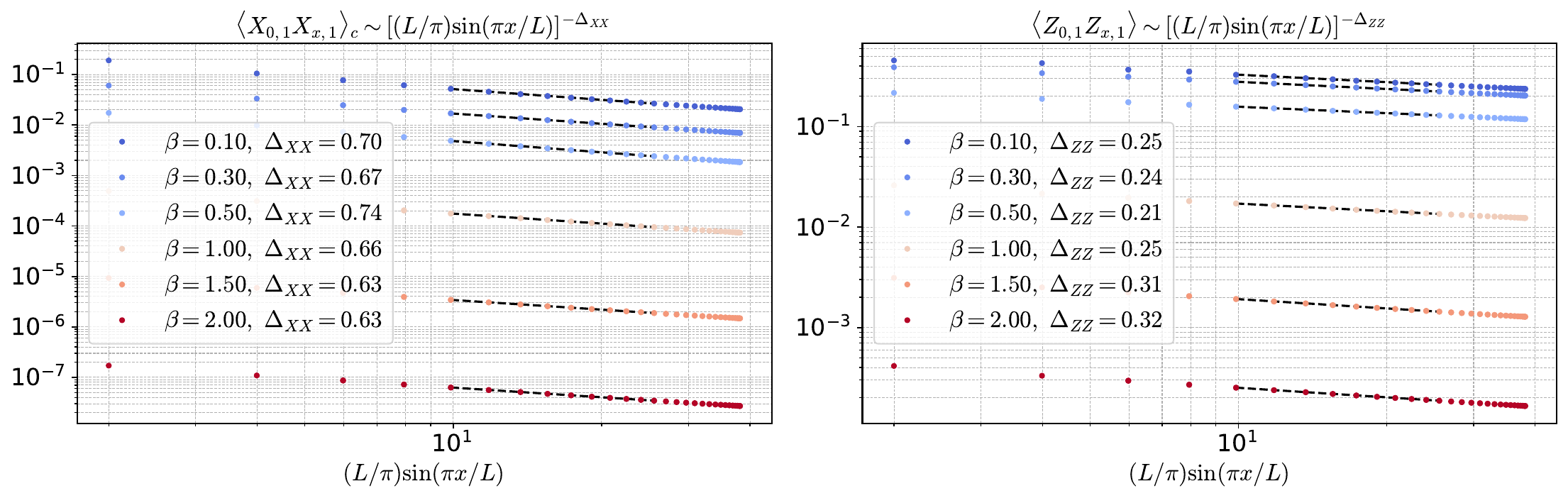}
    \caption{Correlation functions of $e^{\beta \sum_j (X_{j,1}+X_{j,2})} \ket{\psi_\Delta}$ with Luttinger parameter $K = 1.5$. The gapless parent state $\ket{\psi_\Delta}$ with length $L=60$ and periodic boundary conditions is obtained using DMRG with bond dimension $\chi=2000$. In the large $\beta$ limit, the theoretical power-law exponents of $\braket{X_{0,1} X_{x,1}}_c$ and $\braket{Z_{0,1} Z_{x,1}}$ are $\frac{1}{K} \approx 0.67$ and $\frac{1}{2K} \approx 0.33$ respectively, which are close to the numerical fittings: $0.63$ and $0.32$ (see the black dashed lines).}
    \label{fig:measX1andX2_corr}
\end{figure}
\begin{figure}[h]
    \centering
    \includegraphics[width=0.6\linewidth]{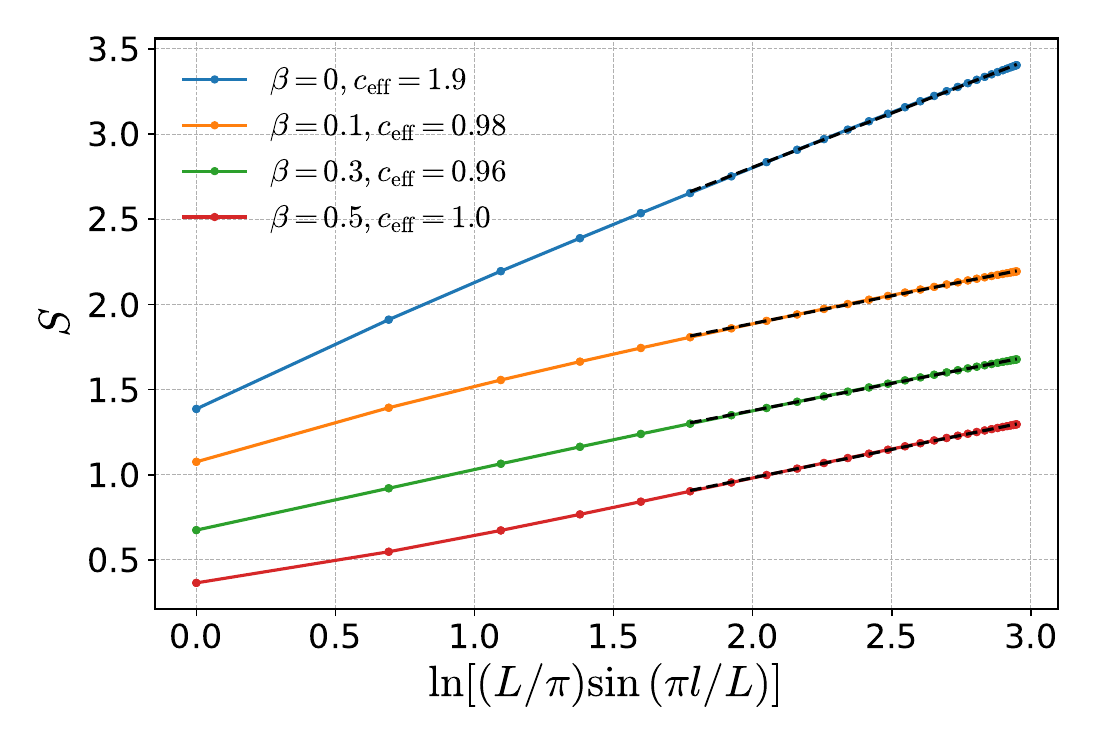}
    \caption{The entanglement entropy $S$ as a function of $\ln[(L/\pi) \sin(\pi l/L)]$ where $l$ is the subsystem size, after weak measurement of $e^{\beta \sum_j (X_{j,1} + X_{j,2})}$ in the gapless parent of the cluster state of size $L=60$ with periodic boundary conditions. The effective central charge $c_{\rm eff}$ is extracted by fitting the data with $S = \frac{c_{\rm eff}}{3} \ln[(L/\pi) \sin(\pi l/L)] + const.$ (see the black dashed lines) for different measurement strength $\beta$.
    }
    \label{fig:measX1andX2_EE}
\end{figure}

At the transition, the post-measurement wavefunction is $e^{\frac{\beta}{\sqrt{2}} \sum_j (X_{j,1}+X_{j,2})} \ket{\psi_\Delta}$, and we now compute its correlation functions. Since the measurement induces a relevant perturbation in the RG sense, we focus on the large $\beta$ limit. Using the bosonization form, we can compute the $XX$ correlation function,
\begin{equation}
    \begin{aligned}
        \braket{X_{0,1} X_{x,1}}_c &= \braket{\mathcal{X}_0 \mathcal{\tilde{X}}_0 \mathcal{X}_x \mathcal{\tilde{X}}_x}_c= \braket{\cos{\theta(0)} \cos{\tilde\theta(0)} \cos{\theta(x)} \cos{\tilde\theta(x)}}_c\\
        & \sim \braket{\cos\theta_+(0) \cos\theta_+(x)}_c + \braket{\cos\theta_-(0) \cos\theta_-(x)}_c\sim x^{-\frac{1}{K}} , \\
    \end{aligned}
\end{equation}
where in the last line we have used the fact that the $\theta_-$ field is pinned at $\tau = 0$, while the $\theta_+$ field remains gapless. Using the analogy with Eq.~\eqref{eq:corr_truncatdWF}, we conclude that the strange correlator decays as
\begin{equation}
    a_{0x} - a_0 a_x \sim x^{-\frac{1}{2K}},
\end{equation}
and correspondingly, for large $\beta$, the correlators of $Y$ and $Z$ Pauli operators are
\begin{equation}
    \braket{Y_{0,1} Y_{x,1}} \sim \braket{Z_{0,1} Z_{x,1}} \sim x^{-\frac{1}{2K}}.
\end{equation}
We numerically check the above results in Fig.~\ref{fig:measX1andX2_corr} by DMRG. Besides, in Fig.~\ref{fig:measX1andX2_EE} we check that the transition at $\omega = \pi/4$ has an effective central charge $c_{\rm eff} = 1$ which corresponds to the gapless $\theta_+$ mode.

\subsection{Power-law exponents of specific operators}\label{app:powerlaw}

In this short section, we compute power-law exponents of several operators in the gapless parent state after $e^{\beta \sum_j X_{j,1}}$ measurement. Then we use the duality argument to find the exponents of $e^{\beta \sum_j Z_{j,1}}$ measurement.

For the $e^{\beta \sum_j X_{j,1}}$ measurement, it is shown in Sec.~\ref{sub:xbasis} that the post-measurement correlator $\braket{Z_{0,2} Z_{x,2}}_{\rm uni}$ is long-range ordered while $\braket{X_{0,2} X_{x,2}}_{\rm uni} \sim \braket{\sin\theta(0) \sin\tilde\theta(0) \sin\theta(x) \sin\tilde\theta(x)}_{\rm uni}\sim x^{-\min(4,4K)}$. Note that $\braket{X_{0,1} X_{x,1}}_{\rm uni}$ has a similar bosonization form as $\braket{X_{0,2} X_{x,2}}_{\rm uni}$, so we also have $\braket{X_{0,1} X_{x,1}}_{\bm s} \sim \braket{\cos\theta(0) \cos\tilde\theta(x) \cos\theta(x) \cos\tilde\theta(x)}_{\rm uni, c} \sim x^{-\min(4,4K)}$, where the subscript $c$ denotes the connected part. Meanwhile, since $X $ is nearly projected to $1$ on the first chain, we have $\braket{X_{0,1} X_{0,2} X_{x,1} X_{x,2}}_{\rm uni} \sim \braket{X_{0,2} X_{x,2}}_{\rm uni} \sim x^{-\min(4,4K)}$. Also, $Z_{x, 1} Z_{x+1, 1} \sim \mathcal{\tilde Y}_j \mathcal{Y}_{j+1} \sim \mathcal{\tilde Y}_j \mathcal{Y}_j \sim X_{x,2}$, so we expect $\braket{Z_{0, 1} Z_{1,1} Z_{x,1} Z_{x+1,1}}_{\rm uni} \sim \braket{X_{0,2} X_{x,2}}_{\rm uni} \sim x^{-\min(4,4K)}$ and, similarly, $\braket{Z_{0, 2} Z_{1,2} Z_{x,2} Z_{x+1,2}}_{\rm uni} \sim \braket{X_{0,1} X_{x,1}}_{\rm uni} \sim x^{-\min(4,4K)}$. We summarize the results in Table~\ref{tab:exponents_x}. We remark that when $K>1$, the scaling dimensions of the selected operators are larger than $1$.
\begin{table*}[h!]
    \centering
    \setlength{\extrarowheight}{2pt}
        \caption{Power-law exponents of selected operators (connected part) after $e^{\beta\sum_j X_{j,1}}$ measurement.
    }
    \begin{tabular*}{\linewidth}{@{\extracolsep{\fill}} |c|c|c|c|c|c|c|c|}
    \hline
         Correlators &$\braket{X_{0,1} X_{x,1}}$ & $\braket{Z_{0,1} Z_{x,1}}$ & $\braket{X_{0,2} X_{x,2}}$ & $\braket{Z_{0,2} Z_{x,2}}$ & $\braket{Z_{0,1} Z_{1, 1} Z_{x,1} Z_{x+1,1}}$ & $\braket{Z_{0,2} Z_{1, 2} Z_{x,2} Z_{x+1,2}}$ & $\braket{X_{0, 1} X_{0, 2} X_{x, 1} X_{x, 2}}$\\
         \hline
         Exponents & $\min(4, 4K)$&$2K$&$\min(4, 4K)$&LRO&$\min(4, 4K)$&$\min(4, 4K)$&$\min(4, 4K)$\\
         \hline
    \end{tabular*}
    \label{tab:exponents_x}
\end{table*}

For the $e^{\beta \sum_j Z_{j,1}}$ measurement, using the argument in Sec.~\ref{sub:zbasis}, we expect the nature of this fixed point is similar to the fixed point of the measurement of $e^{\beta \sum_jZ_{j,1} Z_{j+1,1}}$. Since in terms of the bosonic representation, $Z_{j,1} Z_{j+1,1} \sim X_{j,2}$, one can immediately obtain the post-measurement power-law exponents by swapping $1 \leftrightarrow 2$ in the previous paragraph. The only difference is that, since measuring $Z_{\cdot, 1}$ does not result in a ``cat'' state as in the measurement of $Z_{j,1}Z_{j+1,1}$, now the connected part of $\braket{Z_{0,1}Z_{x,1}}_{\rm uni, c}$ does not exhibit long-range order, as opposed to $\braket{Z_{0,2} Z_{x,2}}_{\rm uni,c}\sim const.$ in the $e^{\beta \sum_j X_{j,1}}$ measurement. Using the same arguments as in Appendix~\ref{app:correlators}, we expect that $\braket{Z_{0,1}Z_{x,1}}_{\rm uni, c} \sim x^{-\min(4,4K)}$. We summarize the results in Table~\ref{tab:exponents_z}. Also note that when $K>1$, the scaling dimensions are larger than $1$. Therefore, we expect that both $X$ and $Z$ measurements of the first chain flow to stable fixed points when $K>1$.
\begin{table*}[h!]
    \centering
        \caption{Power-law exponents of select operators (connected part) after $e^{\beta\sum_j Z_{j,1}}$ measurement.}
    \setlength{\extrarowheight}{2pt}
    \begin{tabular*}{\linewidth}{@{\extracolsep{\fill}} |c|c|c|c|c|c|c|c|}
    \hline
         Correlators &$\braket{X_{0,1} X_{x,1}}$ & $\braket{Z_{0,1} Z_{x,1}}$ & $\braket{X_{0,2} X_{x,2}}$ & $\braket{Z_{0,2} Z_{x,2}}$ & $\braket{Z_{0,1} Z_{1, 1} Z_{x,1} Z_{x+1,1}}$ & $\braket{Z_{0,2} Z_{1, 2} Z_{x,2} Z_{x+1,2}}$ & $\braket{X_{0, 1} X_{0, 2} X_{x, 1} X_{x, 2}}$\\
         \hline
         Exponents & $\min(4, 4K)$&$\min(4, 4K)$&$\min(4, 4K)$&$2K$&$\min(4, 4K)$&$\min(4, 4K)$&$\min(4, 4K)$\\
         \hline
    \end{tabular*}
    \label{tab:exponents_z}
\end{table*}

\begin{figure}[h]
    \centering
    \includegraphics[width=0.5\linewidth]{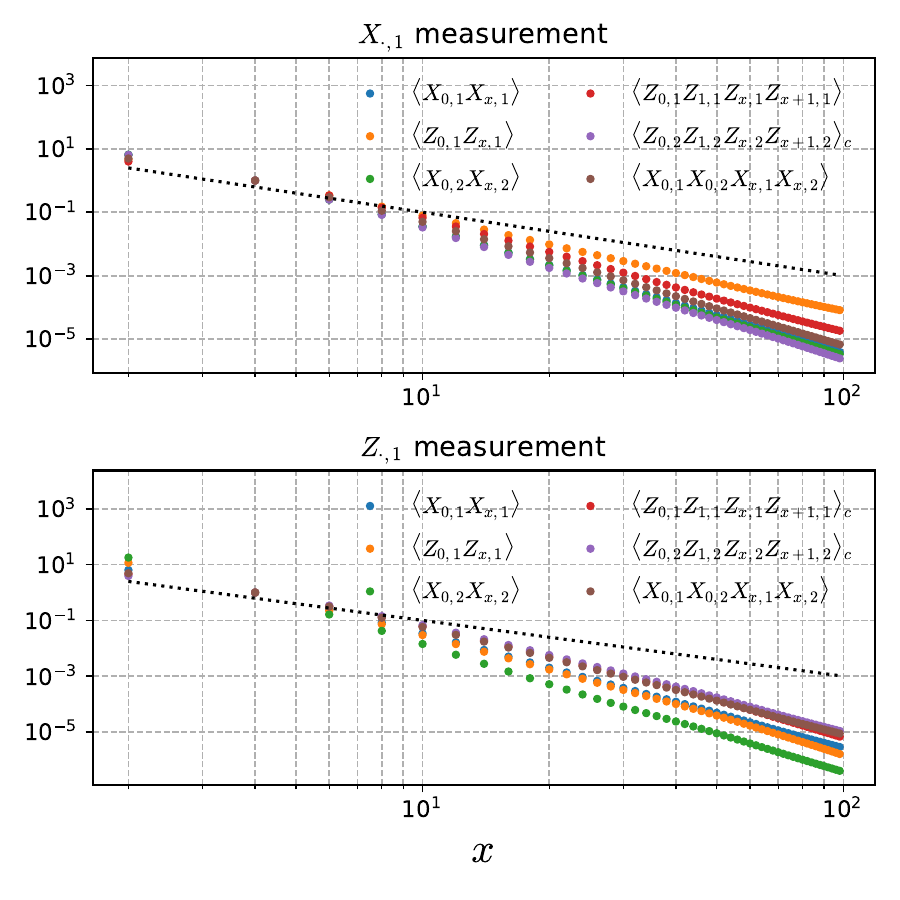}
    \caption{Two-point correlation functions of selected operators in the gapless parent of the cluster state with Luttinger parameter $K=1.5$ after (a) $e^{\beta \sum_j X_{j,1}}$ and (b) $e^{\beta \sum_j Z_{j,1}}$ measurement with $\beta = 1$. The correlators are rescaled for a better visualization of the data. As a guide-line, the black dotted line has exponent $-2$. An operator is irrelevant if the two-point correlator decays faster than the reference line, as in the case presented here.}
    \label{fig:scalingdims}
\end{figure}

We numerically check the irrelevance of selected operators for $K=1.5$ in Fig.~\ref{fig:scalingdims}.
\subsection{Details of DMRG simulations for the entanglement entropy at the intermediate fixed point}\label{app:dmrgconvergence}

In Fig.~\ref{fig:EE_measXandZ}, we investigate the entanglement scaling at the intermediate fixed point. The numerical data are obtained using finite-size DMRG with open boundary conditions and bond dimension $\chi = 1200$. By fitting the post-measurement half-chain entanglement entropy with the system size using $S_{L/2} = \frac{c_{\rm eff}}{6} \log(L) + const.$ for $L = [20, 40, 60, 120, 200]$, we obtain the effective central charge $c_{\rm eff}$ in Fig.~\ref{fig:EE_measXandZ}(c).
The error bars in Fig.~\ref{fig:EE_measXandZ}(c) is three times the standard deviation of $c_{\rm eff}$ from the linear fitting $S_{L/2}$-$\log(L)$.

\begin{figure}[h]
    \centering
    \includegraphics[width=\linewidth]{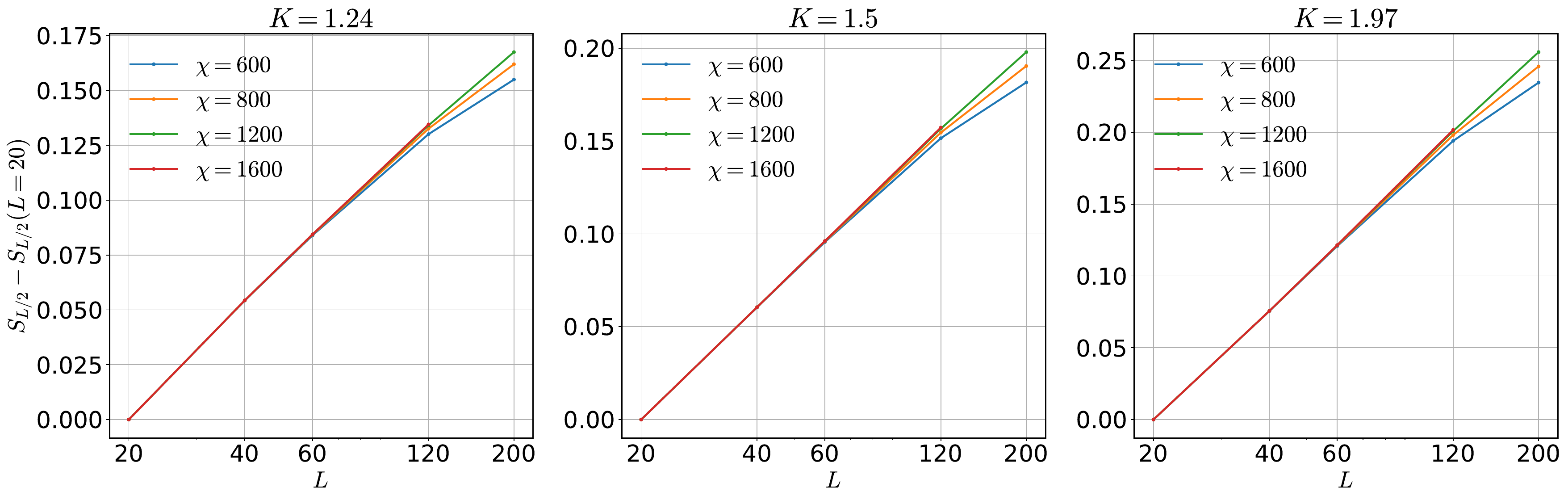}
    \caption{Entanglement entropy obtained by DMRG with different bond dimensions $\chi = 600, 800, 1200, 1600$.}
    \label{fig:DMRGconvergence}
\end{figure}

To check whether $\chi=1200$ is enough for DMRG convergence, we plot in Fig.~\ref{fig:DMRGconvergence} entanglement entropy at the transition for $K=1.24, 1.5, 1.97$ obtained using bond dimension $\chi = 600, 800, 1200, 1600$. Due to lack of computer memory, we do not have data for $L=200, \chi=1600$. From Fig.~\ref{fig:DMRGconvergence} we conclude that $\chi=1200$ is enough for convergence at least for $L\lesssim120$.

\subsection{Replica model with different values of $K$}\label{app:replica_numerics}
\begin{figure}[h]
    \centering
    \includegraphics[width=\linewidth]{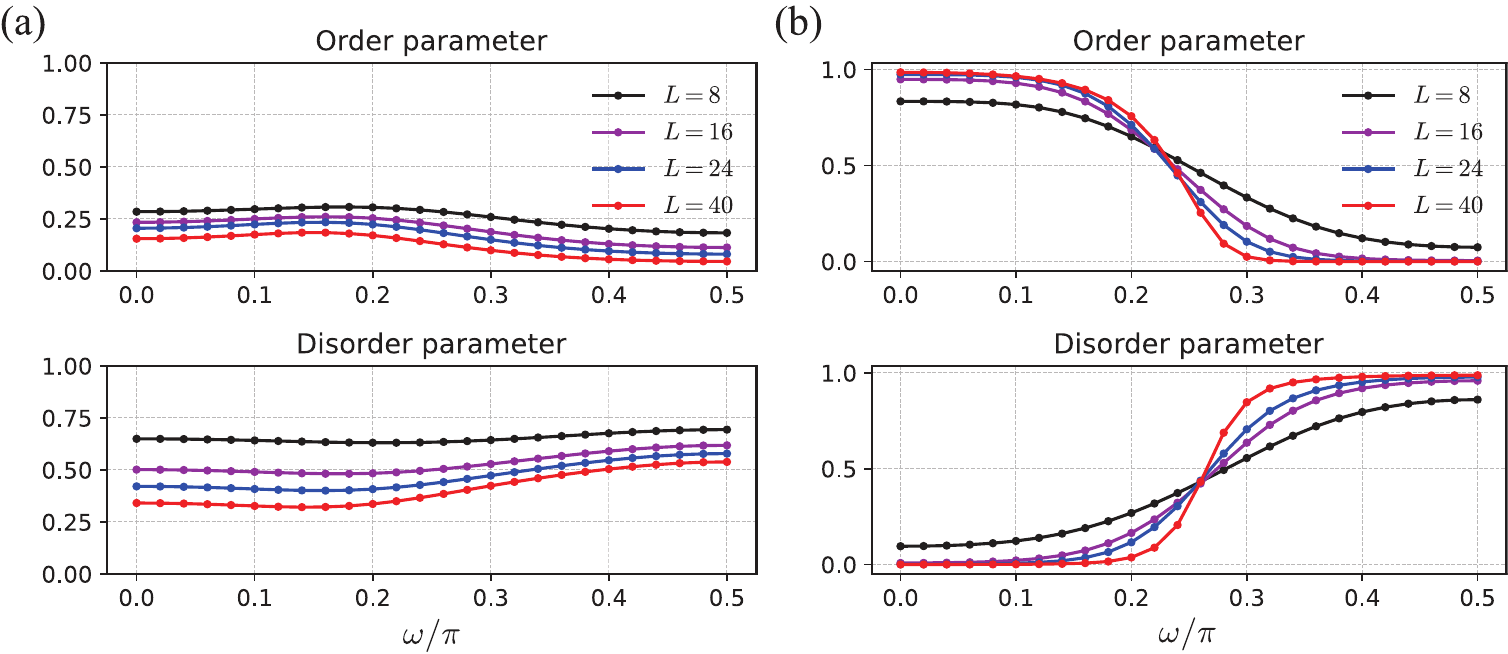}
    \caption{Order and disorder parameters as a function of the tilting angle for (a) $K = 0.75$ and (b) $K = 2.44$.}
    \label{fig:replica_app}
\end{figure}

Since it is difficult to study analytically the replica model of Section \ref{sec:nonlinear}, we add more numerical results for the tilted measurement basis with different values of $K$. In Fig.~\ref{fig:replica_app} we plot the order parameter , $\langle\langle Z_{\frac{L}{4},2} Z_{\frac{3L}{4},2}\rangle\rangle$, and disorder parameter, $\langle\langle{\prod_{j = L/4}^{3L/4} X_{j,2}}\rangle\rangle$ for $K=0.75$ and $K=2.44$, as an addition to the case shown in Fig.~\ref{fig:replica} for $K=1.5$.

For $K=0.75$ in \sfigref{fig:replica_app}{a}, $X_{j,1}$ is irrelevant so at $\omega = 0$ the long-range physics is still described by the pristine Luttinger liquid theory, and we do not find a significant sign of phase transition in the order and disorder parameters. For $K=2.44$ in \sfigref{fig:replica_app}{b}, however, we find the crossing point in the order and disorder parameter curves with different system sizes, which, similar to Fig.~\ref{fig:replica}, signals the presence of the measurement-induced boundary transition discussed in Sec.~\ref{sec:nonlinear}.

\section{$\mathbb{Z}_2 \times \mathbb{Z}_2$ SPT phase with interaction reflection symmetry and its relation with a spin-1 chain} \label{app:sptwithreflection}

In Sec.~\ref{sec:symcluster}, we have shown that the cluster state SPT can be obtained by breaking the interchain reflection symmetry of the gapless parent state, i.e., taking $\alpha\neq \pi/4$ in Eq.~\eqref{eq:SymmetricCluster}. 
A natural question to ask is whether it is possible to obtain an inter-chain-reflection symmetric $\mathbb{Z}_2 \times \mathbb{Z}_2$ SPT by gapping out the gapless parent state. We will show that the answer is yes.

We begin with the $\mathcal{XY}$ representation of the gapless parent of the cluster state with central charge $c=2$,
\begin{equation}
    \begin{aligned}
        H =- \sum_j\big[ & \mathcal{X}_j \mathcal{X}_{j+1}+\mathcal{Y}_j \mathcal{Y}_{j+1} +\tilde{\mathcal{X}}_j \tilde{\mathcal{X}}_{j+1}+\tilde{\mathcal{Y}}_j \tilde{\mathcal{Y}}_{j+1}\big].\\
    \end{aligned}
\end{equation}
For simplicity, we only focus on the $K=1$ case. The interchain-reflection symmetry transforms as follows,
\begin{equation}
    \mathcal{X}_j \to \mathcal{Y}_j G_2, \quad \mathcal{\tilde{X}}_j \to \mathcal{\tilde{Y}}_j G_2
\end{equation}
where the $\mathbb{Z}_2$ symmetry generators $G_{1,2}$ are defined in Eq.~\ref{Z2generators}. The ground state $\ket{\psi_\Delta}$ satisfies $G_{1,2} = 1$ and the reflection transformation is just $\mathcal{X}_j \leftrightarrow \mathcal{Y}_j$ and $\mathcal{\tilde X}_j \leftrightarrow \mathcal{\tilde Y}_j$. 

Now consider adding interchain coupling terms to the Hamiltonian,
\begin{equation}
\begin{aligned}
    H_{\perp} &= -\lambda \sum_j(\mathcal{X}_j \mathcal{\tilde X}_j + \mathcal{Y}_j \mathcal{\tilde Y}_j + \mathcal{Z}_j \mathcal{\tilde Z}_j), \quad (\lambda > 0)\\
    &= -\lambda \sum_j(X_{j,1} + X_{j,2} - X_{j,1}X_{j,2}) \quad \text{(in the $ZXZ$ basis)}.\\
\end{aligned}
\end{equation}
In terms of boson fields, $H + H_{\perp}$ can be decomposed using symmetric ($+$) and antisymmetric ($-$) fields $\phi_\pm = \frac{1}{\sqrt{2}} (\phi \pm \tilde \phi)$ and $\theta_\pm = \frac{1}{\sqrt{2}} (\theta \pm \tilde \theta)$~\cite{Giamarchi2003}, 
\begin{equation}
\begin{aligned}
    H + H_{\perp} &= H_+ + H_- \\
    H_+ &= \int_x \frac{1}{2\pi} \left[ K_+ (\nabla \theta_+)^2 + \frac{1}{K_+} (\nabla \phi_+)^2 \right] +g_+ \int_x \cos(\sqrt{8} \phi_+)\\
    H_- &= \int_x \frac{1}{2\pi} \left[ K_- (\nabla \theta_-)^2 + \frac{1}{K_-} (\nabla \phi_-)^2 \right] +g^{(1)}_{-} \int_x \cos(\sqrt{8} \phi_-) + g^{(2)}_{-} \int_x \cos(\sqrt{2} \theta_-)\\
\end{aligned}
\end{equation}
where $g_+, g^{(1,2)}_{-} \propto \lambda$, and
\begin{equation}
    K_\pm = K \left( 1 \mp \frac{K \lambda a}{2 \pi u} \right)
\end{equation}
with $a$ being the lattice spacing and $u$ the Fermi velocity.
The antisymmetric part is always gapped once $H_{\perp}$ is added. For the symmetric part, $\cos(\sqrt{8}\phi_+)$ is relevant when $K_+<1$. At the $\mathcal{XY}$ point ($K=1$), we have $K_+>1$ when $\lambda>0$. Therefore the symmetric part remains gapless and one should expect an effective central charge $c=1$.

\begin{figure}[h!]
    \centering
    \includegraphics[width=\linewidth]{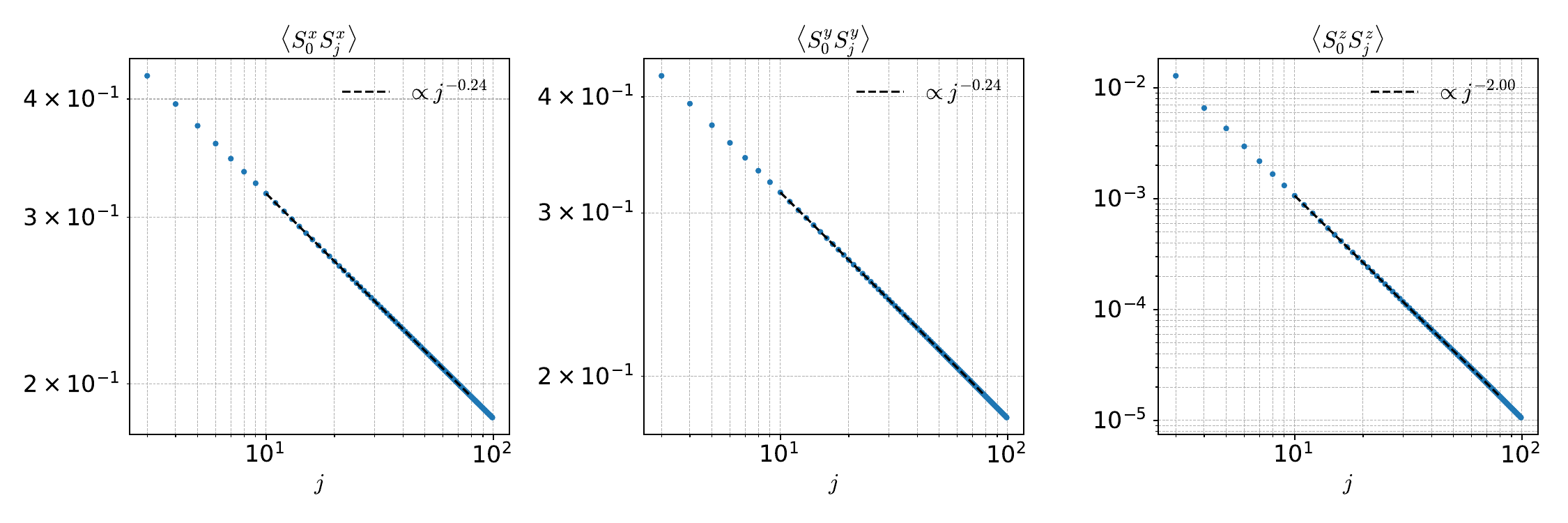}
    \caption{The correlation functions of the ground state of a spin-1 $xy$ chain. The power-law exponents are obtained by fitting the data in log-log scale (see the black dashed lines). The wavefunction is obtained by iDMRG with bond dimension $\chi=450$.}
    \label{fig:Spin1XYModel_corr}
\end{figure}

Another useful way to understand the effect of $H_\perp$ is to consider the $\lambda\to\infty$ limit, where $H_{\perp}$ projects the two spin-1/2's on the same rung, $(\mathcal{X,Y,Z})_j$ and $(\mathcal{\tilde X,\tilde Y,\tilde Z})_j$, onto the triplet subspace with spin-1 operators ${\bf S}_j = \frac{1}{2}[(\mathcal{X}_j + \tilde{\mathcal{X}}_j){\bf \hat{x}} + (\mathcal{Y}_j + \tilde{\mathcal{Y}}_j){\bf \hat{y}} + (\mathcal{Z}_j + \tilde{\mathcal{Z}}_j){\bf \hat{z}})]$.
Within the triplet subspace, the Hamiltonian $H + H_{\perp}$ acts like a spin-1 $xy$ chain,
\begin{equation} \label{eq:spin1xy}
    H_1 = - \sum_j (S^x_j S^x_{j+1} + S^y_j S^y_{j+1}),
\end{equation}
which is also gapless but with central charge $c=1$. The interchain reflection transforms as $x \leftrightarrow y$. The correlations in the ground state of $H_1$ are (see Fig.~\ref{fig:Spin1XYModel_corr} for DMRG simulations),
\begin{equation}
    \braket{S^x_j S^x_k}= \braket{S^y_j S^y_k} \sim \frac{1}{|j-k|^{\frac{1}{4}}}, ~~ \braket{S^z_j S^z_k}\sim \frac{1}{|j-k|^2}.
\end{equation}

The interchain-reflection symmectric $\mathbb{Z}_2\times\mathbb{Z}_2$ SPT phase of the spin-1/2 ladder corresponds to the ferromagnetically ordered phase in the effective spin-1 chain with $\braket{S^x_j S^x_k}\sim \braket{S^y_{j} S^y_{k}}\sim const.$, which can be obtained by further adding a next-nearest-neighbor perturbation to the spin-1 $xy$ chain, e.g.,
\begin{equation}
    \begin{aligned}
       H' &= J' \sum_j \sum_{\mu=x,y} S^\mu_{j-1}[2(S^\mu_j)^2-1]S^\mu_{j+1} 
    \end{aligned}
\end{equation}
with $J'>0$.
In the $\mathcal{XXZ}$ basis, the perturbation term can be written as
\begin{equation}
    H' = J' \sum_j (\mathcal{X}_j \mathcal{X}_{j+1} \mathcal{\tilde X}_{j+1} \mathcal{\tilde X}_{j+2} + \mathcal{Y}_j \mathcal{Y}_{j+1} \mathcal{\tilde Y}_{j+1} \mathcal{\tilde Y}_{j+2}),
\end{equation}
and in the original $ZXZ$ basis, $H'$ represents next-nearest-neighbor interactions,
\begin{equation}
    H' = J' \sum_j (X_{j,1} Z_{j,2} X_{j+2,1} Z_{j+2,2} + Z_{j,1} X_{j,2} Z_{j+2,1} X_{j+2,2}).
\end{equation}
In Ref.~\onlinecite{Liu2023}, it is shown that such next-nearest-neighbor couplings yield the interchain-reflection symmetric $\mathbb{Z}_2 \times \mathbb{Z}_2$ SPT phase.

One can also study the measurement problem in the spin-1 represeatation. Numerically, DMRG simulation of the spin-1 chain requires less computational resource than that of the gapless parent of the cluster state.
Within the triplet subspace, $X_{j,1} =\mathcal{X}_j \mathcal{\tilde X}_j = 2(S^x_j)^2 - 1$, the uniform projective measurement with $X_{j,1}=+1$ is equivalent to applying $\prod_j \frac{1+X_{j,1}}{2} = \prod_j (S^x_j)^2$ to the ground state.

\begin{figure}[h!]
    \centering
    \includegraphics[width=\linewidth]{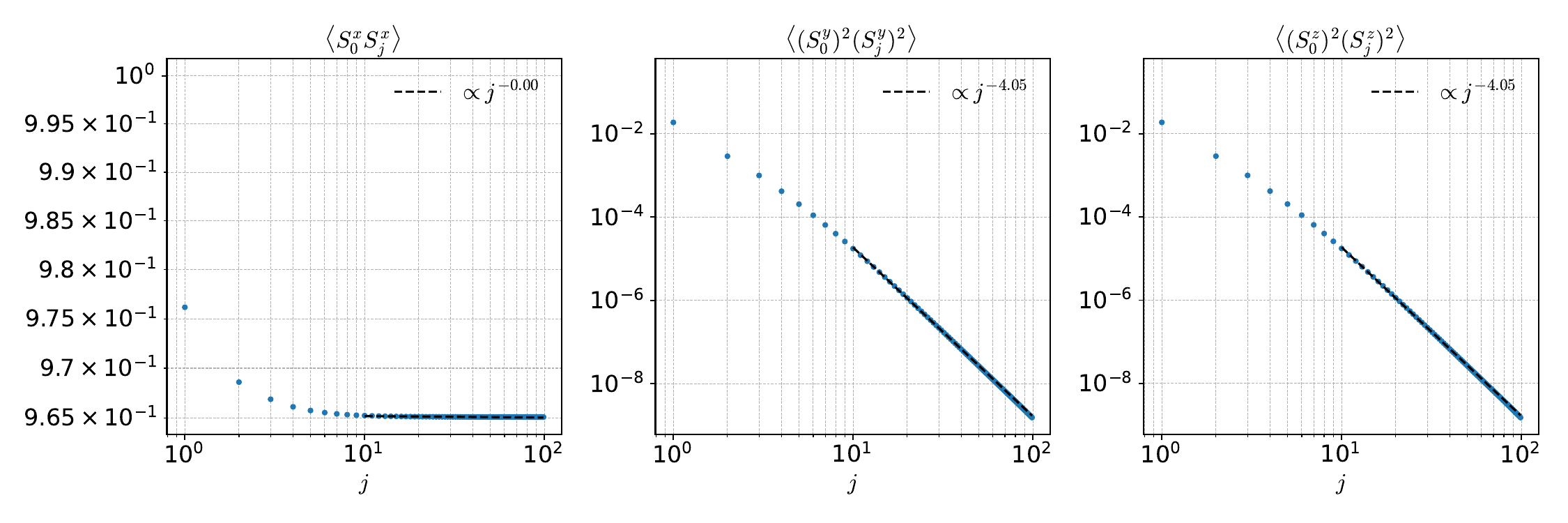}
    \caption{Select correlation functions of the spin-1 $xy$ chain after applying $\prod_j (S^x_j)^2$. The power-law exponents are obtained by fitting the data in log-log scale (see the black dashed lines). The wavefunction is obtained by infinite DMRG with bond dimension $\chi=450$.}
    \label{fig:Spin1XYModel_corr_postmeas}
\end{figure}

The $Z_{j,2} Z_{k,2}$ correlation within the triplet subspace is equivalent to the string operator,
\begin{equation} \label{eq:spin1zz}
\begin{aligned}
    \Braket{Z_{j,2} Z_{k,2}} & \sim (-1)^{k-j-1} \Braket{S^x_j \left[\prod_{l=j+1}^{k} \exp(i \pi S^x_l) \right] S^x_k} \\
\end{aligned}
\end{equation}
which is short-ranged in the ferromagnetic phase of the spin-1 chain and decays algebraically at the gapless point.
Once the uniform outcome ${\boldsymbol{s}}=\{X_{j,1}=+1\}$ is post-selected, one can simplify Eq.~\eqref{eq:spin1zz} by $-\exp(i \pi S^x_l) =  2(S^x_l)^2 - 1 = 1$ and obtain,
\begin{equation}
    \Braket{Z_{j,2} Z_{k,2}}_{\rm uni} \sim \Braket{S^x_j  S^x_k}_{\rm uni},
\end{equation}
which is analogous to Eq.~\eqref{eq:ZZ2}.
Also, the counterpart of Eq.~\eqref{eq:2XYcorrelations_1} is
\begin{equation}
    \Braket{X_{j,2} X_{k,2}}_{\rm uni} \sim \Braket{(S^y_j)^2 (S^y_k)^2}_{\rm uni}.
\end{equation}
From DMRG calculations (see Fig.~\ref{fig:Spin1XYModel_corr_postmeas}), we see that
\begin{equation}
    \Braket{Z_{j,2} Z_{k,2}}_{\rm uni} \sim const, \quad  \Braket{X_{j,2} X_{k,2}}_{\rm uni} \sim |j-k|^{-4},
\end{equation}
which are consistent with Eq.~\eqref{eq:ZZ2} and Eq.~\eqref{eq:2XYcorrelations_1} at $K=1$.

\section{Boundary fixed points and RG flows}\label{app:BCFT}

In BCFT, the conformal invariant boundary conditions (BCs) and the corresponding boundary fixed points are characterized by Cardy states \cite{Cardy1989}. The operator content and conformal spectrum can be obtained by the fusion rules of the primary operators corresponding to the Cardy states. As an example, in the 2D Ising CFT there are three primary fields, $I$ (identity), $\varepsilon$ (energy), and $\sigma$ (spin). One can associate the free BC (denoted by ``0'') with a Cardy state corresponding to the spin field, $\ket{0}\equiv\ket{\tilde \sigma}$. The operator content can be obtained by the fusion rule, $\sigma \times \sigma = I + \varepsilon$. There are also two fixed BCs obtained by applying an external boundary field $h_b$ that breaks the $\mathbb{Z}_2$ symmetry. We denote them by ``$+$'' and ``$-$'', depending on the sign of $h_b$. The Cardy states of these two BCs are $\ket{+}\equiv \ket{\tilde I}$ and $\ket{-}\equiv\ket{\tilde \varepsilon}$, and in their operator contents only the identity $I$ appears, because both $I \times I$ and $\varepsilon \times \varepsilon$ fuse to $I$. Therefore, the conformal spectrum of free BC is $\{0, \frac{1}{2}, \frac{3}{2}, 2, \frac{5}{2}, 3, \frac{7}{2}, 4, 4, \frac{9}{2}, \frac{9}{2}, \dots\}$, and those of fixed BCs are $\{0, 2, 3, 4, 4, 5, 5, \dots\}$. 

The scaling dimension of local boundary operators, such as the spin in the Ising model example, can be extracted from the conformal spectrum and generally differs from that of the same operator when located far from the boundary. In the measurement problem discussed in the main text, the scaling dimension of local operators in the post-measurement wavefunction should also take value in the conformal spectrum, at least when the relevant measurement operator flows to a specific BCFT. For example, let $\ket{\psi_0}$ be the ground state of a critical transverse-field Ising chain $H = -\sum_j (Z_j Z_{j+1} + X_j)$. Then the post-measurement state $e^{\beta \sum_j X_j} \ket{\psi_0}$, $\beta \to \infty$, corresponds to the free BC and the post-measurement scaling dimension $[Z]_{\rm uni} = 1/2$ can be read off from the free BC conformal spectrum. Similarly, $e^{\beta \sum_j Z_j} \ket{\psi_0}$ corresponds to fixed BC and $[Z]_{\rm uni} = 2$ is also consistent with the first gap in the fixed BC conformal spectrum. 
After measurements of relevant operators, an alternative way to find these scaling dimensions is by exploiting the power-law decay of the strange correlator $a_{jk} - a_{j} a_{k}$ defined in Eq.~\eqref{eq:ajk}.

\begin{table}[ht]
    \centering
        \caption{The conformally invariant boundary conditions, Cardy states, and corresponding operator contents of a 2D tricritical Ising CFT. The boundary conditions are labeled by symbols in the parentheses: $0$, $+$, $-$, etc.}
    \setlength{\extrarowheight}{2pt}
    \begin{tabular}{c|c|c|c}
    \hline
        Boundary condition & Cardy state & Operator content & Conformal spectrum \\
        \hline
        \hline
        Free ($0$) & $\ket{0} \equiv \ket{\tilde{\sigma'}}$ & $I + \varepsilon''$ & $\{0, \frac{3}{2}, 2, \frac{5}{2}, 3, \frac{7}{2}, \frac{7}{2}, \dots\}$\\
        \hline
        Fixed ($+$) & $\ket{+} \equiv \ket{\tilde{I}}$ & $I$ & $\{0, 2, 3, 4, 4, 5, 5, \dots\}$\\
        \hline
        Fixed ($-$) & $\ket{-} \equiv \ket{\tilde{\varepsilon''}}$ & $I$ & $\{0, 2, 3, 4, 4, 5, 5, \dots\}$\\
        \hline
        Partially polarized ($0+$) & $\ket{0+} \equiv \ket{\tilde{\varepsilon}}$ & $I + \varepsilon'$ & $\{0, \frac{3}{5}, \frac{8}{5}, 2, \frac{13}{5}, \frac{13}{5}, 3,\dots\}$\\
        \hline
        Partially polarized ($0-$) & $\ket{0-} \equiv \ket{\tilde{\varepsilon'}}$ & $I + \varepsilon'$ & $\{0, \frac{3}{5}, \frac{8}{5}, 2, \frac{13}{5}, \frac{13}{5}, 3,\dots\}$\\
        \hline
        degenerate ($\rm d$) & $\ket{\rm d} \equiv \ket{\tilde{\sigma}}$& $I + \varepsilon + \varepsilon' + \varepsilon''$ & $\{0, \frac{1}{10}, \frac{3}{5}, \frac{11}{10}, \frac{3}{2}, \frac{8}{5}, 2,\dots\}$\\
        \hline
        Fixed ($+\&-$) & $\ket{+\&-} \equiv \ket{+} + \ket{-}$ & $2(I + \varepsilon'')$ & $\{0, 0, \frac{3}{2}, \frac{3}{2}, 2, 2, \frac{5}{2}, \frac{5}{2},\dots\}$\\
        \hline
    \end{tabular}
    \label{tab:TCI_BC}
\end{table}

The boundary fixed points and Cardy states of the 2D tricritical Ising CFT are more complicated. As we have already reviewed in the main text, there are six primary fields in this case. Four of them preserve the $\mathbb{Z}_2$ spin-inversion symmetry, $I$, $\varepsilon$, $\varepsilon$, $\varepsilon'$ and $\varepsilon''$, while two of them break it, $\sigma$ and $\sigma'$. We list the Cardy states and corresponding BCs in Table.~\ref{tab:TCI_BC}.
The RG flow between the boundary fixed points has been studied in Refs.~\cite{Chim1996,Affleck2000}, and is summarized in Fig.~\ref{fig:TCI}(a).

\begin{figure}[h]
    \centering
    \includegraphics[width=\linewidth]{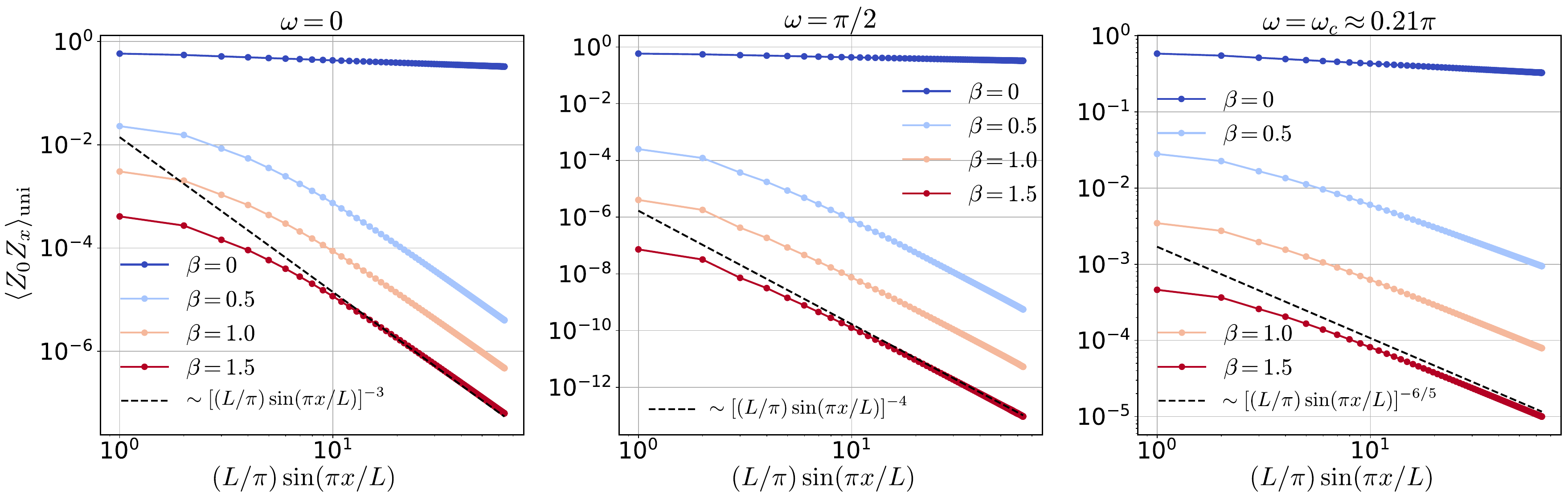}
    \caption{Post-measurement correlation functions of $e^{\beta\sum_j [\cos(\omega)X_j + \sin(\omega)Z_j]} \ket{\psi_0}$, where $\ket{\psi_0}$ is the ground state of Eq.~\eqref{eq:TCI} at the tricritical Ising point. Data obtained using DMRG with system size $L=200$, bond dimension $\chi=1000$ and periodic boundary conditions. In this figure, $\omega = 0$ and $\omega = \pi/2$ correspond to free ($0$) and fixed ($+$) BC respectively, and $\omega = \omega_c \approx 0.21\pi$ corresponds to the intermediate fixed point --- partially polarized ($0+$) BC.}
    \label{fig:TCI_fixedpoints}
\end{figure}

In the measurement problem, if $\ket{\psi_0}$ is the ground state of Eq.~\eqref{eq:TCI}, the weak measurement $e^{\beta \sum_j X_j} \ket{\psi_0}$, $\beta>0$ flows to the free BC, and the strange correlator decays as $a_{jk} - a_j a_k \sim |j-k|^{-3}$, whose exponent is just twice the first gap of the conformal spectrum we can read from the first line of Table \ref{tab:TCI_BC}. This result justifies the power-law decay of Eq.~\eqref{eq:TCI_measX}. The measurement $e^{\beta \sum_j Z_j}$, $\beta>0$, flows to the fixed $+$ BC, and one immediately gets Eq.~\eqref{eq:TCI_measZ} by looking at the second line of Table \ref{tab:TCI_BC}.
When the measurement angle is tilted between the $X$ and $Z$ axes, it will go across a transition corresponding to the partially polarized $0+$ BC, where the strange correlator decays as $a_{jk} - a_j a_k \sim |j-k|^{-6/5}$. We numerically analyze these results in Fig.~\ref{fig:TCI_fixedpoints}: in the left panel ($\omega=0$), we verify that the order parameter decays as $x^{-3}$, as we would expect from free BCs, in the middle panel ($\omega=\pi/2$) we show the decay of the order parameter as $x^{-4}$, consistently with the flow to fixed BCs. Finally, the right panel ($\omega=\omega_c$) corresponds to the measurement-induced boundary transition described by the intermediate fixed point $0+$, in agreement with the order parameter decay $x^{-6/5}$. As discussed in Appendix~\ref{app:pert}, the post-measurement wavefunction is expected to obey an area law entanglement since the exponent $6/5$ exceeds $1$.

\begin{figure}[h]
    \centering
    \includegraphics[width=\linewidth]{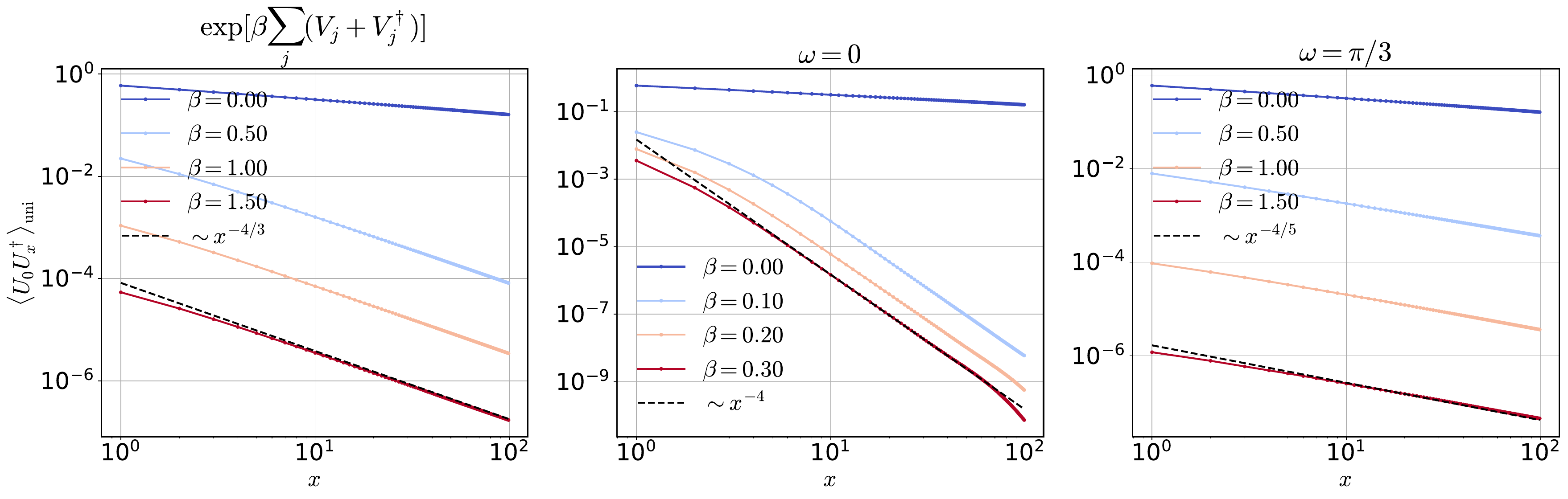}
    \caption{Correlation functions of three-state Potts model after measurement of $e^{\beta\sum_j [V_j + V^\dagger_j]}$ (left panel) and measurement in Eq.~\ref{eq:measPotts} with $\omega = 0$ (middle panel) or $\omega = \pi/3$ (right panel). Data obtained using iDMRG with bond dimension $\chi=500$. }
    \label{fig:Potts_corr}
\end{figure}

We conclude this Appendix by observing that also the three-state Potts model has six primary fields. In the main text, we have mainly focused on $\sigma$ and $\varepsilon$, whose scaling dimensions are respectively $2/15$ and $4/5$. We have observed that the measurement $e^{\beta\sum_j (V_j+V^{\dagger}_j)}$, with $\beta>0$, flows to the free boundary condition, and the power-law exponent of $\braket{U_0U^{\dagger}_j}_{\rm uni}$ can be read off from the boundary conformal spectrum, and it is equal to $4/3$, as we also show in the left panel of Fig.~\ref{fig:Potts_corr}. Then, we have shown how the measurement operator in Eq. \eqref{eq:measPotts} is responsible for a flow to fixed boundary conditions, $A,B$ or $C$, while by tuning $\omega=\pi/3,\pi,5/\pi$, the measurement flows to mixed boundary conditions, $CA,BC$ or $AB$. The first energy gap of the BCFT with fixed boundary conditions suggests that $[U+U^{\dagger}]_{\rm uni}=2$, while for mixed boundary conditions $[U+U^{\dagger}]_{\rm uni}=2/5$. These predictions are supported by the middle and right panels of Fig.~\ref{fig:Potts_corr}.

\end{document}